\documentclass[journal,twocolumn]{IEEEtran} \twocolumn 
\usepackage{epsfig,algorithm,amsmath,amssymb,color,subfigure,url}
\usepackage{epstopdf}
\usepackage{caption}
\usepackage{tikz}
\usetikzlibrary{decorations.pathreplacing}
\usetikzlibrary{decorations.markings}
\usetikzlibrary{fadings}

\author{
\authorblockN{Jinchun Zhan and Namrata Vaswani \\
}
\authorblockA{
Dept. of Electrical and Computer Engineering \\
Iowa State University, Ames, IA 50011, USA \\
Email: jzhan@iastate.edu, namrata@iastate.edu%
}
\thanks{This work was supported by NSF grants ECCS-0725849 and CCF-0917015. Parts of this work were presented at Allerton 2010 \cite{stability_allerton} and at ISIT 2013 \cite{stab_jinchun}. The results in this paper are a significant generalization of both conference papers and include all the proofs. 
}
}

\title{Time Invariant Error Bounds for Modified-CS based Sparse Signal Sequence Recovery} 

\begin{document}
\def\x{0.5}
\def\y{1}
\def\a{2}
\def\ab{1.5}
\def\w{4}
\setlength{\arraycolsep}{0.03cm}
\newcommand{\xhat}{\hat{x}}
\newcommand{\xpred}{\hat{x}_{t|t-1}}
\newcommand{\Ppred}{P_{t|t-1}}
\newcommand{\ty}{\tilde{y}_t}
\newcommand{\tty}{\tilde{y}_{t,\text{res}}}
\newcommand{\tw}{\tilde{w}_t}
\newcommand{\ttw}{\tilde{w}_{t,f}}
\newcommand{\betahat}{\hat{\beta}}

\newcommand{\ypast}{y_{1:t-1}}
\newcommand{\sone}{S_{*}}
\newcommand{\sinf}{{S_{**}}}
\newcommand{\smax}{S_{\max}}
\newcommand{\smin}{S_{\min}}
\newcommand{\samax}{S_{a,\max}}
\newcommand{\Nhat}{{\hat{\cal N}}}

\newcommand{\sgn}{\text{sgn}}

\newcommand{\Dnum}{D_{num}}
\newcommand{\pss}{p^{**,i}}
\newcommand{\fr}{f_{r}^i}

\newcommand{\A}{{\cal A}}
\newcommand{\Z}{{\cal Z}}
\newcommand{\B}{{\cal B}}
\newcommand{\R}{{\cal R}}
\newcommand{\reg}{{\cal G}}
\newcommand{\const}{\mbox{const}}

\newcommand{\trace}{\mbox{tr}}

\newcommand{\hsim}{{\hspace{0.0cm} \sim  \hspace{0.0cm}}}
\newcommand{\he}{{\hspace{0.0cm} =  \hspace{0.0cm}}}

\newcommand{\vect}[2]{\left[\begin{array}{cccccc}
     #1 \\
     #2
   \end{array}
  \right]
  }

\newcommand{\matr}[2]{ \left[\begin{array}{cc}
     #1 \\
     #2
   \end{array}
  \right]
  }
\newcommand{\vc}[2]{\left[\begin{array}{c}
     #1 \\
     #2
   \end{array}
  \right]
  }

\newcommand{\gdot}{\dot{g}}
\newcommand{\Cdot}{\dot{C}}
\newcommand{\re}{\mathbb{R}}
\newcommand{\n}{{\cal N}}  
\newcommand{\N}{{\overrightarrow{\bf N}}}  
\newcommand{\chat}{\tilde{C}_t}
\newcommand{\chati}{\chat^i}

\newcommand{\cmin}{C^*_{min}}
\newcommand{\twi}{\tilde{w}_t^{(i)}}
\newcommand{\twj}{\tilde{w}_t^{(j)}}
\newcommand{\wi}{{w}_t^{(i)}}
\newcommand{\twio}{\tilde{w}_{t-1}^{(i)}}

\newcommand{\tWi}{\tilde{W}_n^{(m)}}
\newcommand{\tWj}{\tilde{W}_n^{(k)}}
\newcommand{\Wi}{{W}_n^{(m)}}
\newcommand{\tWio}{\tilde{W}_{n-1}^{(m)}}

\newcommand{\ds}{\displaystyle}

\newcommand{\SAR}{S$\!$A$\!$R }
\newcommand{\MAR}{MAR}
\newcommand{\MMRF}{MMRF}
\newcommand{\AR}{A$\!$R }
\newcommand{\GMRF}{G$\!$M$\!$R$\!$F }
\newcommand{\DTM}{D$\!$T$\!$M }
\newcommand{\MSE}{M$\!$S$\!$E }
\newcommand{\RCS}{R$\!$C$\!$S }
\newcommand{\uomega}{\underline{\omega}}
\newcommand{\lu}{\mu}
\newcommand{\g}{g}
\newcommand{\bft}{{\bf t}}
\newcommand{\refmap}{{\cal R}}
\newcommand{\totrefl}{{\cal E}}
\newcommand{\beq}{\begin{equation}}
\newcommand{\eeq}{\end{equation}}
\newcommand{\bdm}{\begin{displaymath}}
\newcommand{\edm}{\end{displaymath}}
\newcommand{\hatz}{\hat{z}}
\newcommand{\hatu}{\hat{u}}
\newcommand{\tilz}{\tilde{z}}
\newcommand{\tilu}{\tilde{u}}
\newcommand{\hhatz}{\hat{\hat{z}}}
\newcommand{\hhatu}{\hat{\hat{u}}}
\newcommand{\tilc}{\tilde{C}}
\newcommand{\hatc}{\hat{C}}
\newcommand{\tim}{n}

\newcommand{\ssp}{\renewcommand{\baselinestretch}{1.0}}
\newcommand{\defd}{\mbox{$\stackrel{\mbox{$\triangle$}}{=}$}}
\newcommand{\goes}{\rightarrow}
\newcommand{\tends}{\rightarrow}
\newcommand{\defn}{\triangleq} 
\newcommand{\se}{&=&}
\newcommand{\sdefn}{& \defn  &}
\newcommand{\sle}{& \le &}
\newcommand{\sge}{& \ge &}
\newcommand{\plusminus}{\stackrel{+}{-}}
\newcommand{\Ey}{E_{Y_{1:t}}}
\newcommand{\ey}{E_{Y_{1:t}}}

\newcommand{\equivto}{\mbox{~~~which is equivalent to~~~}}
\newcommand{\nonzero}{i:\pi^n(x^{(i)})>0}
\newcommand{\nonzeroc}{i:c(x^{(i)})>0}

\newcommand{\supn}{\sup_{\phi:\|\phi\|_\infty \le 1}}

\newtheorem{theorem}{Theorem}[section]
\newtheorem{lemma}[theorem]{Lemma}
\newtheorem{corollary}[theorem]{Corollary}
\newtheorem{definition}[theorem]{Definition}
\newtheorem{remark}[theorem]{Remark}
\newtheorem{example}[theorem]{Example}
\newtheorem{ass}[theorem]{Assumption}
\newtheorem{proposition}[theorem]{Proposition}

\newtheorem{fact}[theorem]{Fact}
\newtheorem{heuristic}[theorem]{Heuristic}
\newcommand{\eps}{\epsilon}
\newcommand{\bd}{\begin{definition}}
\newcommand{\ed}{\end{definition}}
\newcommand{\udq}{\underline{D_Q}}
\newcommand{\td}{\tilde{D}}
\newcommand{\epsinv}{\epsilon_{inv}}
\newcommand{\al}{\mathcal{A}}

\newcommand{\bfx} {\bf X}
\newcommand{\bfy} {\bf Y}
\newcommand{\bfz} {\bf Z}
\newcommand{\ddas}{\mbox{${d_1}^2({\bf X})$}}
\newcommand{\ddbs}{\mbox{${d_2}^2({\bfx})$}}
\newcommand{\dda}{\mbox{$d_1(\bfx)$}}
\newcommand{\ddb}{\mbox{$d_2(\bfx)$}}
\newcommand{\xinc}{{\bfx} \in \mbox{$C_1$}}
\newcommand{\eqa}{\stackrel{(a)}{=}}
\newcommand{\eqb}{\stackrel{(b)}{=}}
\newcommand{\eqe}{\stackrel{(e)}{=}}
\newcommand{\leqc}{\stackrel{(c)}{\le}}
\newcommand{\leqd}{\stackrel{(d)}{\le}}

\newcommand{\leqa}{\stackrel{(a)}{\le}}
\newcommand{\leqb}{\stackrel{(b)}{\le}}
\newcommand{\leqe}{\stackrel{(e)}{\le}}
\newcommand{\leqf}{\stackrel{(f)}{\le}}
\newcommand{\leqg}{\stackrel{(g)}{\le}}
\newcommand{\leqh}{\stackrel{(h)}{\le}}
\newcommand{\leqi}{\stackrel{(i)}{\le}}
\newcommand{\leqj}{\stackrel{(j)}{\le}}

\newcommand{\halpha}{\hat{\alpha}}
\newcommand{\hsigma}{\hat{\sigma}}
\newcommand{\slmax}{\sqrt{\lambda_{max}}}
\newcommand{\slmin}{\sqrt{\lambda_{min}}}
\newcommand{\lmax}{\lambda_{max}}
\newcommand{\lmin}{\lambda_{min}}

\newcommand{\da} {\frac{\alpha}{\sigma}}
\newcommand{\chka} {\frac{\check{\alpha}}{\check{\sigma}}}
\newcommand{\sumo}{\sum _{\underline{\omega} \in \Omega}}
\newcommand{\distance}{d\{(\hatz _x, \hatz _y),(\tilz _x, \tilz _y)\}}
\newcommand{\col}{{\rm col}}
\newcommand{\rcs}{\sigma_0}
\newcommand{\CalR}{{\cal R}}
\newcommand{\df}{{\delta p}}
\newcommand{\dq}{{\delta q}}
\newcommand{\dZ}{{\delta Z}}
\newcommand{\pprime}{{\prime\prime}}

\newcommand{\vn}{N}

\newcommand{\bv}{\begin{vugraph}}
\newcommand{\ev}{\end{vugraph}}
\newcommand{\bi}{\begin{itemize}}
\newcommand{\ei}{\end{itemize}}
\newcommand{\ben}{\begin{enumerate}}
\newcommand{\een}{\end{enumerate}}
\newcommand{\be}{\protect\[}
\newcommand{\ee}{\protect\]}
\newcommand{\bean}{\begin{eqnarray*} }
\newcommand{\eean}{\end{eqnarray*} }
\newcommand{\bea}{\begin{eqnarray} }
\newcommand{\eea}{\end{eqnarray} }
\newcommand{\nn}{\nonumber}
\newcommand{\ba}{\begin{array} }
\newcommand{\ea}{\end{array} }
\newcommand{\ep}{\mbox{\boldmath $\epsilon$}}
\newcommand{\epp}{\mbox{\boldmath $\epsilon '$}}
\newcommand{\Lep}{\mbox{\LARGE $\epsilon_2$}}
\newcommand{\und}{\underline}
\newcommand{\pdif}[2]{\frac{\partial #1}{\partial #2}}
\newcommand{\odif}[2]{\frac{d #1}{d #2}}
\newcommand{\dt}[1]{\pdif{#1}{t}}
\newcommand{\urho}{\underline{\rho}}

\newcommand{\spc}{{\cal S}}
\newcommand{\tspc}{{\cal TS}}

\newcommand{\uv}{\underline{v}}
\newcommand{\us}{\underline{s}}
\newcommand{\uc}{\underline{c}}
\newcommand{\utheta}{\underline{\theta}^*}
\newcommand{\ualpha}{\underline{\alpha^*}}

\newcommand{\uxy}{\underline{x}^*}
\newcommand{\uxyj}{[x^{*}_j,y^{*}_j]}
\newcommand{\arcl}[1]{arclen(#1)}
\newcommand{\one}{{\mathbf{1}}}

\newcommand{\uxyjt}{\uxy_{j,t}}
\newcommand{\E}{\mathbb{E}}

\newcommand{\rhomat}{\left[\begin{array}{c}
                        \rho_3 \ \rho_4 \\
                        \rho_5 \ \rho_6
                        \end{array}
                   \right]}
\newcommand{\deltat}{\tau} 
\newcommand{\deltatt}{\Delta t_1}
\newcommand{\ceil}[1]{\ulcorner #1 \urcorner}

\newcommand{\xxi}{x^{(i)}}
\newcommand{\txi}{\tilde{x}^{(i)}}
\newcommand{\txj}{\tilde{x}^{(j)}}

\newcommand{\mi}[1]{{#1}^{(m,i)}}

\setlength{\arraycolsep}{0.05cm}
\newcommand{\rest}{{T_\text{rest}}}
\newcommand{\zetahat}{\hat{\zeta}}
\newcommand{\tDelta}{{\tilde{\Delta}}}
\newcommand{\tDeltae}{{\tilde{\Delta}_e}}
\newcommand{\tT}{{\tilde{\cal T}}}
\newcommand{\add}{\text{add}}
\newcommand{\rem}{{\cal R}}
\newtheorem{sigmodel}{Signal Change Assumptions}

\newcommand{\modcs}{{\text{modcs}}}
\newcommand{\thr}{{\text{thr}}}
\newcommand{\delthr}{{\text{del-thr}}}
\newcommand{\delbound}{{b}}
\newcommand{\err}{{\text{err}}}
\newcommand{\Q}{{\cal Q}}

\newcommand{\del}{{\text{del}}}
\newcommand{\CSres}{{\text{CSres}}}
\newcommand{\diff}{{\text{diff}}}
\newcommand{\Section}[1]{ \vspace{-0.13in}  \section{#1} \vspace{-0.12in} } 
\newcommand{\Subsection}[1]{  \vspace{-0.12in} \subsection{#1}  \vspace{-0.08in} } 
\newcommand{\Subsubsection}[1]{   \subsubsection{#1} } 

\newcommand{\Aset}{{\cal A}}
\newcommand{\Bset}{{\cal B}}
\newcommand{\Rset}{{\cal R}}
\newcommand{\Iset}{{\cal I}}
\newcommand{\Dset}{{\cal D}}
\newcommand{\Sset}{{\cal S}}
\newcommand{\Lset}{{\cal L}}
\newcommand{\Tset}{{\cal T}}
\newcommand{\Nset}{{\cal N}}
\newcommand{\SDset}{{\cal SD}}
\newcommand{\Inc}{\text{Inc}}
\newcommand{\Dec}{\text{Dec}}
\newcommand{\Con}{\text{Con}}
\newcommand{\sm}{e}  

\newcommand{\Ahat}{\hat{\cal A}}
\newcommand{\Rhat}{\hat{\cal R}}
\newcommand{\Shat}{\hat{\cal S}}
\newcommand{\Mhat}{\hat{\cal M}}

\date{}
\maketitle

\begin{abstract}
In this work, we obtain performance guarantees for modified-CS and for its improved version, modified-CS-Add-LS-Del, for recursive reconstruction of a time sequence of sparse signals from a reduced set of noisy measurements available at each time. Under mild assumptions, we show that the support recovery error of both algorithms is bounded by a time-invariant and small value at all times. The same is also true for the reconstruction error.  Under a slow support change assumption, (i) the support recovery error bound is small compared to the support size; and (ii) our results hold under weaker assumptions on the number of measurements than what $\ell_1$ minimization for noisy data needs.
We first give a general result that only assumes a bound on support size, number of support changes and number of small magnitude nonzero entries at each time.
Later, we specialize the main idea of these results for two sets of signal change assumptions that model the class of problems in which a new element that is added to the support either gets added at a large initial magnitude or its magnitude slowly increases to a large enough value within a finite delay. Simulation experiments are shown to back up our claims.


\end{abstract}

\section{Introduction}

The static sparse reconstruction problem has been studied for a while \cite{mallat,bpdn,bdrao_focuss,bresler1996new,bresler1996spectrum}. The papers on compressive sensing (CS) from 2005 \cite{donoho,candes,donoho_large,decodinglp,candesnoise,candes_rip} (and many other more recent works) provide the missing theoretical guarantees -- conditions for exact recovery and error bounds when exact recovery is not possible. In more recent works, the problem of recursively recovering a time sequence of sparse signals, with slowly changing sparsity patterns has also been studied \cite{kfcsicip,just_lscs,modcsjp,ibm,schniter_track,zhang_rao,romberg_ciss11,dyn_supp_track}. By ``recursive" reconstruction, we mean that we want to use only the current measurements' vector and the previous reconstructed signal to recover the current signal. This problem occurs in many applications such as real-time dynamic magnetic resonance imaging (MRI); single-pixel camera based real-time video imaging; recursively separating the region of the brain that is activated in response to a stimulus from brain functional MRI (fMRI) sequences \cite{rrpcp_allerton11} and recursively extracting sparse foregrounds (e.g. moving objects) from slow-changing (low-dimensional) backgrounds in video sequences \cite{rrpcp_allerton}. For other potential applications, see \cite{rice_cs,nuit_blanche}.

An important assumption introduced and empirically verified in \cite{kfcsicip,just_lscs} is that for many natural signal/image sequences, the sparsity pattern (support set of its projection into the sparsity basis) changes slowly over time. In \cite{modcsjp}, the authors exploited this fact to reformulate the above problem as one of sparse recovery with partially known support and introduced a solution approach called modified-CS. Given the partial support knowledge $\Tset$, modified-CS tries to find a signal that is sparsest outside of $\Tset$ among all signals that satisfy the data constraint. Exact recovery conditions were obtained for modified-CS and it was argued that these are weaker than those for simple $\ell_1$ minimization (basis pursuit) under the slow support change assumption. Related ideas for support recovery with prior knowledge about the support entries, that appeared in parallel, include \cite{hassibi}, \cite{camsap07}. All of \cite{modcsjp}, \cite{hassibi} and \cite{camsap07} studied the noise-free measurements' case. Later work includes \cite{friedlander, on_partial_sparse}.
Error bounds for modified-CS for noisy measurements were obtained in \cite{jacques}, \cite{regmodbpdn}, \cite{stability_allerton}. When modified-CS is used for recursive reconstruction, these bounds tell us that the reconstruction error bound at the current time is proportional to the support recovery error  (misses and extras in the support estimate) from the previous time. Unless we impose extra conditions, this support error can keep increasing over time, in which case the bound is not useful. Thus, for recursive reconstruction, the important question is, under what conditions can we obtain time-invariant bounds on the support error (which will, in turn, imply time-invariant bounds on the reconstruction error)? In other words, when can we ensure ``stability" over time?
Notice that, even if we did nothing, i.e. we set $\xhat_t=0$, the support error will be bounded by the support size. If the support size is bounded, then this is a naive stability result too, but is not useful. Here, we look for results in which the support error bound is small compared to the support size. 

\begin{figure}
\centerline{
\begin{tabular}{cc}
\epsfig{file = 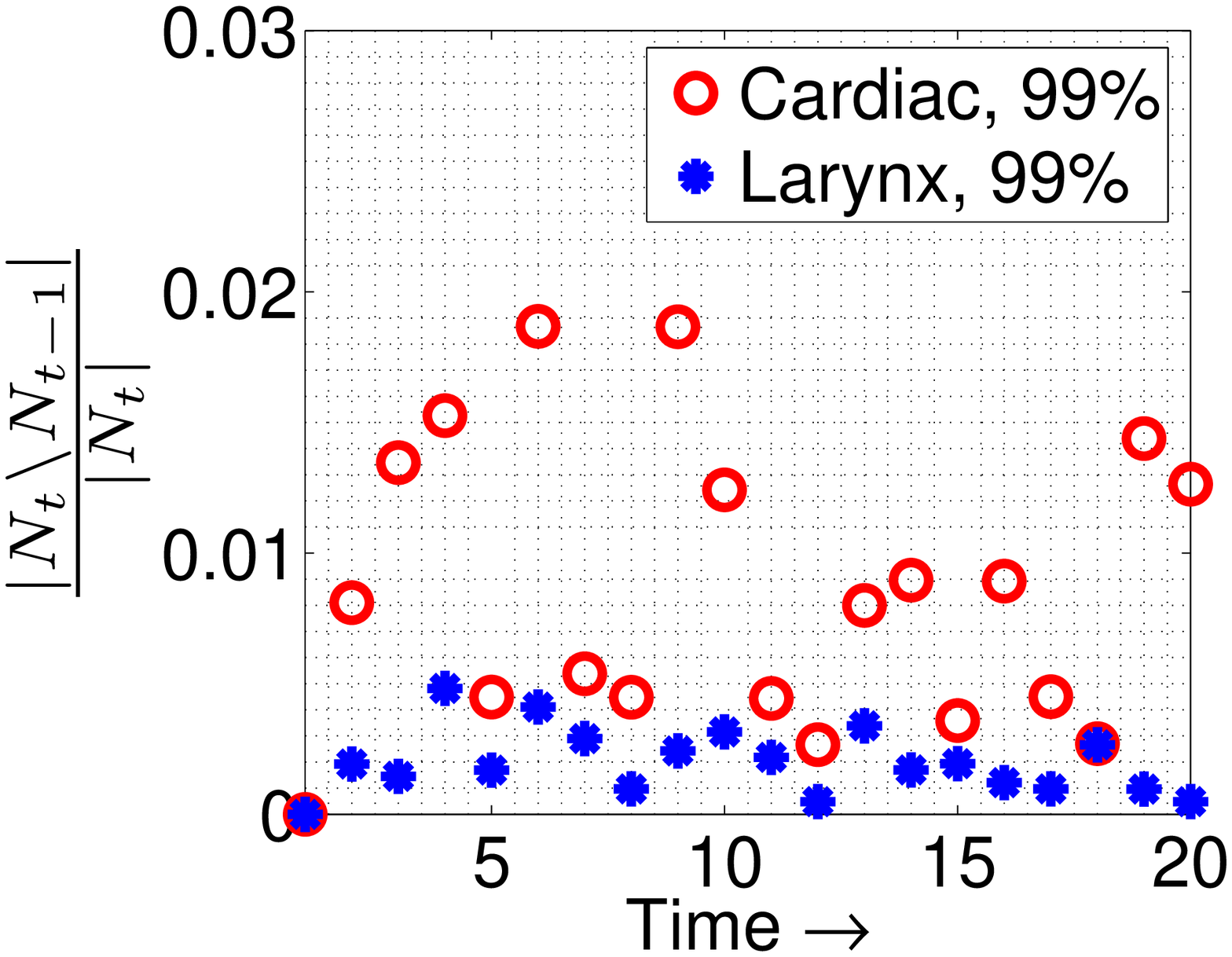,height=2.5cm, width = 4cm} & 
\epsfig{file = 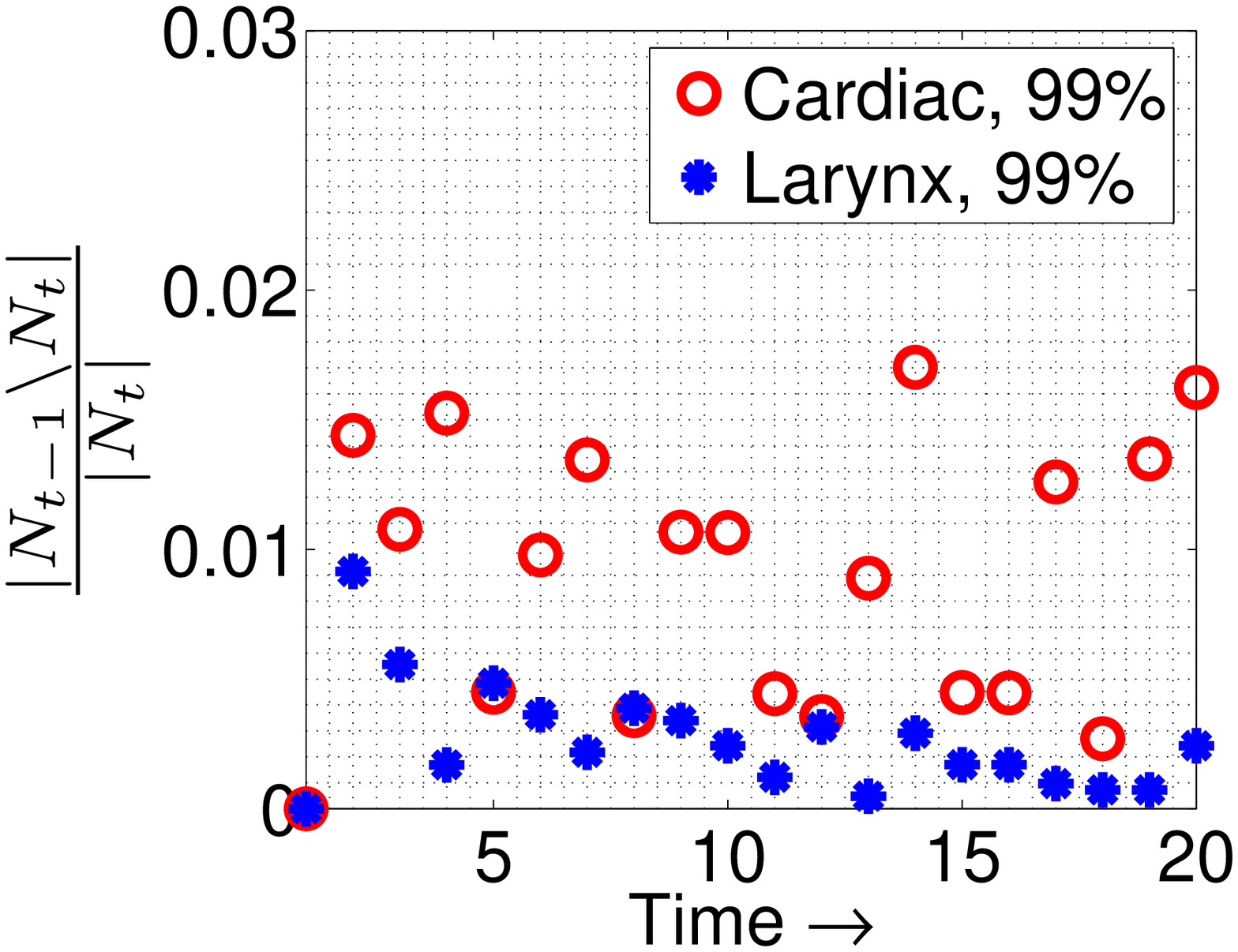,height=2.5cm, width = 4cm}
\end{tabular}
}
\caption{\small{Slow support change in medical image sequences.
The two-level Daubechies-4 2D discrete wavelet transform (DWT) served as the sparsity basis.
Since real image sequences are only approximately sparse, we use $N_t$ to denote the 99\%-energy support of the DWT of these sequences. The support size, $|N_t|$, was 6-7\% of the image size for both sequences. 
We plot the number of additions (left) and the number of removals (right) as a fraction of $|N_t|$. {\em Notice that all changes are less than 2\% of the support size.}
}}
\vspace{-0.1in}
\label{suppchange}
\end{figure}

Stability over time has not been studied much for recursive recovery of sparse signal sequences. To the best of our knowledge, it has only been addressed in \cite{just_lscs}, and in very recent work \cite{dyn_supp_track}. 
The result of  \cite{dyn_supp_track} is for exact dynamic support recovery in the noise-free case and it studies a different problem: the multiple measurement vector (MMV) version of the recursive recovery problem.
The result from \cite{just_lscs} for Least Squares CS-residual (LS-CS) stability) holds under mostly mild assumptions; its one limitation is that it assumes that support changes occur every $p$ frames. But from testing the slow support change assumption for real data (medical image sequences), it has been observed that support changes usually occur at {\em every} time, e.g. see Fig. \ref{suppchange}. {\em This important case is the focus of the current work.} We explain the differences of our results w.r.t. the LS-CS result in detail later in Sec \ref{compare_lscs}.


\subsection{Contributions}
In this work, we introduce modified-CS-add-LS-del which is a modified-CS based algorithm for recursive recovery with an improved support estimation step and we explain how to set its parameters in practice. The main contribution of this work is to obtain conditions for stability of modified-CS and modified-CS-add-LS-del  for recursive recovery of a time sequence of sparse signals.
%
%
Under mild assumptions, we show that the support recovery error and the reconstruction error of both algorithms is bounded by a time-invariant value at all times. The support error bound is proportional to the maximum allowed support change size. Under slow support change, this bound is small compared to the support size, making our result meaningful. Similar arguments can be made for the reconstruction error also. The assumptions we need are: weaker restricted isometry property (RIP) conditions \cite{decodinglp} on the measurement matrix than what $\ell_1$ minimization for noisy data (henceforth referred to as {\em noisy $\ell_1$}) needs; bounded cardinality of the support and support change; all but a small number of existing nonzero entries are above a threshold in magnitude; appropriately set support estimation thresholds; and a special start condition. Here and elsewhere in the paper noisy $\ell_1$ (or simple CS) refers to the solution of (\ref{simpcs}).

A second main contribution of this work is to show two examples of signal change assumptions under which the required conditions hold and prove stability results for these. The first case is a simple signal change model that helps to illustrate the key ideas and allows for easy comparison of the results. The second set of assumptions is realistic, but more complicated to state.  We use MRI image sequences to demonstrate that these assumptions are indeed valid for real data. The essential requirement in both cases is that, for any new element that is added to the support,  either its initial magnitude is large enough, or for the first few time instants, its magnitude increases at a large enough rate; and a similar assumption for magnitude decrease and removal from the support.

Let $S$ be the bound on the maximum support size and $S_a$ the bound on the maximum number of support additions or removals. All our results require $s$-RIP to hold with $s = S+k S_a$ where $k$ is a constant. On the other hand, noisy $\ell_1$ needs $s$-RIP for $s=2S$ \cite{candes_rip} which is a stronger requirement when $S_a \ll S$ (slow support change). 

\subsection{Notation}
\label{notn}
We let $[1,m]:=[1,2,\dots m]$. We let $\emptyset$ denote an empty set. We use $\Tset^c$ to denote the complement of a set $\Tset$ w.r.t. $[1,m]$, i.e. $\Tset^c := \{i \in [1,m]: i \notin \Tset \}$. We use $|\Tset|$ to denote the cardinality of $\Tset$. Also, $\emptyset$ denotes the empty set. The set operations $\cup$, $\cap$, $\setminus$ have their usual meanings (recall that ${\cal A} \setminus {\cal B} : = {\cal A} \cap {\cal B}^c$). If two sets ${\cal B}$, ${\cal C}$ are disjoint, we just write ${\cal D }\cup {\cal B} \setminus {\cal C}$ instead of writing $({\cal D} \cup {\cal B}) \setminus {\cal C}$.

For a vector, $v$, and a set, $\Tset$, $v_\Tset$ denotes the $|\Tset|$ length sub-vector containing the elements of $v$ corresponding to the indices in the set $\Tset$. $\| v \|_k$ denotes the $\ell_k$ norm of a vector $v$. {\em If just $\|v\|$ is used, it refers to $\|v\|_2$.} Similarly, for a matrix $M$, $\|M\|_k$ denotes its induced $k$-norm, while just $\|M\|$ refers to $\|M\|_2$. $M'$ denotes the transpose of $M$ and $M^\dag$ denotes the Moore-Penrose pseudo-inverse of $M$ (when $M$ is full column rank, $M^\dag:=(M'M)^{-1} M'$). Also, $M_\Tset$ denotes the sub-matrix obtained by extracting the columns of $M$ corresponding to indices in $\Tset$.

We refer to the left (right) hand side of an equation or inequality as LHS (RHS).

\subsection{Problem Definition}
We assume the following observation model: 
\bea
y_t = A_t x_t + w_t,  \ \|w_t\| \le \eps   
\label{obsmod}
\eea
where $x_t$ is an $m$ length sparse vector with support set ${\cal N}_t$, i.e. ${\cal N}_t:= \{i: (x_t)_i \neq 0\}$; $A_t$ is a $n_t \times m$ measurement matrix; $y_t$ is the $n_t$ length observation vector at time $t$ (with $n_t<m$); and $w_t$ is the observation noise. For $t>0$, we fix $n_t = n$.

Our goal is to recursively estimate $x_t$ using $y_1, \dots y_t$. By {\em recursively}, we mean, use only $y_t$ and the estimate from $t-1$, $\xhat_{t-1}$, to compute the estimate at $t$.

\begin{remark}[Why bounded noise]
All results for bounding $\ell_1$ minimization error in noise, and hence all results for bounding modified-CS error in noise, either assume a deterministic noise bound and then bound $\|\xhat-x\|$,  e.g., \cite{candes_rip}, \cite{cosamp}, \cite{jacques,modcsicassp10}, \cite{cai_tsp10}; or assume unbounded, e.g. Gaussian, noise and then bound $\|\xhat-x\|$ with ``large" probability, e.g. \cite{dantzig}, \cite[Sec IV]{tropp}, \cite[Section III-A]{just_lscs}, \cite{cai_tsp10}. The latter approach is not useful for recovering a time sequence of sparse signals because the error bound will hold for all times $0 \le t < \infty$ with probability zero.

One way to get a meaningful error stability result with unbounded, e.g. Gaussian noise, is to compute or bound the expected value of the error at each time, i.e. compute $\E[(\xhat_t -x_t)(\xhat_t -x_t)']$ or bound some norm of it. This is possible to do, for example, for a Kalman filter applied to a linear system model with additive Gaussian noise; and hence in that case, one can assume Gaussian noise and still get a time-invariant bound on the expected value of the error under mild assumptions. However, for $\ell_1$ minimization based methods, such as modified-CS, there is no easy way to compute or bound the expected value of the error. Moreover, even if one could do this for a given time, it would not tell us anything about the support recovery error (for the given noise sequence realization) and hence would not be useful for analyzing modified-CS. 

As a sidenote, we should point out that, in most applications, the noise is typically bounded (because of finite sensing power available). One often chooses to model the noise as Gaussian because it simplifies performance analysis.
%
\end{remark}

\subsection{Other Related Work}
``Recursive sparse reconstruction" also sometimes refers to homotopy methods, e.g. \cite{romberg_jp}, whose goal is to use the past reconstructions and homotopy to speed up the current optimization, but not to achieve accurate recovery from fewer measurements than what noisy $\ell_1$ needs.
The goals in the above works are quite different from ours.

Iterative support estimation approaches (using the recovered support from the first iteration for a second weighted $\ell_1$ step and doing this iteratively) have been studied in recent work \cite{reweighted_cs,trunc_bp,iterative_support,hassibi_twostep}. This is done for iteratively improving the recovery of a {\em single} signal.


\subsection{Paper Organization}
The rest of the paper is organized as follows. The algorithms -- modified-CS and  modified-CS-add-LS-del -- are introduced in Sec \ref{algos}. This section also includes definitions for certain quantities and sets used later in the paper.
In Sec \ref{nomodel}, we provide stability results for modified-CS and modified-CS-add-LS-del that do not assume anything about signal change over time except a bound on the number of small magnitude nonzero coefficients and a bound on maximum number of support additions or removals per unit time.
In Sec \ref{simpmodel}, we give a simple set of signal change assumptions and give stability results for both algorithms under these and other simple assumptions. In Sec \ref{genmodel}, we do the same for a realistic signal change model. The results are discussed in Sec \ref{simp_disc} and \ref{gen_disc} respectively. In Sec \ref{model_veri}, we demonstrate that the signal model assumptions of Sec \ref{genmodel} are indeed valid for medical imaging data. In Sec \ref{sims}, we explain how to set the algorithm parameters automatically for both modified-CS and  modified-CS-add-LS-del. In this section, we also give simulation experiments that back up some of our discussions from earlier sections. Conclusions and future work are given in Sec \ref{conc}.

\section{Modified-CS and Modified-CS-add-LS-del for Recursive Reconstruction} \label{algos}

\subsection{Modified-CS}
Modified-CS was first proposed in \cite{modcsjp} as a solution to the problem of sparse reconstruction with partial, and possibly erroneous, knowledge of the support. Denote this ``known" support by $\Tset$. Modified-CS tries to find a signal that is sparsest outside of the set $\Tset$ among all signals satisfying the data constraint. In the noisy case, it solves $\min_\beta  \|(\beta)_{\Tset^c}\|_1 \ \text{s.t.} \ \| y_t - A \beta \| \le \eps$. For recursively reconstructing a time sequence of sparse signals, we use the support estimate from the previous time, $\Nhat_{t-1}$, as the set $\Tset$. The simplest way to estimate the support is by thresholding the output of modified-CS. We summarize the complete algorithm in Algorithm \ref{modcsalgo}.

At the initial time, $t=0$, we let $\Tset$ be the empty set, $\emptyset$, i.e. we solve noisy $\ell_1$. Alternatively, as explained in \cite{modcsjp}, we can use prior knowledge of the initial signal's support as the set $\Tset$ at $t=0$, e.g. for wavelet sparse images with no (or a small) black background, the set of indices of the approximation coefficients can form the set $\Tset$. This prior knowledge is usually not as accurate. 

We explain how the parameter $\alpha$ can be set in practice in Sec \ref{param}.


\begin{algorithm}[h!]
\caption{{\bf \small Modified-CS}}
For $t \ge 0$, do
\ben
\item {\em Noisy $\ell_1$. } If $t = 0$, set $\Tset_t = \emptyset$ and compute $\xhat_{t,modcs}$ as the solution of
\bea
\min_\beta  \|(\beta)\|_1 \ \text{s.t.} \ \| y_0 - A_0 \beta \| \le \eps
\label{simpcs}
\eea

\item {\em Modified-CS. } If $t>0$, set $\Tset_t = \Nhat_{t-1}$ and compute $\xhat_{t,modcs}$ as the solution of
\label{step1_0}
\bea
\min_\beta  \|(\beta)_{\Tset_t^c}\|_1 \ \text{s.t.} \ \| y_t - A_t \beta \| \le \eps
\label{modcs}
\eea

\item {\em Estimate the Support. } Compute $\tT_t$ as
\label{supp_estim}
\bea
\tT_t=\{i \in [1,m] : |(\xhat_{t,modcs})_i| > \alpha \}
\label{supp_est}
\eea

\item Set $\Nhat_t = \tT_t$. Output $\hat{x}_{t,modcs}$. Feedback $\Nhat_t$.
\een
\label{modcsalgo}
\end{algorithm}

\subsection{Limitation: Biased solution}
Modified-CS uses single step thresholding for estimating the support $\Nhat_t$. The threshold, $\alpha$, needs to be large enough to ensure that all (or most) removed elements are correctly deleted and there are no (or very few) false detections. But this means that the new additions to the support set will either have to be added at a large value, or their magnitude will need to increase to a large value quickly enough to ensure correct detection within a small delay. This issue is further exaggerated by the fact that $\xhat_{t,modcs}$ is a biased estimate of $x_t$. Along $\Tset_t^c$, the values of $\xhat_{t,modcs}$ will be biased toward zero (because we minimize $\|(\beta)_{\Tset_t^c}\|_1$), while, along $\Tset_t$, they may be biased away from zero. This will create the following problem. The set $\Tset_t$ contains the set $\Delta_{e,t}$ which needs to be deleted. Since the estimates along $\Delta_{e,t}$ may be biased away from zero, one will need a higher threshold to delete them. But that would make detection more difficult, especially since the estimates along $\Delta_t \subseteq \Tset_t^c$ will be biased towards zero. A similar issue for noisy CS, and a possible solution (Gauss-Dantzig selector), was first discussed in \cite{dantzig}.

%
\subsection{Modified-CS with Add-LS-Del}
\label{addLSdel_modcs}
The bias issue can be partly addressed by replacing the support estimation step of Modified-CS by a three step Add-LS-Del procedure summarized in Algorithm \ref{modcsalgo_2}. It involves a support addition step (that uses a smaller threshold - $\alpha_{\add}$), as in (\ref{addstep}), followed by LS estimation on the new support estimate, $\Tset_{\add,t}$, as in (\ref{lsstep}), and then a deletion step that thresholds the LS estimate, as in (\ref{deletestep}). This can be followed by a second LS estimation using the final support estimate, as in (\ref{finalls}), although this last step is not critical. The addition step threshold, $\alpha_{\add}$, needs to be just large enough to ensure that the matrix used for LS estimation, $A_{\Tset_{\add,t}}$ is well-conditioned. If $\alpha_{\add}$ is chosen properly and if $n$ is large enough, the LS estimate on $\Tset_{\add,t}$ will have smaller error and will be less biased than the modified-CS output. As a result, deletion will be more accurate when done using this estimate. This also means that one can use a larger deletion threshold, $\alpha_{\del}$, which will ensure quicker deletion of extras.

Related ideas were introduced in our older work \cite{just_lscs,kfcsicip} for KF-CS and LS-CS, and in \cite{subspacepursuit,cosamp} for a greedy algorithm for static sparse reconstruction.

We explain how to automatically set the parameters for both modified-CS-add-LS-del and modified-CS in Sec \ref{param}.

\begin{algorithm}[h]
\caption{{\bf \small Modified-CS-Add-LS-Del}}
For $t \ge 0$, do
\ben
\item {\em Noisy $\ell_1$. } If $t = 0$, set $\Tset_t = \emptyset$ and compute $\xhat_{t,modcs}$ as the solution of (\ref{simpcs}).

\item {\em Modified-CS. } If $t>0$, set $\Tset_t = \Nhat_{t-1}$ and compute $\xhat_{t,modcs}$ as the solution of (\ref{modcs}).

\label{step1}

\item {\em Additions / LS.} Compute $\Tset_{\add,t}$ and the LS estimate using it:%
\label{addls}
\bea
\label{addstep}
\Ahat_t \se \{i: |(\xhat_{t,modcs})_i| > \alpha_{\add}  \} \nn \\
\Tset_{\add,t} \se \Tset_t \cup \Ahat_t \ \ \ \  \\
(\xhat_{t,\add})_{\Tset_{\add,t}} \se {A_{\Tset_{\add,t}}}^\dag y_t, \ \ (\xhat_{t,\add})_{\Tset_{\add,t}^c} = 0
\label{lsstep}
\eea

\item {\em Deletions / LS.} Compute $\tT_t$ and LS estimate using it:%
\label{delete}
\bea
\label{deletestep}
\Rhat_t\se \{i\in \Tset_{\add,t}: |(\xhat_{t,\add})_i| \le \alpha_{\del} \} \nn \\
\tT_t \se  \Tset_{\add,t} \setminus \Rhat_t  \\
(\xhat_{t})_{\tT_t} \se {A_{\tT_t}}^\dag  y_t, \ \ (\xhat_{t})_{\tT_t^c} = 0  
\label{finalls}
\eea

\item Set $\Nhat_t = \tT_t$. Feedback $\Nhat_t$.  Output $\xhat_{t}$.

\een
\label{modcsalgo_2}
\end{algorithm}

\subsection{Some Definitions}
\bd
For any matrix, $A$, the left $S$-restricted isometry constant (left-RIC) $\delta_{S, \text{left}}(A)$ and right $S$-restricted isometry constant (right-RIC) $\delta_{S, \text{right}}(A)$ are the smallest real numbers satisfying
\bea
(1- \delta_{S, \text{left}}(A)) \|c\|^2 \le \|A_\Tset c\|^2 \le (1 + \delta_{S, \text{right}}(A)) \|c\|^2
\label{def_delta}
\eea
for all sets $\Tset \subset [1,m]$ of cardinality $|\Tset| \le S$ and all real vectors $c$ of length $|\Tset|$.
The restricted isometry constant (RIC)\cite{decodinglp} is the larger of the two, i.e., $$\delta_S = \max\{\delta_{S, \text{left}}(A), \delta_{S, \text{right}}(A)\}.$$
\ed
\bd
The restricted orthogonality constant (ROC) \cite{decodinglp}, $\theta_{S_1,S_2}(A)$, is the smallest real number satisfying
\bea
| {c_1}'{A_{\Tset_1}}'A_{\Tset_2} c_2 | \le \theta_{S_1,S_2} \|c_1\| \ \|c_2\|
\label{def_theta}
\eea
for all disjoint sets $\Tset_1, \Tset_2 \subset [1,m]$ with $|\Tset_1| \le S_1$, $|\Tset_2| \le S_2$ and $S_1+S_2 \le m$, and for all vectors $c_1$, $c_2$ of length $|\Tset_1|$, $|\Tset_2|$ respectively.
\ed

In this work, we need the same condition on the RIC and ROC of all measurement matrices $A_t$ for $t>0$. Thus, in the rest of this paper, we let
$$\delta_S := \max_{t>0} \delta_S (A_t), \ \text{and} \ \theta_{S_1,S_2}:= \max_{t>0} \theta_{S_1,S_2}(A_t). $$
If we need the RIC of ROC of any other matrix, then we specify it explicitly. 


As seen above, we use $\alpha$ to denote the support estimation threshold used by modified-CS and we use $\alpha_{\add}, \alpha_{\del}$ to denote the support addition and deletion thresholds used by modified-CS-add-LS-del. 
We use $\Nhat_t$ to denote the support estimate at time $t$. 

\bd[$\Tset_t$, $\Delta_t$, $\Delta_{e,t}$]
We use $\Tset_t  : = \Nhat_{t-1}$ to denote the support estimate from the previous time. This serves as the predicted support at time $t$.
We use $\Delta_t := {\cal N}_t \setminus \Tset_t$ to denote the unknown part of $\Nset_t$ and $\Delta_{e,t} := \Tset_t \setminus {\cal N}_t$ to denote the ``erroneous" part of $\Nset_t$. 
\ed
With the above definition, clearly,
${\cal N}_t = \Tset_t \cup \Delta_t \setminus \Delta_{e,t}.$

\bd[$\tT_t$, $\tDelta_t$, $\tDelta_{e,t}$]
We use $\tT_t := \Nhat_t$ to denote the final estimate of the current support; $\tDelta_t : = {\cal N}_t \setminus \tT_t$ to denote the ``misses" in $\Nhat_t$ and $\tDelta_{e,t} := \tT_t \setminus {\cal N}_t$ to denote the ``extras". 
\ed

\bd [Define $\Tset_{\add,t},\Delta_{\add,t}, \Delta_{e,\add,t}$]
The set $\Tset_{\add,t}$ is the support estimate obtained after the support addition step in Algorithm \ref{modcsalgo_2} (modified-CS-add-LS-del). It is defined in (\ref{addstep}). The set $\Delta_{\add,t}:=\Nset_t \setminus  \Tset_{\add,t}$ denotes the set of missing elements from $\Nset_t$ and the set $\Delta_{e,\add,t}:=\Tset_{\add,t} \setminus \Nset_t$ denotes the set of extras in it. 
\label{defdett}
\ed

\begin{remark}
At certain places in the paper, we remove the subscript $t$ for ease of notation.
\end{remark}



\subsection{Modified-CS error bound at time $t$ }
By adapting the approach of \cite{candes_rip}, the error of modified-CS can be bounded as a function of $|\Tset_t|=|\Nset_t|+|\Delta_{e,t}|-|\Delta_t|$ and $|\Delta_t|$. This was done in \cite{arxiv}. We state a modified version here.

%

\begin{lemma}[modified-CS error bound]
Assume that $y_t$ satisfies (\ref{obsmod}) and the support of $x_t$ is $\Nset_t$.
Consider step 2 of Algorithm \ref{modcsalgo} or \ref{modcsalgo_2}.
If $\delta_{|\Tset_t|+3|\Delta_t|} =\delta_{|\Nset_t|+ |\Delta_{e,t}|+ 2|\Delta_t|} < (\sqrt{2}-1)/2$, 
then
$$\|x_t - \xhat_{t,modcs}\| \le C_1(|\Tset_t|+3|\Delta_t|) \eps \le 7.50 \eps, \  C_1(S) \defn \frac{4 \sqrt{1+\delta_S}}{1 - 2 \delta_S}.$$
\label{modcsbnd_2}
\end{lemma}
For completeness, we provide a proof in Appendix \ref{modcserrorbound_2}. 

Notice that the bound by $C_1(|\Tset_t|+3|\Delta_t|)\eps$ will hold as long as $\delta_{|\Tset_t|+3|\Delta_t|}< 1/2$. By enforcing that $\delta_{|\Tset_t|+3|\Delta_t|} \le1/2 c$ for a $c<1$, we ensure that $C_1(.)$ is bounded by a fixed constant. To state the above lemma we pick $c=\sqrt{2}-1$ and this gives $C_1(.) = 7.50$. We can state a similar result for CS \cite{candes_rip}.

\begin{lemma}[CS error bound \cite{candes_rip}]
Assume that $y_t$ satisfies (\ref{obsmod}) and the support of $x_t$ is $\Nset_t$. Let $\xhat_{t,cs}$ denote the solution of (\ref{modcs}) with $\Tset_t = \emptyset$. If  $\delta_{2|\Nset_t|} < (\sqrt{2}-1)/2$,
then
\bea
\|x_t - \xhat_{t,cs}\| \sle C_1(2|\Nset_t|) \eps \le 7.50 \eps \nn
\eea
\label{cs_bnd}
\end{lemma}

\subsection{LS step error bound at time $t$}
We can claim the following about the LS step error in step 3 of Algorithm \ref{modcsalgo_2}.
\begin{lemma}
Assume that $y_t$ satisfies (\ref{obsmod}) and the support of $x_t$ is $\Nset_t$. Consider step 3 of Algorithm \ref{modcsalgo_2}.
\ben

\item $(x_t - \xhat_{t,\add})_{\Tset_{\add,t}} = ({A_{\Tset_{\add,t}}}'A_{\Tset_{\add,t}})^{-1} [ {A_{\Tset_{\add,t}}}' w_t + {A_{\Tset_{\add,t}}}' A_{\Delta_{\add,t}}  (x_t)_{\Delta_{\add,t}}]$, $(x_t - \xhat_{t,\add})_{\Delta_{\add,t}} = (x_t)_{\Delta_{\add,t}}$, and $(x_t - \xhat_{t,\add})_{i} = 0, \ \text{if} \ i \notin \Tset_{\add,t} \cup \Delta_{\add,t}.$

\item  
\ben
\item
$\|(x_t - \xhat_{t,\add})_{\Tset_{\add,t}}\| \le \frac{1}{\sqrt{1-\delta_{|\Tset|}}} \eps  + \frac{\theta_{|\Tset_{\add,t}|,|\Delta_{\add,t}|}}{1-\delta_{|\Tset|}} \|(x_t)_{\Delta_{\add,t}}\|$. 
\item $\|(x_t - \xhat_{t,\add})\| \le \frac{1}{\sqrt{1-\delta_{|\Tset|}}} \eps + (1+ \frac{\theta_{|\Tset_{\add,t}|,|\Delta_{\add,t}|}}{1-\delta_{|\Tset|}} ) \|(x_t)_{\Delta_{\add,t}}\|$.
\een

\een
\label{errls1}
\end{lemma}

Proof: The first claim follows directly from the expression for $\xhat_{t,\add}$. The second claim uses the first claim and the facts that
 $||{A_{\Tset}}^\dag||_2 \le 1/\sqrt{1-\delta_{|\Tset|}}$, $||({A_{\Tset}}'A_{\Tset})^{-1}|| \le 1/{(1-\delta_{|\Tset|})}$ and $||A_{\Tset\cup\Delta}||_2 \le \theta_{|\Tset|,|\Delta|}$ \cite{just_lscs}.

\section{Stability over time results without signal value change assumptions}
\label{nomodel}
As suggested by an anonymous reviewer, we begin by first stating a stability over time result for modified-CS and modified-CS-add-LS-del without assuming any model on how the signal changes. This result is quite general and is applicable to various types of signal change models. In Sections \ref{simpmodel} and \ref{genmodel}, we specialize the proof technique to get stability results for two sets of signal change assumptions.

\subsection{Stability Over Time Result for Modified-CS}
The following facts are immediate from Algorithm \ref{modcsalgo}.
\begin{proposition}[simple facts] Consider Algorithm \ref{modcsalgo}.
\ben
\item An $i \in \Nset_t$ will definitely get detected in step \ref{supp_estim} if $|(x_t)_i|  > \alpha + \|x_t - \xhat_{t,modcs}\|_\infty$.
\label{det1}

\item Similarly, all $i \in \Delta_{e,t}$ (the zero elements of $\Tset_t$) will definitely get deleted in step \ref{supp_estim} if $\alpha \ge \|x_t - \xhat_{t, modcs}\|_\infty$.

\een
\label{prop0}
\end{proposition}

Using the above facts and Lemma \ref{modcsbnd_2} and an induction argument, we get the following result.
\begin{theorem}
\label{theorem_modcs_simplified}
Consider Algorithm \ref{modcsalgo}. Assume that the support size of $x_t$ is bounded by $S$ and there are at most $S_a$ additions and removals at all times. Assume that $y_t$ satisfies (\ref{obsmod}). 
If the following hold
\ben
\item {\em (support estimation threshold) } set $\alpha =  7.50 \eps$, 

\item {\em (number of measurements)} 
$\delta_{S + 6S_a}  \le 0.207$,
\label{measmodel_simplified}

\item {\em (number of small magnitude entries)} $|\Bset_t|\leq S_a$, where $\Bset_t= \{i\in \Nset_t:  |(x_t)_i| \leq \alpha+ 7.50 \eps\},$
\label{minmag_simplified}
\item {\em (initial time)} at $t=0$, $n_0$ is large enough to ensure that $|\tDelta_t|=0$, $\tDelta_{e,t}=0$.
\label{initass_modcs_simplified}
\een
then for all $t$,
\ben
\item $|\tDelta_t|\leq S_a$, $|\tDelta_{e,t}|=0$, $|\tT_t| \leq S$,
\item $|\Delta_t|\leq 2S_a$, $|T_t|\leq S$, $|\Delta_{e,t}| \leq S_a$,
\item and $\|x_t-\hat{x}_t\|\leq 7.50\epsilon$.
\een
\end{theorem}

The proof is provided in Appendix \ref{proof_modcs_simplified}.

\subsection{Stability over time Result for Modified-CS-add-LS-del}
A result similar to the one above can also be proved for modified-CS-add-LS-del.

\begin{theorem}
\label{theorem_modcsls_simplified}
Consider Algorithm \ref{modcsalgo_2}. Assume that the support size of $x_t$ is bounded by $S$ and there are at most $S_a$ additions and removals at all times. Assume that $y_t$ satisfies (\ref{obsmod}).
If the following hold
\ben
\item {\em (addition and deletion thresholds) }
\ben
\label{threshes_simplified}
\item $\alpha_\add$ is large enough so that at most $f$ false additions per unit time, 
\label{addthresh_simplified}
\item $\alpha_\del=1.12\eps + 0.261\sqrt{S_a}(\alpha_\add + 7.50\eps)$,
\label{delthresh_simplified}
\een

\item {\em (number of measurements)} 
$\delta_{S + 6S_a}  \le 0.207$,
$\delta_{S + 2S_a + f}  \le 0.207$,
\label{measmodel_modcsls_simplified}

\item {\em (number of small magnitude entries)} $|\Bset_t|\leq S_a$, where $\Bset_t= \{i\in \Nset_t:  |(x_t)_i| \leq \max \{\alpha_\add+7.50 \eps, 2\alpha_\del\}\},$
\label{minmag_modcsls_simplified}
\item {\em (initial time)} at $t=0$, $n_0$ is large enough to ensure that $|\tDelta_t|=0$, $\tDelta_{e,t}=0$.
\label{initass_simplified}
\een
then for all $t$,
\ben
\item $|\tDelta_t|\leq S_a$, $\tDelta_{e,t}=0$, $|\tT_t|\leq S$,
\item $|\Delta_t|\leq 2S_a$, $|\Delta_{e,t}| \leq S_a$, $|T_t|\leq S$,
\item $|\Delta_{\add,t}|\leq S_a$, $|\Delta_{e,\add,t}|\leq S_a+f$, $|\Tset_{\add,t}|\leq S+S_a+f$,
\item $\|x_t-\hat{x}_{t,modcs}\|\leq 7.50\epsilon$
\item and $\|x_t-\hat{x}_t\|\leq 1.12\epsilon+1.261\sqrt{2\alpha_\del S_a}$.
\een
\end{theorem}

Proof is provided in Appendix \ref{proof_modcsls_simplified}.

\subsection{Discussion} \label{simp_disc}
Notice that the support error bound in both results above is $2 S_a$. Under slow support change, $S_a \ll S$, this bound is small compared to the support size $S$, making the result a meaningful one. Also, the reconstruction error is upper bounded by a constant times $\eps$. Under a high enough signal-to-noise ratio (SNR), this bound is also small compared to the signal power.

If $f = S_a$ in Theorem \ref{theorem_modcsls_simplified}, both Modified-CS and Modified-CS-add-LS-del need $\delta_{S + 6S_a} \le 0.207$. Consider noisy $\ell_1$, i.e. (\ref{simpcs}).  Since it is not a recursive approach (each time instant is handled separately), Lemma \ref{cs_bnd} is also a stability result for it. From Lemma \ref{cs_bnd}, it needs $\delta_{2S} \le 0.207$ to get the same error bound. When $S_a \ll S$, clearly it requires a stronger condition than either of the modified-CS algorithms.

\begin{remark} 
Consider the noise-free case, i.e. the case when $\eps=0$, $y_t=A_tx_t$, with the number of support additions and removals per unit time at most $S_a$. In this case, our results say the following: as long as the signal change assumptions hold, $\delta_{S+kS_a}< 0.207$ is sufficient for both algorithms. It is easy to show that $\delta_{S+S_a,\text{left}} < 1$ is also necessary for both algorithms.
We give a proof for this in Appendix \ref{modcs_noise_free_rip_proof}. Thus the sufficient condition that our results need are of the same order in both $S$ and $S_a$ as the necessary condition and hence these results cannot be improved much. Thus, for example, RIP of order $S+k\sqrt{S_a}$ or  $\sqrt{S+kS_a}$ will not work. This remark is inspired by a concern of an anonymous reviewer.
\label{modcs_noise_free_rip}
\end{remark}

\section{Stability results: Simple but restrictive signal change assumptions} \label{simpmodel}
In this section, we assume a very simple but restrictive signal change model that allows for slow nonzero coefficient magnitude increase after a new coefficient is added and slow decrease in magnitude before a coefficient is removed. 


\subsection{Simple but restrictive signal change assumptions}
\label{signalmodel_simple}
We use a single parameter, $r$, for the newly added elements' magnitude and for the magnitude increase and decrease rate of all elements at all times. We also fixes the number of support additions and removals to be $S_a$.

\begin{figure*}
\centering
\begin{tikzpicture}[xscale=1,yscale=1,line width=0.8pt]
\draw (2*\w,3*\a+\ab+\x+\y) rectangle (2*\w+\x,3*\a+\ab+2*\x+\y)node[pos=.5] {dr};
\filldraw(2*\w+0.5*\x,3*\a+\ab+\x+0.25*\y) circle (2pt);
\filldraw(2*\w+0.5*\x,3*\a+\ab+\x+.5*\y) circle (2pt);
\filldraw(2*\w+0.5*\x,3*\a+\ab+\x+.75*\y) circle (2pt);
\draw (2*\w,3*\a+\ab) rectangle (2*\w+\x,3*\a+\ab+\x)node[pos=.5] {dr};
\draw[decorate,decoration={brace,mirror,raise=2pt}] (2*\w+\x,3*\a+\ab+0.1*\x) -- (2*\w+\x,3*\a+\ab+2*\x+\y-0.1*\x);
\node (d) at (2*\w+1.1*\x,3*\a+\ab+\x+.5*\y)[right,align=left]{$\mathcal{N}_{t+1}\setminus\mathcal{S}_{t+1}$,\\ $|\mathcal{N}_{t+1}\setminus\mathcal{S}_{t+1}|\geq S-(2d-2)S_a$};

\filldraw(2*\w+0.5*\x,3*\a+0.5*\x+0.25*\y) circle (2pt);
\filldraw(2*\w+0.5*\x,3*\a+0.5*\x+.5*\y) circle (2pt);
\filldraw(2*\w+0.5*\x,3*\a+0.5*\x+.75*\y) circle (2pt);

\draw (2*\w,2*\a+\x+\y) rectangle (2*\w+\x,2*\a+2*\x+\y)node[pos=.5] {2r};
\filldraw(2*\w+0.5*\x,2*\a+\x+0.25*\y) circle (2pt);
\filldraw(2*\w+0.5*\x,2*\a+\x+.5*\y) circle (2pt);
\filldraw(2*\w+0.5*\x,2*\a+\x+.75*\y) circle (2pt);
\draw (2*\w,2*\a) rectangle (2*\w+\x,2*\a+\x)node[pos=.5] {2r};
\draw[decorate,decoration={brace,mirror,raise=2pt}] (2*\w+\x,2*\a+0.1*\x) -- (2*\w+\x,2*\a+2*\x+\y-0.1*\x);
\node (c) at (2*\w+1.1*\x,2*\a+\x+.5*\y)[right,align=left]{$\mathcal{I}_{t+1}(2)\cup\mathcal{D}_{t+1}(2)$,\\ $|\mathcal{I}_{t+1}(2)|=|\mathcal{D}_{t+1}(2)|=S_a$};

\draw (2*\w,\a+\x+\y) rectangle (2*\w+\x,\a+2*\x+\y)node[pos=.5] {r};
\filldraw(2*\w+0.5*\x,\a+\x+0.25*\y) circle (2pt);
\filldraw(2*\w+0.5*\x,\a+\x+.5*\y) circle (2pt);
\filldraw(2*\w+0.5*\x,\a+\x+.75*\y) circle (2pt);
\draw (2*\w,\a) rectangle (2*\w+\x,\a+\x)node[pos=.5] {r};
\draw[decorate,decoration={brace,mirror,raise=2pt}] (2*\w+\x,\a+0.1*\x) -- (2*\w+\x,\a+2*\x+\y-0.1*\x);
\node (b) at (2*\w+1.1*\x,\a+\x+.5*\y)[right,align=left]{$\mathcal{I}_{t+1}(1)\cup\mathcal{D}_{t+1}(1)$,\\ $|\mathcal{I}_{t+1}(1)|=|\mathcal{D}_{t+1}(1)|=S_a$};

\draw (2*\w,\x+\y) rectangle (2*\w+\x,2*\x+\y)node[pos=.5] {0};
\filldraw(2*\w+0.5*\x,\x+0.25*\y) circle (2pt);
\filldraw(2*\w+0.5*\x,\x+.5*\y) circle (2pt);
\filldraw(2*\w+0.5*\x,\x+.75*\y) circle (2pt);
\draw (2*\w,0) rectangle (2*\w+\x,\x)node[pos=.5] {0};
\draw[decorate,decoration={brace,mirror,raise=2pt}] (2*\w+\x,0.1*\x) -- (2*\w+\x,2*\x+\y-0.1*\x);
\node (a) at (2*\w+1.1*\x,\x+.5*\y)[right,align=left]{$\mathcal{N}_{t+1}^c$,\\ $|\mathcal{N}_{t+1}^c|=m-S$};

\draw (2*\w,0) rectangle (2*\w+\x,3*\a+\ab+2*\x+\y);


\draw (\w,3*\a+\ab+\x+\y) rectangle (\w+\x,3*\a+\ab+2*\x+\y)node[pos=.5] {dr};
\filldraw(\w+0.5*\x,3*\a+\ab+\x+0.25*\y) circle (2pt);
\filldraw(\w+0.5*\x,3*\a+\ab+\x+.5*\y) circle (2pt);
\filldraw(\w+0.5*\x,3*\a+\ab+\x+.75*\y) circle (2pt);
\draw (\w,3*\a+\ab) rectangle (\w+\x,3*\a+\ab+\x)node[pos=.5] {dr};
\draw[decorate,decoration={brace,raise=2pt}] (\w,3*\a+\ab+0.1*\x) -- (\w,3*\a+\ab+2*\x+\y-0.1*\x);
\node (d) at (\w-.1*\x,3*\a+\ab+\x+.5*\y)[left,align=right]{$\mathcal{N}_{t}\setminus\mathcal{S}_{t}$,\\ $|\mathcal{N}_{t}\setminus\mathcal{S}_{t}|=S-(2d-2)S_a$};

\filldraw(\w+0.5*\x,3*\a+0.5*\x+0.25*\y) circle (2pt);
\filldraw(\w+0.5*\x,3*\a+0.5*\x+.5*\y) circle (2pt);
\filldraw(\w+0.5*\x,3*\a+0.5*\x+.75*\y) circle (2pt);

\draw (\w,2*\a+\x+\y) rectangle (\w+\x,2*\a+2*\x+\y)node[pos=.5] {2r};
\filldraw(\w+0.5*\x,2*\a+\x+0.25*\y) circle (2pt);
\filldraw(\w+0.5*\x,2*\a+\x+.5*\y) circle (2pt);
\filldraw(\w+0.5*\x,2*\a+\x+.75*\y) circle (2pt);
\draw (\w,2*\a) rectangle (\w+\x,2*\a+\x)node[pos=.5] {2r};
\draw[decorate,decoration={brace,raise=2pt}] (\w,2*\a+0.1*\x) -- (\w,2*\a+2*\x+\y-0.1*\x);
\node (c) at (\w-.1*\x,2*\a+\x+.5*\y)[left,align=right]{$\mathcal{I}_{t}(2)\cup\mathcal{D}_{t}(2)$,\\ $|\mathcal{I}_{t}(2)\cup\mathcal{D}_{t}(2)|=S_a$};

\draw (\w,\a+\x+\y) rectangle (\w+\x,\a+2*\x+\y)node[pos=.5] {r};
\filldraw(\w+0.5*\x,\a+\x+0.25*\y) circle (2pt);
\filldraw(\w+0.5*\x,\a+\x+.5*\y) circle (2pt);
\filldraw(\w+0.5*\x,\a+\x+.75*\y) circle (2pt);
\draw (\w,\a) rectangle (\w+\x,\a+\x)node[pos=.5] {r};
\draw[decorate,decoration={brace,raise=2pt}] (\w,\a+0.1*\x) -- (\w,\a+2*\x+\y-0.1*\x);
\node (b) at (\w-.1*\x,\a+\x+.5*\y)[left,align=right]{$\mathcal{I}_{t}(1)\cup\mathcal{D}_{t}(1)$,\\ $|\mathcal{I}_{t}(1)\cup\mathcal{D}_{t}(1)|=S_a$};

\draw (\w,\x+\y) rectangle (\w+\x,2*\x+\y)node[pos=.5] {0};
\filldraw(\w+0.5*\x,\x+0.25*\y) circle (2pt);
\filldraw(\w+0.5*\x,\x+.5*\y) circle (2pt);
\filldraw(\w+0.5*\x,\x+.75*\y) circle (2pt);
\draw (\w,0) rectangle (\w+\x,\x)node[pos=.5] {0};
\draw[decorate,decoration={brace,raise=2pt}] (\w,0.1*\x) -- (\w,2*\x+\y-0.1*\x);
\node (a) at (\w-.1*\x,\x+.5*\y)[left,align=right]{$\mathcal{N}_t^c$,\\ $|\mathcal{N}_t^c|=m-S$};

\draw (\w,0) rectangle (\w+\x,3*\a+\ab+2*\x+\y);

\node (left) at (\w+0.5*\x,-0.8*\x){\LARGE $x_t$};
\node (left) at (2*\w+0.5*\x,-0.8*\x){\LARGE $x_{t+1}$};

\draw[dashed] (-\x,\a) rectangle (\w+1.5*\x,3*\a+\ab+2.1*\x+\y);
\node (right) at (2*\x,2*\a+3*\x+\y){\Large $\mathcal{N}_t$};

\draw[dashed] (2*\w-0.5*\x,\a) rectangle (3*\w+3*\x,3*\a+\ab+2.1*\x+\y);
\node (right) at (3*\w-2*\x,2*\a+3*\x+\y){\Large $\mathcal{N}_{t+1}$};


\draw[red,dash pattern=on 10pt off 10pt,decoration={markings,mark=at position 1 with {\arrow[scale=2]{>}}},postaction={decorate},shorten >=0.4pt](\w+\x,\x+.5*\y)--(2*\w,\x+.5*\y)node[pos=.5,above]{$m-S-S_a$};
\draw[red,dash pattern=on 10pt off 10pt,decoration={markings,mark=at position 1 with {\arrow[scale=2]{>}}},postaction={decorate},shorten >=0.4pt](\w+\x,\x+.5*\y)--(2*\w,\a+\x+.5*\y)node[pos=.3,above,sloped]{$S_a$};

\draw[blue,decoration={markings,mark=at position 1 with {\arrow[scale=2]{>}}},postaction={decorate},shorten >=0.4pt](\w+\x,\a+\x+.5*\y)--(2*\w,\x+.5*\y)node[pos=.3,above,sloped]{$S_a$};
\draw[blue,decoration={markings,mark=at position 1 with {\arrow[scale=2]{>}}},postaction={decorate},shorten >=0.4pt](\w+\x,\a+\x+.5*\y)--(2*\w,2*\a+\x+.5*\y)node[pos=.3,above,sloped]{$S_a$};

\draw[line width=1.5pt,dotted,decoration={markings,mark=at position 1 with {\arrow[scale=1.5]{>}}},postaction={decorate},shorten >=0.4pt](\w+\x,2*\a+\x+.5*\y)--(2*\w,\a+\x+.5*\y)node[pos=.3,above,sloped]{$S_a$};
\draw[line width=1.5pt,dotted,decoration={markings,mark=at position 1 with {\arrow[scale=1.5]{>}}},postaction={decorate},shorten >=0.4pt](\w+\x,2*\a+\x+.5*\y)--(2*\w,2.7*\a+\x+.5*\y)node[pos=.3,above,sloped]{$S_a$};

\draw[cyan,dash pattern=on 5pt off 3pt on 2pt off 3pt,decoration={markings,mark=at position 1 with {\arrow[scale=2]{>}}},postaction={decorate},shorten >=0.4pt](\w+\x,3*\a+\ab+\x+.5*\y)--(2*\w,3*\a+\ab+\x+.5*\y) node[pos=.5,above,sloped]{$S-(2d-1)S_a$};
\draw[cyan,dash pattern=on 5pt off 3pt on 2pt off 3pt,decoration={markings,mark=at position 1 with {\arrow[scale=2]{>}}},postaction={decorate},shorten >=0.4pt](\w+\x,3*\a+\ab+\x+.5*\y)--(2*\w,3*\a+\x+.5*\y)node[pos=.4,above,sloped]{$S_a$};

\end{tikzpicture}
\caption{Signal Change Assumptions 1 (Values inside rectangular denote magnitudes.)}
\label{fig_sigmod2}
\end{figure*}
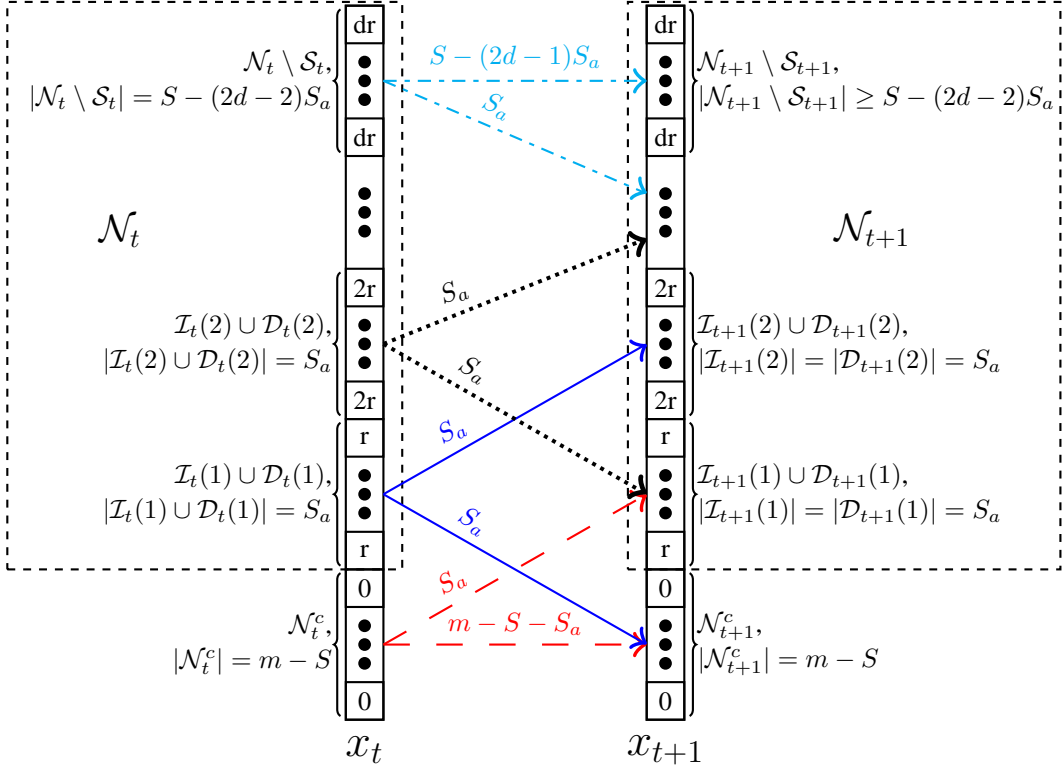

\begin{sigmodel} Assume the following.
\ben

\item {\em (addition and increase) } At each $t>0$, $S_a$ new coefficients get added to the support at magnitude $r$. Denote this set by $\Aset_t$. At each $t>0$, the magnitude of $S_a$ coefficients out of all those which had magnitude $(j-1)r$ at $t-1$ increases to $jr$. This occurs for all $2 \le j \le d$. Thus the maximum magnitude reached by any coefficient is $M:=dr$.

\item {\em (decrease and removal) } At each $t>0$, the magnitude of $S_a$ coefficients out of all those which had magnitude $(j+1)r$ at $t-1$ decreases to $jr$. This occurs for all $1 \le j \le (d-2)$. At each $t>0$, $S_a$ coefficients out of all those which had magnitude $r$ at $t-1$ get removed from the support (magnitude becomes zero). Denote this set by $\Rset_t$.

\item {\em (initial time) } At $t=0$, the support size is $S$ and it contains $2S_a$ elements each with magnitude $r,2r, \dots (d-1)r$, and $(S-(2d-2)S_a)$ elements with magnitude $M$.%
\een
\label{sigmod2}
\end{sigmodel}
Fig. \ref{fig_sigmod2} illustrates the above signal change assumptions.

To understand its implications, define the following sets.
For $0 \le j \le d-1$, let $$\Dset_{t}(j): = \{i: |x_{t,i}| = jr, \ |x_{t-1,i}| = (j+1)r \}$$ denote the set of elements that {\em decrease} from $(j+1)r$ to $jr$ at time, $t$.
For $1 \le j \le d$, let $$\Iset_{t}(j): = \{i: |x_{t,i}| = jr, \ |x_{t-1,i}| = (j-1)r \}$$ denote the set of elements that {\em increase} from $(j-1)r$ to $jr$ at time, $t$.
For $1 \le j \le d-1$, let  $$\Sset_t(j):= \{i:  0 < |x_{t,i}| < j r \}$$ denote the set of {\em small but nonzero} elements, with smallness threshold $jr$.
Clearly, the newly added set, $\Aset_t= \Iset_t(1)$ and the newly removed set, $\Rset_t= \Dset_t(0)$. Also, $|\Iset_{t}(j)|=S_a$, $|\Dset_{t}(j)|=S_a$, $|\Sset_t(j)| = 2(j-1)S_a$.

Consider a $1 < j \le d$. From Signal Change Assumptions \ref{sigmod2}, it is clear that at any time, $t$, $S_a$ elements enter the small elements' set, $\Sset_t(j)$, from the bottom (set $\Aset_t$) and $S_a$ enter from the top (set $\Dset_{t}(j-1)$). Similarly $S_a$ elements leave  $\Sset_t(j)$ from the bottom  (set $\Rset_t$) and $S_a$ from the top (set $\Iset_{t}(j)$). Thus,
\bea
\Sset_t(j) = \Sset_{t-1}(j)  \cup (\Aset_t \cup \Dset_{t}(j-1)) \setminus (\Rset_t \cup \Iset_{t}(j)) \ \ \
\label{sseteq}
\eea
%
Since $\Aset_t, \Rset_t, \Dset_{t}(j-1),\Iset_{t}(j)$ are mutually disjoint, $\Rset_t \subseteq \Sset_{t-1}(j)$ and $\Iset_{t}(j) \subseteq \Sset_{t-1}(j)$, thus, (\ref{sseteq}) implies that
\bea
\Sset_{t-1}(j)  \cup \Aset_t  \setminus \Rset_t = \Sset_t(j) \cup \Iset_{t}(j) \setminus \Dset_{t}(j-1)
\label{sseteq_2}
\eea
Also, clearly,
\bea
\Nset_t \se \Nset_{t-1} \cup \Aset_t \setminus \Rset_t
\eea

\subsection{Stability result for modified-CS}
\label{stab_modcs_simple}

The first step is to find sufficient conditions for a certain set of large coefficients to definitely get detected, and for the elements of $\Delta_e$ to definitely get deleted. These are obtained in Lemma \ref{lemma_modcs} by using Lemma \ref{modcsbnd_2} and the following simple facts. Next, we use Lemma \ref{lemma_modcs} to ensure that all new additions to the support get detected within a finite delay, and all removals from the support get deleted immediately.

In general, for any vector $z$, $\|z\|_\infty \le \|z\|$ with equality holding only if $z$ is one-sparse (exactly one element of $z$ is nonzero). If the energy of $z$ is more spread out, $\|z\|_\infty$ will be smaller than $\|z\|$.  Typically the error $x_t - \xhat_{t,modcs}$ will not be one-sparse, but will be more spread out. The assumption below states this. 
\begin{ass}
Consider Algorithm \ref{modcsalgo}. Assume that the Modified-CS reconstruction error is spread out enough so that
\bea
\|x_t - \hat{x}_{t,modcs}\|_\infty \leq \frac{\zeta_M}{\sqrt{S_a}}\|x_t-\hat{x}_{t,modcs}\| \nn
\eea
for some $\zeta_M\leq \sqrt{S_a}$.
\label{spread1}
\end{ass}

Combining Proposition \ref{prop0} and the above assumption with Lemma \ref{modcsbnd_2}, we get the following lemma.
\begin{lemma}
Consider Algorithm \ref{modcsalgo}. Assume Assumption \ref{spread1}.
Assume that $|\Nset_t| = S_{\Nset_t}$, $|\Delta_{e,t}| \le S_{\Delta_{e,t}}$ and $|\Delta_t| \le S_{\Delta_t}$.
\ben
\item All elements of the set $\{i \in \Nset_t: |(x_t)_i| \ge b_1 \}$ will get detected in step \ref{supp_estim} if
\bi
\item $\delta_{S_{\Nset_t} + S_{\Delta_{e,t}} + 2S_{\Delta_t}}  \le 0.207$, and $b_1 > \alpha +  \frac{\zeta_M}{\sqrt{S_a}}7.50 \eps$.
\ei

\item In step \ref{supp_estim}, there will be no false additions, and all the true removals from the support (the set $\Delta_{e,t}$) will get deleted at the current time, if
\bi
\item $\delta_{S_{\Nset_t} + S_{\Delta_{e,t}} + 2S_{\Delta_t}}  \le 0.207$, and $\alpha \ge \frac{\zeta_M}{\sqrt{S_a}}7.50 \eps$. 
\ei

\een
\label{lemma_modcs}
\end{lemma}

We use the above lemma to obtain sufficient conditions to ensure the following: for some $d_0 \le d$, at all times, $t$, (i) only coefficients with magnitude less than $d_0r$ are part of the final set of misses, $\tDelta_t$ and (ii) the final set of extras, $\tDelta_{e,t}$, is an empty set. In other words, we find conditions to ensure that $\tDelta_t \subseteq \Sset_{t}(d_0)$ and $|\tDelta_{e,t}|=0$. Using Signal Change Assumptions \ref{sigmod2}, $|\Sset_{t}(d_0)|=2(d_0-1)S_a$ and thus $\tDelta_t \subseteq \Sset_{t}(d_0)$ will imply that $|\tDelta_t| \le 2(d_0-1)S_a$.

\begin{theorem}[Stability of modified-CS]
Consider Algorithm \ref{modcsalgo}. Assume Signal Change Assumptions \ref{sigmod2} on $x_t$. Also assume that $y_t$ satisfies (\ref{obsmod}). Assume that Assumption \ref{spread1} holds.
If, for some $d_0 \le d$, the following hold
\ben
\item {\em (support estimation threshold) } set $\alpha =  \frac{\zeta_M}{\sqrt{S_a}}7.50 \eps$ 
\label{threshes_simple}

\item {\em (number of measurements)} 
$\delta_{S + (2k_1+1)S_a}  \le 0.207$,
\label{measmodel_simple}

\item {\em (new element increase rate) } $r \ge G$, where
\label{add_del_simple}
\bea
G \sdefn \frac{ \alpha + \frac{\zeta_M}{\sqrt{S_a}}7.50 \eps}{d_0} \eps
\eea

\item {\em (initial time)} at $t=0$, $n_0$ is large enough to ensure that $\tDelta_0  \subseteq \Sset_0(d_0)$, $|\tDelta_0| \le 2(d_0-1)S_a$,  $|\tDelta_{e,0}| =0$ and $|\tT_0| \le S$
\label{initass_simple}
\een
where
\bea
k_1 \sdefn \max(1,2d_0-2)
\eea
then,
\ben
\item at all $t \ge 0$, $|\tT_t| \le S$, $|\tDelta_{e,t}| =0$, $\tDelta_t \subseteq \Sset_t(d_0)$ and so $|\tDelta_t| \le 2(d_0-1)S_a$,
\item at all $t > 0$, $|\Tset_t| \le S$, $|\Delta_{e,t}| \le S_a$, and $|\Delta_t| \le k_1S_a$,
\item at all $t > 0$,  $\|x_t - \xhat_{t,modcs}\| \le 7.50 \eps$ 
\een
\label{stabres_simple_modcs}
\end{theorem}

{\em Proof: } The proof is given in Appendix \ref{proof_simple_modcs}. It follows using induction. 

%

\begin{remark}
The condition \ref{initass} is not restrictive. It is easy to see that this will hold if $n_0$ is large enough to ensure that $\delta_{2S}(A_0)  \le 0.207$.
\label{initass_remark}
\end{remark}

\subsection{Stability result for Modified-CS with Add-LS-Del}
\label{stab_modcsald_simple}

The first step to show stability is to find sufficient conditions for (a) a certain set of large coefficients to definitely get detected, and (b) to definitely not get falsely deleted, and (c) for the zero coefficients in $\Tset_\add$ to definitely get deleted. These can be obtained using  Lemma \ref{modcsbnd_2} and simple facts similar to Proposition \ref{prop0}.

As explained before, we can assume that the modified-CS reconstruction error is not one-sparse but is more spread out. The same assumption should also be valid for the LS step error. We state these next.
\begin{ass}
Consider Algorithm \ref{modcsalgo_2}. Assume that the Modified-CS reconstruction error is spread out enough so that Assumption \ref{spread1} holds and
assume that the LS step error along ${\Tset_{\add,t}}$ is spread out enough so that
$$\|(x_t - \xhat_{\add,t})_{\Tset_{\add,t}}\|_\infty \le \frac{\zeta_L}{\sqrt{S_a}} \|(x_t - \xhat_{\add,t})_{\Tset_{\add,t}}\|$$ 
at all times, $t$, for some $\zeta_L \leq \sqrt{S_a}$.
\label{spread2}
\end{ass}

Combining the above assumption with  Lemmas \ref{modcsbnd_2} and \ref{errls1}, we get the following lemmas.

\begin{lemma}[Detection condition]
Consider Algorithm \ref{modcsalgo_2}. Assume Assumption \ref{spread2}.
 Assume that $|\Nset_t| = S_{\Nset_t}$, $|\Delta_{e,t}| \le S_{\Delta_{e,t}}$, $|\Delta_t| \le S_{\Delta_t}$.
 Pick a $b_1 > 0$.
All elements of 
the set $\{i \in \Delta: |(x_t)_i| \ge b_1 \}$
will get detected in step \ref{addls} if
\bi
\item  $\delta_{S_{\Nset_t} + S_{\Delta_{e,t}} + 2S_{\Delta_t}}  \le 0.207$, and $b_1 > \alpha_{\add} + \frac{\zeta_M}{\sqrt{S_a}}7.50\eps$.
\ei
\label{detectcond_modcs}
\end{lemma}


\begin{lemma}[Deletion and No false-deletion condition]
Consider Algorithm \ref{modcsalgo_2}.  Assume Assumption \ref{spread2}.  Assume that $|\Tset_{\add,t}| \le S_{\Tset_{\add,t}}$ and $|\Delta_{\add,t}| \le S_{\Delta_{\add,t}}$.
\ben

\item Pick a $b_1 > 0$. No element of the set $\{i \in \Tset_{\add,t}: |(x_t)_i| \ge b_1\}$
will get (falsely) deleted in step \ref{delete} if
\bi
\item $\delta_{S_{\Tset_{\add,t}}} < 1/2$ and $b_1 > \alpha_{\del} + \frac{\zeta_L}{\sqrt{S_a}}(\sqrt{2} \eps + 2{\theta_{S_{\Tset_{\add,t}},S_{\Delta_{\add,t}}}} \|(x_t)_{\Delta_{\add,t}}\|)$.
\ei
%
%

\item All elements of $\Delta_{e,\add}$ will get deleted in step \ref{delete} if
\bi
\item $\delta_{S_{\Tset_{\add,t}}} < 1/2$ and $\alpha_{\del} \ge \frac{\zeta_L}{\sqrt{S_a}}(\sqrt{2} \eps + 2{\theta_{S_{\Tset_{\add,t}},S_{\Delta_{\add,t}}}} \|(x_t)_{\Delta_{\add,t}}\|)$.
\ei

\een
\label{nofalsedels_truedels_cond}
\end{lemma}


Using the above lemmas, we can obtain sufficient conditions to ensure that, for some $d_0 \le d$, at each time $t$, $\tDelta_t \subseteq \Sset_t(d_0)$ (so that $|\tDelta_t| \le (2d_0-2)S_a$) and $|\tDelta_{e,t}|=0$.

\begin{theorem}[Stability of modified-CS with add-LS-del]
Consider Algorithm \ref{modcsalgo_2}. Assume Signal Change Assumptions \ref{sigmod2} on $x_t$. Also assume that $y_t$ satisfies (\ref{obsmod}). Assume that Assumption \ref{spread2} holds.
If, for some $1 \le d_0 \le d$, the following hold%
\ben
\item {\em (addition and deletion thresholds) }
\ben
\item $\alpha_{\add}$ is large enough so that there are at most $f$ false additions per unit time,
\label{addthresh}

\item $\alpha_{\del}  = \sqrt{\frac{2}{S_a}}\zeta_L \eps +  2  k_3  \theta_{S+S_a+f,k_2S_a}\zeta_L r  $,
\label{delthresh}
\een

\item {\em (number of measurements)} 
\ben
\item $\delta_{S + S_a(1 + 2k_1) }  \le 0.207$,
\label{measmodald_delta}
\item $\delta_{S+S_a + f} < 1/2$,
\label{measmodald_delta2}
\item $\theta_{S+S_a+f,k_2S_a} < \frac{1}{2} \frac{d_0}{4k_3 \zeta_L}$,
\label{measmodald_theta}
\een
\label{measmodel}

\item {\em (new element increase rate) } $r \ge \max({G}_1,{G}_2)$, where
\label{add_del}
\bea
{G}_1 \sdefn \frac{ \alpha_{\add} + \frac{\zeta_M}{\sqrt{S_a}}7.50\eps }{d_0}  \nn \\ 
{G}_2 \sdefn \frac{2\sqrt{2} \zeta_L \eps}{\sqrt{S_a} (d_0 -  4k_3 \theta_{S+S_a+f,k_2S_a}\zeta_L) }  \ \ \ \ \ \ \
\eea
\item {\em (initial time)} $n_0$ is large enough to ensure that $\tDelta_0  \subseteq \Sset_0(d_0)$, $|\tDelta_0| \le (2d_0-2)S_a$,  $|\tDelta_{e,0}| =0$, $|\tT_0| \le S$,
\label{initass}
\een
where
\bea
k_1 \sdefn \max(1,2d_0-2) \nn \\
k_2 \sdefn \max(0,2d_0-3) \nn \\
k_3 \sdefn \sqrt{ \sum_{j=1}^{d_0-1} j^2 +  \sum_{j=1}^{d_0-2} j^2 }  
\eea
then,  at all $t \ge 0$,
\ben
\item $|\tT_t| \le S$, $|\tDelta_{e,t}| =0$, and $\tDelta_t \subseteq \Sset_t(d_0)$ and so $|\tDelta_t| \le (2d_0-2)S_a$,

\item $|\Tset_t| \le S$, $|\Delta_{e,t}| \le S_a$, and $|\Delta_t| \le k_1 S_a$,

\item $|\Tset_{\add,t}| \le S+S_a+f$, $|\Delta_{e,\add,t}| \le S_a+f$, and $|\Delta_{\add,t}| \le k_2 S_a,$

\item $\|x_t - \xhat_{t,modcs}\| \le C_1(S+S_a + 2k_1 S_a) \eps \le 7.50 \eps,$


\item $\|x_t-\xhat_{t} \| \leq 1.261k_3\sqrt{S_a}r+1.12\epsilon.$
\een
\label{gencase}
\end{theorem}

{\em Proof: } The proof is given in Appendix \ref{proof_addLSdel_modcs}.

\subsection{Discussion} \label{simp_disc}

Notice that, with Signal Change Assumptions \ref{sigmod2},  at all times, $t$, the signals have the same support set size, $|\Nset_t|=S$ and the same signal power, $\|x_t\|^2=(S-(2d-2)S_a)M^2 + 2S_a\sum_{j=1}^{d-1} j^2 r^2$.
%
As in the previous section, here again the support error bound in both results above is proportional to $S_a$. Under slow support change, this means that the support error is small compared to the support size. 
To make the comparison of the above two results simpler, let us fix $d_0 = 2$ and let $f = S_a$ in Theorem \ref{gencase}.
Consider the conditions on the number of measurements. Modified-CS needs $\delta_{S + 5S_a} \le 0.207$.
Modified-CS-add-LS-del needs $\delta_{S + 5S_a} \le 0.207$; $\delta_{S+2S_a} < 0.5$ (this is implied by the first condition) and $\theta_{S+2S_a,S_a} \le \frac{1}{4 \zeta_L}$. Since $\theta_{S+2S_a,S_a} \le \delta_{S + 3S_a}$, the third condition is also implied by the first as long as $\zeta_L \le 1.2$. In simulation tests (described in Sec \ref{gen_disc}) we observed that this was usually true. Then, both modified-CS  and modified-CS-add-LS-del need the same condition on the number of measurements: $\delta_{S + 5S_a} \le 0.207$.
Consider noisy $\ell_1$ i.e. (\ref{simpcs}).  As explained earlier, Lemma \ref{cs_bnd} serves as a stability result for it. From Lemma \ref{cs_bnd}, iy needs $\delta_{2S} \le 0.207$ to get the same error bound which is significantly stronger when $S_a \ll S$.
%

Let us compare the requirement on $r$. In Theorem \ref{gencase} for modified-cs-add-ls-del, since $\theta_{S+S_a+f,k_2S_a} \le \frac{1}{2} \frac{d_0}{4k_3 \zeta_L}$, so $G_2 \le \frac{4\sqrt{2} \zeta_L}{\sqrt{S_a}d_0} \eps < \frac{5.7 \eps}{d_0} < \frac{7.50 \eps}{d_0} < G_1$ and thus $G_1$ is what decides the minimum allowed value of $r$. Thus, it needs $r \ge G_1 = \frac{1}{d_0}[ \alpha_{\add} + \frac{\zeta_M}{\sqrt{S_a}}7.50\eps]$. On the other hand, modified-CS needs $r \ge G = \frac{1}{d_0} [ 2\frac{\zeta_M}{\sqrt{S_a}}7.50\eps]$.
If $\alpha_{\add}$ is close to zero, this means that the minimum magnitude increase rate, $r$, required by Theorem \ref{gencase} is almost half of that required by Theorem \ref{stabres_simple_modcs}. In our simulation experiments, $\alpha_{\add}$ was typically quite small: it was usually close to a small constant times $\eps/\sqrt{n}$ (see Sec \ref{sims}). 

\begin{remark}
From the above results, observe that, if the rate of magnitude change, $r$, is smaller, $r \ge {G}_1$ or $r \ge G$ will hold for a larger value of $d_0$. This means that the support error bound,  $(2d_0-2)S_a$, will be larger. This, in turn, decides what conditions on the RIC and ROC are needed (in other words, how many measurements, $n_t$, needed). Smaller $r$ means a larger $d_0$ is needed which, in turn, means that stronger conditions on the RIC and ROC (larger $n_t$) are needed. Thus, for a given $n_t=n$, as $r$ is reduced, the algorithm will stabilize to larger and larger support error levels (larger $d_0$) and finally become unstable (because the given $n$ does not satisfy the conditions on $\delta,\theta$ for the larger $d_0$).
\label{d0_need}
\end{remark}

%


\section{Stability Results: Realistic Signal Change Assumptions} \label{genmodel}
We introduce the signal change assumptions in the next subsection and then give the results in the following two subsections. The discussion of the results and a comparison with the results of LS-CS \cite{just_lscs} is provided in the two subsequent subsections.

\subsection{Realistic Signal Change Assumptions}
Briefly, we assume the following. At any time the signal vector $x_t$ is a sparse vector with support set ${\cal N}_t$ of size $S$ or less. At most $S_a$ elements get added to the support at each time $t$ and at most $S_a$ elements get removed from it. At time $t={t_j}$, a new element $j$ gets added at an initial magnitude $a_{j}$, and its magnitude increases for the next $d_j \ge d_{\min}$ time units. Its magnitude increase at time $\tau$ (for any ${t_j} < \tau \le t_j+d_j$ is $r_{j,\tau}$. Also, at each time $t$, at most $S_a$ elements out of the ``large elements" set (defined in the signal model) leave the set and begin to decrease. These elements keep decreasing and get removed from the support in at most $b$ time units. In the model as stated above, we are implicitly allowing an element $j$ to get added to the support at most once. In general, $j$ can get added, then removed and then added again. To allow for this, we let $\mathbf{t_j}$ be the {\em set} of time instants at which $j$ gets added; we replace $a_j$ by $a_{j,t}$ and we replace $d_{j}$ by $d_{j,t}$ (both of which are nonzero only for $t \in \mathbf{t_j}$).

As demonstrated in Section \ref{model_veri}, the above assumptions are practically valid for MRI sequences. 

\begin{sigmodel}
Assume the following.
\ben
\item At the initial time, $t=0$, the support set, ${\cal N}_0$, contains $S_0$ nonzero elements, i.e. $|{\cal N}_0| = S_0$.

\item At time $t$, $S_{a,t}$ elements are added to the support set. Denote this set by $\Aset_t$. At time $t$, a new element $j$ gets added to the support at an initial magnitude $a_{j,t}$ and its magnitude increases for at least the next $d_{\min}>0$ time instants. At time $\tau$ (for $t < \tau \le t+ d_{\min}$), the magnitude of element $j$ increases by $r_{j,\tau} \geq 0$.
    \bi
    \item $a_{j,t}$ is nonzero only if element $j$ got added at time $t$, for all other times, we set it to zero. 
    \ei

\item We define the ``large set" as $$\Lset_t:=\{j \notin \cup_{\tau=t-d_{\min}+1}^t \Aset_\tau: |(x_t)_j| \ge \ell \},$$ for a given constant $\ell$. Elements in $\Lset_{t-1}$ either remain in $\Lset_{t}$ (while increasing or decreasing or remaining constant) or decrease enough to leave $\Lset_{t}$.

\item At time $t$, $S_{d,t}$ elements out of $\Lset_{t-1}$ decrease enough to leave $\Lset_{t-1}$. Denote this set $\Bset_t$. All these elements continue to keep decreasing and become zero (removed from support) within at most $b$ time units. Also, at time $t$, $S_{r,t}$ elements out of these decreasing elements are removed from the support. Denote this set by $\Rset_t$.

\item At all times $t$, $0 \le S_{a,t} \le S_a$, $0 \le S_{d,t} \le \min\{S_a,|\Lset_{t-1}|\}$, $0\leq S_{r,t}\leq S_a$  and the support size, $S_t:=|{\cal N}_t| \le S$ for constants $S$ and $S_a$ such that $S + S_a \le m$.
\een
\label{sigmodgen}
\end{sigmodel}

\begin{figure*}
\centering
\begin{tikzpicture}[xscale=1.5,yscale=1.5, line width=0.8pt]

\draw (2*\w,2*\a+\x+\y) rectangle (2*\w+\x,2*\a+2*\x+\y)node[pos=.5] {$\geq \ell$};
\filldraw(2*\w+0.5*\x,2*\a+\x+0.25*\y) circle (2pt);
\filldraw(2*\w+0.5*\x,2*\a+\x+.5*\y) circle (2pt);
\filldraw(2*\w+0.5*\x,2*\a+\x+.75*\y) circle (2pt);
\draw (2*\w,2*\a) rectangle (2*\w+\x,2*\a+\x)node[pos=.5] {$\geq \ell$};
\draw[decorate,decoration={brace,mirror,raise=2pt}] (2*\w+\x,2*\a+0.1*\x) -- (2*\w+\x,2*\a+2*\x+\y-0.1*\x);
\node (c) at (2*\w+1.1*\x,2*\a+\x+.5*\y)[right, align=left]{$\mathcal{L}_{t+1}$,\\ $|\mathcal{L}_{t+1}|\geq S_0 - (\frac{b+1}{2}+d_0)S_a$};

\draw (2*\w,\a+\x+\y) rectangle (2*\w+\x,\a+2*\x+\y)node[pos=.5] {$>0$};
\filldraw(2*\w+0.5*\x,\a+\x+0.25*\y) circle (2pt);
\filldraw(2*\w+0.5*\x,\a+\x+.5*\y) circle (2pt);
\filldraw(2*\w+0.5*\x,\a+\x+.75*\y) circle (2pt);
\draw (2*\w,\a) rectangle (2*\w+\x,\a+\x)node[pos=.5] {$>0$};
\draw[decorate,decoration={brace,mirror,raise=2pt}] (2*\w+\x,\a+0.1*\x) -- (2*\w+\x,\a+2*\x+\y-0.1*\x);
\node (b) at (2*\w+1.1*\x,\a+\x+.5*\y)[right, align=left]{$\mathcal{N}_{t+1} \setminus \mathcal{L}_{t+1},$\\ $  |\mathcal{N}_{t+1} \setminus \mathcal{L}_{t+1}|\leq (\frac{b+1}{2}+d_0)S_a$};

\draw (2*\w,\x+\y) rectangle (2*\w+\x,2*\x+\y)node[pos=.5] {0};
\filldraw(2*\w+0.5*\x,\x+0.25*\y) circle (2pt);
\filldraw(2*\w+0.5*\x,\x+.5*\y) circle (2pt);
\filldraw(2*\w+0.5*\x,\x+.75*\y) circle (2pt);
\draw (2*\w,0) rectangle (2*\w+\x,\x)node[pos=.5] {0};
\draw[decorate,decoration={brace,mirror,raise=2pt}] (2*\w+\x,0.1*\x) -- (2*\w+\x,2*\x+\y-0.1*\x);
\node (a) at (2*\w+1.1*\x,\x+.5*\y)[right]{$\mathcal{N}_{t+1}^c, |\mathcal{N}_{t+1}^c|= m-S_{t+1}$};

\draw (2*\w,0) rectangle (2*\w+\x,2*\a+2*\x+\y);


\draw (\w,2*\a+\x+\y) rectangle (\w+\x,2*\a+2*\x+\y)node[pos=.5] {$\geq \ell$};
\filldraw(\w+0.5*\x,2*\a+\x+0.25*\y) circle (2pt);
\filldraw(\w+0.5*\x,2*\a+\x+.5*\y) circle (2pt);
\filldraw(\w+0.5*\x,2*\a+\x+.75*\y) circle (2pt);
\draw (\w,2*\a) rectangle (\w+\x,2*\a+\x)node[pos=.5] {$\geq \ell$};
\draw[decorate,decoration={brace,raise=2pt}] (\w,2*\a+0.1*\x) -- (\w,2*\a+2*\x+\y-0.1*\x);
\node (c) at (\w-.1*\x,2*\a+\x+.5*\y)[left, align=right]{$\mathcal{L}_t,$ \\ $  |\mathcal{L}_t|\geq S_0 -(\frac{b+1}{2}+d_0)S_a$};

\draw (\w,\a+\x+\y) rectangle (\w+\x,\a+2*\x+\y)node[pos=.5] {$>0$};
\filldraw(\w+0.5*\x,\a+\x+0.25*\y) circle (2pt);
\filldraw(\w+0.5*\x,\a+\x+.5*\y) circle (2pt);
\filldraw(\w+0.5*\x,\a+\x+.75*\y) circle (2pt);
\draw (\w,\a) rectangle (\w+\x,\a+\x)node[pos=.5] {$>0$};
\draw[decorate,decoration={brace,raise=2pt}] (\w,\a+0.1*\x) -- (\w,\a+2*\x+\y-0.1*\x);
\node (b) at (\w-.1*\x,\a+\x+.5*\y)[left, align=right]{$\mathcal{N}_t \setminus \mathcal{L}_t,$\\ $ |\mathcal{N}_t \setminus \mathcal{L}_t|\leq (\frac{b+1}{2}+d_0)S_a$};

\draw (\w,\x+\y) rectangle (\w+\x,2*\x+\y)node[pos=.5] {0};
\filldraw(\w+0.5*\x,\x+0.25*\y) circle (2pt);
\filldraw(\w+0.5*\x,\x+.5*\y) circle (2pt);
\filldraw(\w+0.5*\x,\x+.75*\y) circle (2pt);
\draw (\w,0) rectangle (\w+\x,\x)node[pos=.5] {0};
\draw[decorate,decoration={brace,raise=2pt}] (\w,0.1*\x) -- (\w,2*\x+\y-0.1*\x);
\node (a) at (\w-.1*\x,\x+.5*\y)[left]{$\mathcal{N}_t^c, |\mathcal{N}_t^c|= m-S_t$};

\draw (\w,0) rectangle (\w+\x,2*\a+2*\x+\y);

\node (left) at (\w+0.5*\x,-0.5*\x){\huge$x_t$};
\node (left) at (2*\w+0.5*\x,-0.5*\x){\huge$x_{t+1}$};

\draw[dashed] (\w-5.5*\x,\a) rectangle (\w+1.5*\x,2*\a+2*\x+\y);
\node (right) at (\w-3.5*\x,2*\a){\Large $\mathcal{N}_t$};

\draw[dashed] (2*\w-0.5*\x,\a) rectangle (2*\w+7.3*\x,2*\a+2*\x+\y);
\node (right) at (2*\w+5*\x,2*\a){\Large $\mathcal{N}_{t+1}$};


\draw[red,dash pattern=on 10pt off 10pt,decoration={markings,mark=at position 1 with {\arrow[scale=3]{>}}},postaction={decorate},shorten >=0.4pt](\w+\x,\x+.5*\y)--(2*\w,\x+.5*\y)node[pos=.5,above]{$m-S_t-S_{a,t+1}$};
\draw[red,dash pattern=on 10pt off 10pt,decoration={markings,mark=at position 1 with {\arrow[scale=3]{>}}},postaction={decorate},shorten >=0.4pt](\w+\x,\x+.5*\y)--(2*\w,\a+\x+.5*\y)node[pos=.3,above,sloped,align=left]{$S_{a,t+1}$};

\draw[blue,decoration={markings,mark=at position 1 with {\arrow[scale=3]{>}}},postaction={decorate},shorten >=0.4pt](\w+\x,\a+\x+.5*\y)--(2*\w,\x+.5*\y)node[pos=.3,above,sloped]{};
\draw[blue,decoration={markings,mark=at position 1 with {\arrow[scale=3]{>}}},postaction={decorate},shorten >=0.4pt](\w+\x,\a+\x+.5*\y)--(2*\w,\a+\x+.5*\y)node[pos=.3,above,sloped]{};
\draw[blue,decoration={markings,mark=at position 1 with {\arrow[scale=3]{>}}},postaction={decorate},shorten >=0.4pt](\w+\x,\a+\x+.5*\y)--(2*\w,2*\a+\x+.5*\y)node[pos=.3,above,sloped]{};

\draw[line width=2pt,dotted,decoration={markings,mark=at position 1 with {\arrow[scale=1.5]{>}}},postaction={decorate},shorten >=0.4pt](\w+\x,2*\a+\x+.5*\y)--(2*\w,2*\a+\x+.5*\y)node[pos=.5,above,sloped]{$|\mathcal{L}_t|-S_{d,t+1}$};
\draw[line width=2pt,dotted,decoration={markings,mark=at position 1 with {\arrow[scale=1.5]{>}}},postaction={decorate},shorten >=0.4pt](\w+\x,2*\a+\x+.5*\y)--(1.5*\w,2*\a+.5*\y) node[pos=.7,above,sloped,align=left]{$S_{d,t+1}$};
\draw[line width=2pt,dotted,decoration={markings,mark=at position 1 with {\arrow[scale=1.5]{>}}},postaction={decorate},shorten >=0.4pt](1.5*\w,2*\a+.5*\y)--(2*\w,\a+\x+.5*\y)node[pos=.3,above,sloped]{};
\draw[line width=2pt,dotted,decoration={markings,mark=at position 1 with {\arrow[scale=1.5]{>}}},postaction={decorate},shorten >=0.4pt](1.5*\w,2*\a+.5*\y)--(2*\w,\x+.5*\y)node[pos=.3,above,sloped]{};

\end{tikzpicture}
\caption{Signal Change Assumptions \ref{sigmodgen} (Values inside rectangular denote magnitudes.)}
\label{fig_sigmodgen}
\end{figure*}
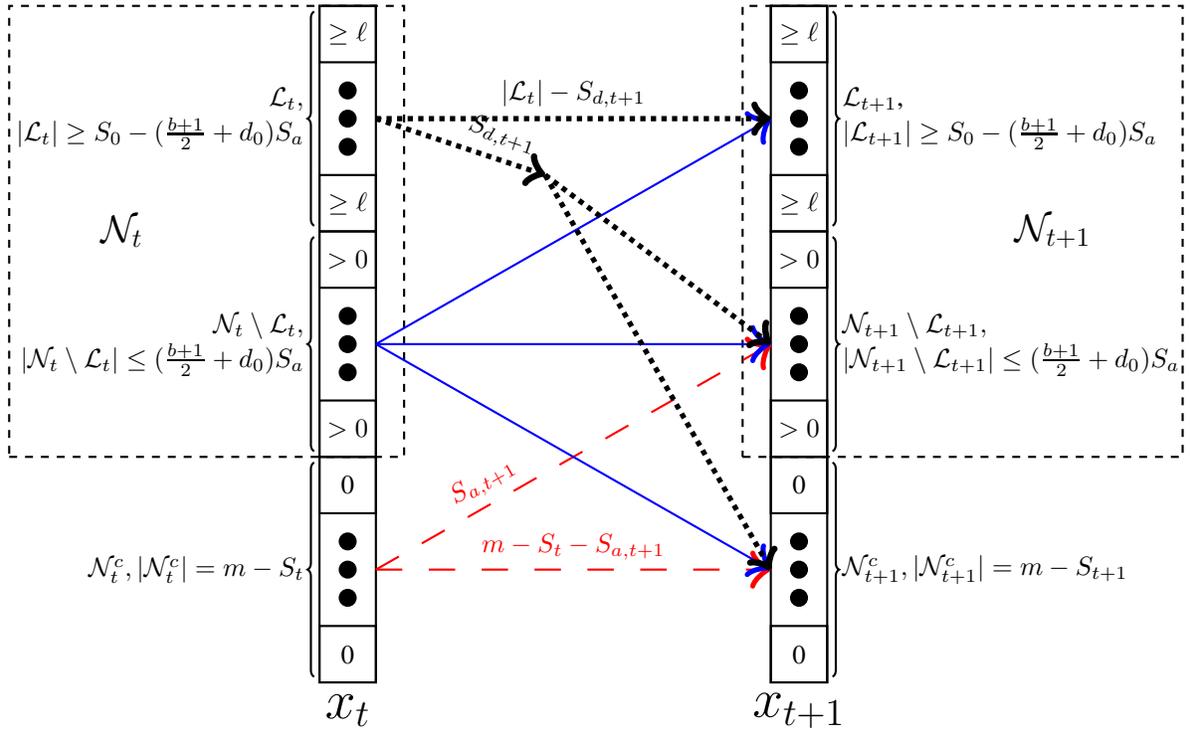

Fig.\ref{fig_sigmodgen} illustrates the above assumptions. We should reiterate that the above is not a generative model. It is only a set of assumptions on signal change. One possible generative model that satisfies these assumptions is given in Appendix \ref{sigmodgengen}.

\begin{remark}
It is easy to see that Signal Change Assumptions \ref{sigmod2} are a special case of Signal Change Assumptions \ref{sigmodgen} with $a_{j,t}=r_{j,t}=r$, $d_{\min}=d$, $b=d$, $S_0=S$, $S_{a,t}=S_{d,t}=S_{r,t}=S_a$, $\ell = dr$.
\label{special}
\end{remark}

From the above assumptions, the newly added elements' set $\Aset_t: = \Nset_t \setminus \Nset_{t-1}$; the newly removed elements' set $\Rset_t: = \Nset_{t-1} \setminus \Nset_{t}$; the set of elements that begin to start decreasing at $t$, $\Bset_t: = \Lset_{t-1} \setminus \Lset_t$.
Define the following sets:
the set of increasing (actually non-decreasing) elements at $t$,
$$\Iset_t:= \{j \in \Nset_t:|(x_t)_j| \geq |(x_{t-1})_j| \};$$
and the set of small and decreasing elements,
$$\SDset_t: = \Lset_t^c \cap |\{i  \in \Nset_t: 0 < |(x_{t})_i| < |(x_{t-1})_i| \}|.$$
Notice that $\Iset_t$ also includes $j$ if its magnitude does not change from $t-1$ to $t$. 


Condition 2 of the above model implies that (i) $|\Aset_t| = S_{a,t}$; (ii) if $j \in \Aset_{t-t_0}$ (i.e. if $j$ is added at $t-t_0$) for a $t_0 \le d_{\min}$, then  $|(x_t)_j| = a_{j,t-t_0} + \sum_{\tau = t-t_0+1}^{t} r_{j,\tau}$; and (iii) $\Aset_t \subseteq \Iset_t \cap \Iset_{t+1} \dots \cap \Iset_{t+d_{\min}}$ (all newly added elements increase for at least $d_{\min}$ time instants).

Condition 3 implies that $\Lset_{t-1} \subseteq \Lset_t \cup \SDset_t$. It also implies that $ ( \cup_{\tau=t-d_{\min}+1}^t\Aset_{\tau} ) \cap \Lset_t = \emptyset$. This, along with condition 2 means that  $\cup_{\tau=t-d_{\min}}^t\Aset_{\tau}\subseteq \Iset_t$.

Condition 4 implies that $|\Bset_t|= S_{d,t}$; $\Lset_{t-1}\setminus \Bset_t \subseteq \Lset_t$;  $\SDset_t = \SDset_{t-1} \cup \Bset_t \setminus \Rset_{t}$; $\sum_{\tau=1}^{t} S_{r,\tau} \ge \sum_{\tau=1}^{t-b} S_{d,\tau}$;  $|\SDset_t| \le \sum_{\tau = t-b+1}^t S_{d,\tau}$; and  $|\Rset_t| = S_{r,t}$.

Condition 5, along with the above, implies that $|\SDset_t| \le b S_a$.

Finally, it is easy to see that ${\cal N}_t = \Iset_t \cup \Lset_t \cup \SDset_t$. The sets $\Iset_t$, $\Lset_t$ are not disjoint, but both are disjoint with $\SDset_t$.

The above model tells us the following. Consider an element $j$ that got added at time $t$, i.e. $j \in \Aset_t$. At $\tau = t,t+1,... t+d_{\min}-1$, $j \in \Iset_\tau$ and $j \notin \Lset_\tau$. At $\tau = t+d_{\min}$, $j \in \Iset_\tau$; if $|(x_\tau)_j| \ge \ell$ then $j \in \Lset_\tau$ as well. For $\tau > t+d_{\min}$, what happens depends on $\tau-1$. If $j \in \Lset_{\tau-1}$, then either $j \in \Lset_\tau$ or it decreases enough to enter the small and decreasing set, i.e. $j \in \Bset_\tau \subseteq \SDset_\tau$. If $j \in \SDset_{\tau-1}$, then either it keeps decreasing or gets removed, i.e. either $j \in \SDset_\tau$ or $j \in \Rset_\tau \subseteq \Nset_\tau^c$. If $j \in \Lset_{\tau-1}^c \cap \Iset_{\tau-1}$, then, if $|(x_\tau)_j| \ge \ell$ then $j \in \Lset_\tau \cap \Iset_{\tau}$, else $j \in  \Lset_\tau^c \cap \Iset_{\tau}$.

We now discuss sufficient conditions for condition 5 of the signal model to hold.

\begin{remark}
Since $S_t = S_{t-1} + S_{a,t} - S_{r,t} = S_0 + \sum_{\tau=1}^{t} S_{a,\tau} - \sum_{\tau=1}^{t} S_{r,\tau}$, thus, $S_t \le S$ holds if $S_0 \le S$ and $ \sum_{\tau=1}^{t} S_{a,\tau} \le \sum_{\tau=1}^{t-b} S_{d,\tau}$. 
\label{rem_St}
\end{remark}

Notice that an element $j$ could get added, then removed and added again later. Let
$$\mathbf{t_j}:=\{t: a_{j,t}\neq 0 \}$$
denote the set of time instants at which $j$ gets added. Clearly, $\mathbf{t_j}=\emptyset$  if $j$ never got added.
Let
$$a_{\min}:= \min_{j:\mathbf{t_j}\neq \emptyset} \min_{t \in \mathbf{t_j}, t \neq 0} a_{j,t}$$
denote the minimum of $a_{j,t}$ over all elements $j$ that got added at $t>0$. We are excluding coefficients that never got added and those that got added at $t=0$.
Let
$$r_{\min}(d):= \mathop{\min}_{j: \mathbf{t_j}\neq \emptyset} \min_{t \in \mathbf{t_j}, t \neq 0}\min_{\tau \in[t+1, t+d]} r_{j,\tau}$$
denote the minimum, over all elements $j$ that got added at $t>0$, of the minimum of $r_{j,\tau}$ over the first $d$ time instants after $j$ got added.

Define
\bea
\ell:=a_{\min}+d_{\min}r_{\min}(d_{\min}).
\label{def_ell}
\eea
With $\ell$ defined this way, clearly, $\Nset_t = (\cup_{\tau=t-d_{\min}+1}^t \Aset_\tau)  \cup \Lset_t \cup \SDset_t$ where the three sets are mutually disjoint.

Also, with $\ell$ as above, it is clear that for $t>d_{\min}$, $\Lset_t = \Lset_{t-1} \cup \Aset_{t-d_{\min}-1} \setminus \Bset_t$, and for $t \le d_{\min}$, $\Lset_t = \Lset_{t-1} \setminus \Bset_t$. Here, by definition, $\Lset_{t-1}$ and $\Aset_{t-d_{\min}-1}$ are disjoint and $\Bset_t \subseteq \Lset_{t-1}$.
Thus,
$$|\Lset_t| = |\Lset_0| + \sum_{\tau = 1}^{t-d_{\min}}S_{a,\tau} - \sum_{\tau=1}^t S_{d,\tau}$$
Also notice that $|\Lset_0| \le S_0$. Using these facts and Remark \ref{rem_St}, we can conclude the following.

\begin{remark} 
Let $\ell:=a_{\min}+d_{\min}r_{\min}(d_{\min})$. Then, condition 5 of Signal Change Assumptions \ref{sigmodgen} holds if
\ben
\item $0 \le S_{a,t} \le S_a$ and $0 \le S_{d,t} \le S_a$,
\item $(d_{\min}+b+1) S_a \le |\Lset_0| \le S_0 \le S$,
and
\item $ \sum_{\tau=1}^{t} S_{a,\tau} \le \sum_{\tau=1}^{t-b} S_{d,\tau} \le |\Lset_0| + \sum_{\tau=1}^{t-b-d_{\min}-1} S_{a,\tau}$.
\een
The leftmost lower bound of the second condition ensures that the upper bound of the third condition is not smaller than the lower bound. The upper bound of the third condition ensures that $S_{d,t} \le |\Lset_{t-1}|$ always (it is actually written to ensure $S_{d,t-b} \le  |\Lset_{t-b-1}|$). $S_0 \le S$ and the lower bound of the third condition ensures that $S_t \le S$ (as explained in Remark \ref{rem_St}).
\end{remark}

A simpler sufficient condition is as follows. 
\begin{remark}
Let $\ell:=a_{\min}+d_{\min}r_{\min}(d_{\min})$. Then, condition 5 of Signal Change Assumptions \ref{sigmodgen} holds if $(d_{\min}+b+1) S_a \le |\Lset_0| \le S_0 \le S$; $S_{d,t}=S_a$ for all $t$; and for $1 \le t \le b$, $S_{a,t} = 0$, and for $t>b$, $S_{a,t}=S_a$.
\end{remark}

In the above model, we only assume that all coefficients will get removed in at most $b$ time units. However, it can happen that some coefficients get removed earlier than that and hence it is fair to include this in the signal model.  We do this below.
\begin{sigmodel}
Assume Signal Change Assumptions \ref{sigmodgen} with the following extra assumption.
\bi
\item Out of the $S_{d,t}$ elements that started decreasing at time $t$, at least $\frac{\tau}{b}  S_{d,t}$ of them get removed by $t+\tau$ for $\tau < b$.
\ei
\label{sigmodgen1}
\end{sigmodel}
All implications of the above model are the same as those of Signal Change Assumptions \ref{sigmodgen}, except that now, $|\SDset_t| \le S_{d,t} + \frac{b-1}{b}  S_{d,t-1} + \dots \frac{1}{b}  S_{d,t-b+1} \le \frac{b+1}{2}S_a$; while for Signal Change Assumptions \ref{sigmodgen}, $|\SDset_t| \le b S_a$.

\subsection{Modified-CS Stability Result}
For the above signal model, we can claim the following.
\begin{theorem}
Consider Algorithm \ref{modcsalgo}. Assume Signal Change Assumptions \ref{sigmodgen1} on $x_t$. Also assume that $y_t$ satisfies (\ref{obsmod}). Assume that Assumption \ref{spread1} holds.
 If there exists a $d_0 \le d_{\min}$ such that the following hold:
\ben
\item algorithm parameters
\ben
\item  $\alpha = \frac{\zeta_M}{\sqrt{S_a}} 7.50 \eps$,
\een
\item number of measurements
\ben
\item  $\delta_{S+ 3(\frac{(b+1)}{2} + d_0 + 1)S_a} \le 0.207$,
\een
\item initial magnitude and magnitude increase rate:
\bea
&& \min\{\ell,\min_{j: \mathbf{t_j} \neq \emptyset} \min_{t \in \mathbf{t_j}} (a_{j,t}+\sum_{\tau=t+1}^{t+d_0}r_{j,\tau}) \} \nn \\
&& > \alpha + \frac{\zeta_M}{\sqrt{S_a}} 7.50 \eps, \nn
\eea
\item at $t=0, n_0$ is large enough to ensure that $|\tDelta_t| \leq \frac{b+1}{2}S_a + d_0S_a, |\tDelta_{e,t}|=0$,
\een
then, for all $t$,
\ben
\item $|{\tDelta}_t| \le \frac{(b+1)}{2} S_a + d_0 S_a$, $|{\tDelta}_{e,t}|= 0$, $|\tT_t| \le S$,
\item $|{\Delta}_t| \le \frac{(b+1)}{2} S_a + d_0 S_a + S_a$, $|\Tset_t| \le S$, $|\Delta_{e,t}|\le S_a$,
\item and $\|x_t - \xhat_t\| \le 7.50 \eps$
\een
\label{modcsthm}
\end{theorem}
Proof: See Appendix \ref{modcs_sta_proof}. 

\begin{corollary}  \label{cor_modcsthm}
Under Signal Change Assumptions \ref{sigmodgen}, the result of Theorem \ref{modcsthm} changes in the following way: replace $\frac{(b+1)}{2} S_a$ by $b S_a$ everywhere in the result.
\end{corollary}


\begin{remark}
Condition 4 of the above result is not restrictive. It is easy to see that it will hold if $\delta_{2S}(A_0)\le 0.207$ and if $|\Lset_0| \ge [S_0 - (\frac{(b+1)}{2} S_a + d_0 S_a)]$.
\label{n_0}
\end{remark}

\begin{remark} \label{simp_suff}
A simpler sufficient condition for condition 3 is: $\min(\ell, a_{\min} + d_0 r_{\min}(d_0)) > \alpha + \frac{\zeta_M}{\sqrt{S_a}} 7.50 \eps$.
\end{remark}

\subsection{Modified-CS-Add-LS-Del Stability Result}
Finally we study Modified-CS-Add-LS-Del.

\begin{theorem}
Consider Algorithm \ref{modcsalgo_2}. Assume Signal Change Assumptions \ref{sigmodgen1} on $x_t$. Also assume that $y_t$ satisfies (\ref{obsmod}). Assume that Assumption \ref{spread2} holds.
If there exists a $d_0 \le d_{\min}$ such that the following hold:
\ben
\item algorithm parameters
\ben
\item  $\alpha_{\add}$ is large enough so that there are at most $f$ false adds at time $t$, i.e. $|\Ahat_t\setminus {\cal N}_t| \le f$
\item $\alpha_{\text{del}} = 1.12\frac{\zeta_L}{\sqrt{S_a}} \eps + 0.261 \zeta_L h$,
where $h^2=(\frac{(b+1)}{2}+d_0)(\alpha_{\add} + \frac{\zeta_M}{\sqrt{S_a}}7.50  \eps)^2$
\een

\item number of measurements
\ben
\item $\delta_{S + 3 (\frac{(b+1)}{2} S_a + d_0 S_a + S_a)} \leq 0.207$
\item $\delta_{S + S_a + f} \leq  0.207$
\item $\theta_{S + S_a + f,\frac{(b+1)}{2} S_a + d_0 S_a}\leq 0.207$
\label{nummea}
\een

\item initial magnitude and magnitude increase rate: 
\bea
&& \min\{\ell,\min_{j: \mathbf{t_j} \neq \emptyset} \min_{t \in \mathbf{t_j}} (a_{j,t}+\sum_{\tau=t+1}^{t+d_0}r_{j,\tau}) \} \nn \\
&& > \max\{\alpha_{\add} + \frac{\zeta_M}{\sqrt{S_a}} 7.50 \eps, 2 \alpha_{\text{del}} \}
\label{eqdel}
\eea
\item at $t=0, n_0$ is large enough to ensure that $|\tDelta_t| \leq \frac{b+1}{2}S_a + d_0S_a, |\tDelta_{e,t}|=0$,
\een
then
\ben
\item $\tDelta_t \subseteq \SDset_{t} \cup \Aset_{t} \cup \Aset_{t-1} \dots \Aset_{t-d_0+1}$
\item $|\tDelta_t| \le\frac{(b+1)}{2} S_a + d_0 S_a$, $|\tDelta_{e,t}| = 0 $, $|\tT_t| \le S$
\item $|\Delta_t| \le\frac{(b+1)}{2} S_a + d_0 S_a + S_a$,  $|\Tset_t| \le S$  
\item $\|x_t - \xhat_{t,modcs}\| \le 7.50 \eps,$
\item $\|x_t - \xhat_{t}\| \le 1.12\eps + 1.261 \sqrt{(\frac{(b+1)}{2}+d_0)(\alpha_{\del}+7.50\epsilon)S_a}.$
\een
\label{modcsaldwkthm}
\end{theorem}
Proof: See Appendix \ref{modcsaldthmproof}.


\begin{remark}
Claims similar to Corollary \ref{cor_modcsthm} and Remarks \ref{n_0} and \ref{simp_suff} hold for the above result also.
\end{remark}

\subsection{Discussion} \label{gen_disc}

\begin{remark}
Notice that  Signal Change Assumptions \ref{sigmodgen} or \ref{sigmodgen1} allow for both slow and fast signal magnitude increase or decrease. Slow magnitude increase/decrease would happen, for example, in an imaging problem when one object slowly morphs into another with gradual intensity changes. Or, in case of brain regions becoming ``active" in response to stimuli, the activity level gradually increases from zero to a certain maximum value within a few milliseconds (10-12 frames of fMRI data), and similarly the ``activity" level decays to zero within a few milliseconds. In both of the above examples, a new coefficient will get added to the support at time $t$ at a small magnitude $a_{j,t}$ and increase by $r_{j,\tau}$ per unit time for sometime after that. Similarly for the decay to zero of the brain's activity level. On the other hand, the signal model also allows support changes resulting from motion of objects, e.g. translation. In this case, the signal magnitude changes will typically not be slow. As the object moves, a set of new pixels enter the support and another set leave. The entering pixels may have large enough pixel intensity and their intensity may never change. For our model this means that the pixel enters the support at a large enough initial magnitude  $a_{j,t}$ but its  magnitude never changes i.e. $r_{j,\tau}=0$ for all $\tau$. If all pixels exit the support without their magnitude first decreasing, then $b=1$.

The only thing that the above results (Theorem \ref{modcsthm} and \ref{modcsaldwkthm}) require is that (i) for any element $j$ that is added, either $a_{j,t}$ is large enough or $r_{j,\tau}$ is large enough for the initial few ($d_0$) time instants so that condition 3 holds; and (ii) a decaying coefficient decays to zero within a short delay, $b$.  (i) ensures that every newly added support element gets detected either immediately or within a finite delay; while (ii) ensures removal within finite delay of a decreasing element.  For the moving object case, this translates to requiring that $a_{j,t}$ be large enough. For the first two examples above, this translates to requiring that $r_{j,\tau}$ be large enough for the first few time instants after $j$ gets added and that $b$ be small enough.

Recall that $\delta_S:= \max_{t>0} \delta_S(A_t)$. Other than the above assumption, the results also need that the support estimation thresholds are set appropriately; enough number of measurements, $n_t$, are available at all times $t>0$ so that condition 2 holds (this number depends on the support size, $S$, the support change size, $S_a$ and on $b$); and condition 4 holds.
\end{remark}


For the above results, the support errors are bounded by a constant times $S_a$. Thus, under slow support change, the bound is small compared to the support size, $S_t$, making the above a meaningful result. The reconstruction error is bounded by a constant times $\eps$. Under high enough SNR, this bound is small compared to the signal power. In fact, for Signal Change Assumptions \ref{sigmodgen} or \ref{sigmodgen1}, the signal power is not bounded. 
To compare the results, let us fix some of the parameters. Suppose that
$b = 3$, $f = S_a$, $S_0=S$,  $S_{a,t} = S_{r,t} =S_{d,t}=S_a$. 
Let $d_0=2$. 
The modified-CS result says the following. If 
\ben
\item $\delta_{S + 15 S_a} \le 0.207$, and
\item $\text{LHS of condition 3} > \frac{\zeta_M}{\sqrt{S_a}} 15 \eps$,
\een
then $|\tDelta_t| \le 4 S_a$ and $|\tDelta_{e,t}|=0$ and $\|x_t - \xhat_{t,modcs}\| \le 7.50 \eps$.
The Modified-CS-add-LS-del result says the following. If
\ben
\item $\delta_{S + 15S_a} \le 0.207$(the other two conditions are implied by this), and 
\item $\text{LHS of condition 3} > \max(\alpha_{\add} + \frac{\zeta_M}{\sqrt{S_a}}7.50 \eps, 2.24 \frac{\zeta_L}{\sqrt{S_a}} \eps + 0.522 \zeta_L h )$, where $h^2=4 (\alpha_{\add} + \frac{\zeta_M}{\sqrt{S_a}}7.50 \eps)^2$.
\een
then $|\tDelta_t| \le 4S_a$ and $|\tDelta_{e,t}| = 0$ and $\|x_t - \xhat_{t,modcs}\| \le 7.50 \eps$.

The CS result from Lemma \ref{cs_bnd} says the following. If
\ben
\item $\delta_{2S} \le 0.207$
\een
then $\|x_t - \xhat_{t,cs}\| \le 7.50 \eps$.

Thus, both modified-CS and modified-CS-add-LS-del need the same restricted isometry condition (condition on the number of measurements). Under the slow support change assumption,  $S_a\ll S_t \le S$. In this case, both the modified-CS algorithms hold under a weaker restricted isometry condition (potentially fewer number of measurements required) than what noisy $\ell_1$ needs for the same error bound.
%
Next we compare the lower bounds on the LHS of condition 3 needed by modified-CS and by modified-CS-add-LS-del. This requires knowing $\zeta_M$ and $\zeta_L$. To get an idea of the values of $\zeta_M$ and $\zeta_L$, we did simulations based on Signal Change Assumptions \ref{sigmodgen} with $S=0.1m, S_{a,t} = S_{d,t}=S_{r,t}=S_a=0.01m, b=d_{\min}=3, r_{j,t}=1, a_{j,t}=1$ (we generated it using the generative model given in Appendix A of \cite{stability_allerton}). The measurement matrices $A_t$ were zero mean random Gaussian $n_t\times m$ matrices with columns normalized to unit norm. For $t=0$, $n_0=160$; for $t> 0$, $n_t=n=57$. The measurement noise, $(w_t)_j \sim^{i.i.d.} uniform(-c_t,c_t)$ for $1\leq j\leq m$. For $t=0$, $c_t=0.01266$; for $t > 0$, $c_t=0.1266$. We used the same measurement Gaussian matrix $A$ for $t >0$. We generated 500 realizations respectively with different choices of $m$, and used both algorithms for reconstruction. When $m=200$, we got, $\zeta_M=0.9328\sqrt{S_a}, \zeta_L=0.8734\sqrt{S_a}$; when $m=1000$, $\zeta_M=0.8295\sqrt{S_a}, \zeta_L=0.8628\sqrt{S_a}$; when $m=2000$, $\zeta_M=0.8497\sqrt{S_a}, \zeta_L= 0.8628\sqrt{S_a}$.

For our comparison, we pick the largest values we got from the above experiment: let $\zeta_M=0.9328\sqrt{S_a}$ and $\zeta_L=0.8734\sqrt{S_a}$. With these values, modified-CS needs $\text{LHS of condition 3} > 13.99\epsilon$ and modified-CS-Add-LS-Del needs $\text{LHS of condition 3}  > \max\{\alpha_{\add}+7.00\epsilon, 10.978 \eps + 3.246\alpha_{\add} \}= 10.978\epsilon + 3.246\alpha_{\add}$. With $\alpha_{\add}$ small enough, clearly modified-CS-add-LS-del requires a weaker assumption. As explained earlier and also in \cite{stability_allerton}, $\alpha_{\add}$ is a small threshold that is typically proportional to the noise bound $c, i.e., \eps/\sqrt{n}$. Thus the mod-CS-Add-LS-Del condition is weaker.

The comparison between modified-CS and modified-CS-add-LS-del above is not as clear-cut as that in the simple model case (Signal Change Assumptions \ref{sigmod2}). The reason is that the simple model tells us exactly how many support additions and removals occur at each time; and it also tells us the exact number of elements with a certain magnitude. As a result, it is possible to get a better bound on $\|x_{\Delta_{t,\add}}\|_2$: this is needed to bound the LS step error. The LS error decides the value of $\alpha_\del$ and $\alpha_\del$, in turn, decides the lower bound on the LHS of condition 3. The current Signal Change Assumptions \ref{sigmodgen} or \ref{sigmodgen1} are much more flexible, but this also means that they not give us exact magnitude information. As a result, the bounds are looser and so the advantage of  modified-CS-add-ls-del is not demonstrated as clearly.


\begin{remark}
Finally, we explain why condition 1a of Theorem \ref{modcsaldwkthm} is stated the way it is. Because of how the modified-CS error is bounded, we cannot get a bound on the reconstruction error for the $j^{th}$ coefficient, $|(\xhat_{t})_j - (x_{t})_j|$. We can only bound this error by its infinity norm. Thus, the only way to get an explicit value for $\alpha_{\add}$ is to let it equal the upper bound on $\|\xhat_{t} - x_{t}\|_\infty$ and this will ensure $f=0$ false adds. However, the key point of the add-LS-del procedure is that one can pick an addition threshold that is smaller than this but results in some false adds, $f$. As long as  $f$ is small enough so that $A_{T_{\add}}$ is well conditioned (condition 2b holds), the LS step error will be much smaller. With $\alpha_{\del}$ chosen appropriately, one can still delete all of these false adds (as well as all elements of the removed set) in the deletion step.
\end{remark}

\subsection{Comparison with the LS-CS result of \cite{just_lscs}} \label{compare_lscs}
In \cite{just_lscs}, we obtained a stability result for LS-CS which was a worse algorithm than modified-CS: it required  stronger conditions for exact recovery, and was worse is simulation experiments as shown in \cite{modcsjp, stability_allerton}.
The same signal model and the same strategy as that of \cite{just_lscs} can be used for modified-CS as well and we will, in fact, get a stronger stability result for it: the modified-CS result will not need condition 3b of the LS-CS stability result (Theorem 2 of \cite{just_lscs}). 

The most important difference between the LS-CS result from \cite{just_lscs} and our results is that \cite{just_lscs} assumed $S_a$ support changes every $p$ frames and the result required a lower bound on $p$. With this, one could ensure that all newly added support elements got detected before the next support change time. 
This meant that one could delete the false adds and removals after all new adds got detected, but before the next change time. At this time, the signal recovery is very accurate (because of zero misses) and hence, for the result of \cite{just_lscs}, a very small deletion threshold could suffice. However, as explained earlier (see Fig \ref{suppchange}), support change every so often is not a practically valid assumption in most applications. In this work, we allow the support to change at every time which is more realistic, but is also more difficult to analyze. With this, one always has some misses at each time instant (except in the simplest case where all new elements are added at very large magnitudes). Thus, one cannot wait for all the missed elements to get detected before deleting the false adds and removals and hence one requires a larger deletion threshold.

A third difference is that the signal change model of \cite{just_lscs} fixed the number of support additions and removals at each time to be just $S_a$; it fixed the initial magnitude and the rate of magnitude increase for a new support element $j$ to both be $a_j$ at all times; and, for decreasing coefficients, it assumed a very specific and fixed rate of magnitude decrease. None of these is a very practical assumption. Our realistic signal change models (Signal Change Assumptions \ref{sigmodgen} or \ref{sigmodgen1}) allow all these things to vary with time.

%

\section{Model Verification}
\label{model_veri}
We verified that two different types of MRI image sequences -- a larynx (vocal tract) MRI sequence and a brain functional MRI sequence -- do indeed satisfy Signal Change Assumptions \ref{sigmodgen}. First we describe model verification for the larynx sequence. We used a 10 frame sequence and extracted out a 36x36 region of this sequence selected as the region that includes the part where most of the changes were visible. As shown in earlier work \cite{modcsjp}, this sequence is approximately sparse in the 2D discrete wavelet transform (DWT) domain. A two level db4 wavelet was used there. We computed this 2D DWT, re-arranged it as a vector and computed its 99.9\% energy support set. All elements not in this set were set to zero. This gave us an exactly sparse sequence $x_t$. Its dimension $m = 36^2=1296$. For this sequence, we observed the following.
The support size $\Nset_t$ satisfied $|\Nset_t| \le S = 113$ for all $t$. The number of additions from $t-1$ to $t$ satisfied $|\Nset_t \setminus \Nset_{t-1}| \le 21$ and the number of removals, $|\Nset_{t-1} \setminus \Nset_t| \le  26$. Thus, $S_a = 26$. Also, the initial nonzero value, $a_{j,t}$, ranged from $13$ to $37$, the rate of magnitude increase, $r_{j,t}$, ranged from $1$ to $37$, and the duration for which the increase occurred, $d_{j,t}$, ranged from $0$ to $4$. Also, the maximum delay between the time that a coefficient began to decrease and when it was removed was $b=7$.

\begin{figure*}[!htb]
   \centerline{\subfigure[BOLD]{\includegraphics[height=5cm]{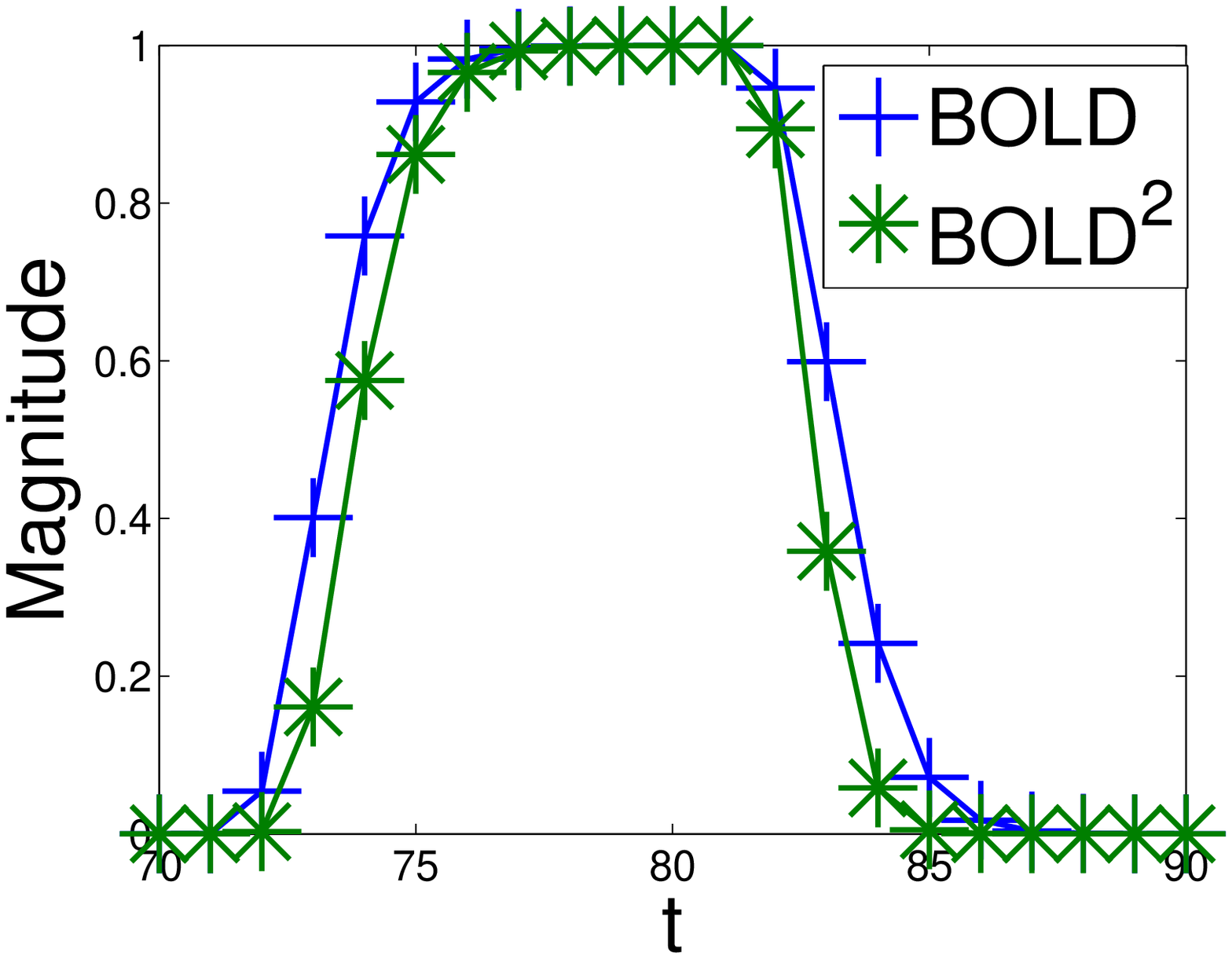}
       \label{BOLD}}
     \hfil
     \subfigure[Active and transient region]{\includegraphics[height=5cm]{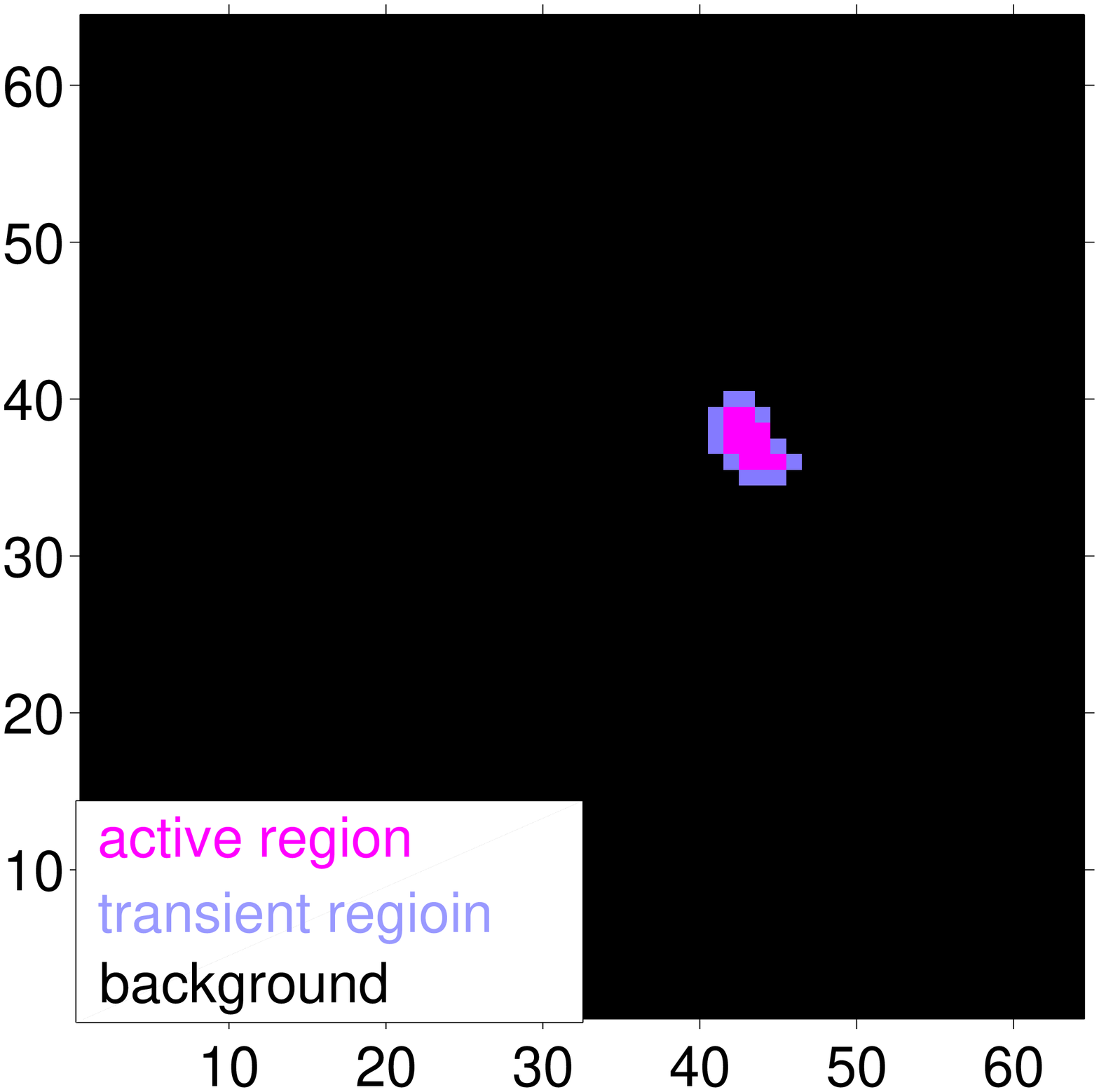}
       \label{a_t_t}}}
   \caption{(a): plot of the BOLD signal and of its square. (b): active, transient and inactive brain regions}
   \label{fig:sim}
 \end{figure*}

Next we consider a 64x64 functional MRI sequence. fMRI is a technique that is used to investigate brain function. The sequence we study here is for the brain responding to a certain type of stimulus (light being turned on and off). This sequence consisted of a rest state brain sequence to which activation was added based on the models suggested in \cite{blindestimation}. The goal is to be able to accurately extract out the activation region from this sequence. As is done in \cite{rrpcp_allerton11}, one can use the undersampled ReProCS algorithm to extract out the sparse activation regions from the low rank background brain image sequence, as long as an initial background brain training sequence is available. 
In our example, the activation started at frame 71. For the purpose of ReProCS, the active region ``image" (the image that is zero everywhere except in the active region), is the sparse signal of interest.
For a 23 pixel region that is known to correspond to the part of the brain that is affected by the above stimulus, the activation was added follows \cite{blindestimation}. The 23 pixel region was split into 2 sub-regions so that the activation intensity was smallest at the boundary of the region and slowly increased as one moved inwards. We show the 2 regions in Fig \ref{a_t_t}. $\Rset_1$ is the innermost region, $\Rset_2$ is the outermost. The activation in these regions satisfied the following model.
For $j\in \Rset_1$, $(x_{t})_j=b(t) M_a$.
For $j\in \Rset_2$, $(x_{t})_j=0.2 b(t)^2 M_a$.
Here $M_a=1783$ is the maximum magnitude in the active region and $b(t)$ is the blood oxygenation level dependent (BOLD) signal taken from \cite{blindestimation}. It is plotted in Fig \ref{BOLD}.  
This image sequence was of size 64x64, i.e. its dimension $m=64^2 = 4096$. We computed its 99.9\% energy support and set all elements not in this set to zero. This gave us our sparse sequence $x_t$. The support size of $x_t$,  $\Nset_t$, satisfied $|\Nset_t| \le S = 23$ for all $t$. The number of additions from $t-1$ to $t$ satisfied $|\Nset_t \setminus \Nset_{t-1}| \le S_a = 13$ and the number of removals, $|\Nset_{t-1} \setminus \Nset_t| \le S_a = 13$. Also, the initial nonzero value, $a_{j,t}$, ranged from $57$ to $97$, the rate of magnitude increase, $r_{j,t}$, ranged from $1$ to $637$, and the duration for which the increase occurred, $d_{j,t}$, ranged from $6$ to $7$. Also, the maximum delay between the time that a coefficient began to decrease and when it was removed was $b=7$.

\section{Setting Algorithm Parameters and Simulation Results}
\label{sims}

\subsection{Setting algorithm parameters automatically}
\label{param}
Algorithm \ref{modcsalgo} has one parameter $\alpha$. Algorithm \ref{modcsalgo_2} has two parameters $\alpha_\add$, $\alpha_\del$. We explain here how to set these thresholds automatically. It is often fair to assume that the noise bound on $\eps$ is known, e.g. it can be estimated using a short initial noise-only training sequence. We assume this here. In cases where it is not known or can change with time, one can approximate it by $\|y_{t-1} - A_{t-1} \xhat_{t-1}\|_2$ (assuming accurate recovery at $t-1$).

Define the minimum nonzero value at time $t$, $x_{\min,t}=\min_{j\in \Nset_t}|(x_t)_j|$. This can be estimated as $\xhat_{\min,t}=\min_{j\in \tT_{t-1}}|(\xhat_{t-1})_j|$.


When setting the thresholds automatically, they will change with time. We set $\alpha_{\add,t}$ using the following heuristic. By Lemma \ref{errls1}, we have $(x_t - \xhat_{t,\add})_{\Tset_{\add,t}} = ({A_{\Tset_{\add,t}}}'A_{\Tset_{\add,t}})^{-1} [ {A_{\Tset_{\add,t}}}' w_t + {A_{\Tset_{\add,t}}}' A_{\Delta_{\add,t}}  (x_t)_{\Delta_{\add,t}}]$. To ensure that this is bounded, we need $\|{A_{\Tset_{\add,t}}}^{\dag}\|$ and $\|({A_{\Tset_{\add,t}}}'A_{\Tset_{\add,t}})^{-1}\|$ to be bounded. Since $\|{A_{\Tset_{\add,t}}}^{\dag}\|=\frac{1}{\sigma_{\min}(A_{\Tset_{\add,t}})}$ and $\|({A_{\Tset_{\add,t}}}'A_{\Tset_{\add,t}})^{-1}\|=\frac{1}{\sigma_{\min}^2(A_{\Tset_{\add,t}})}$, we pick $\alpha_{\add,t}$ as smallest number such that $\sigma_{\min}({A_{\Tset_{\add,t}}})\geq 0.4$.

If one could set $\alpha_\del$ equal to the lower bound on $x_{\min,t}-\|(x_t - \xhat_{t,\add})_{\Tset_{\add,t}}\|_{\infty}$, there will be zero misses. Using this idea, we  let $\alpha_{\del,t}$ be an estimate of the lower bound of this quantity. Notice that
\bea
&&\|(x_t - \xhat_{t,\add})_{\Tset_{\add,t}}\|_{\infty}\le \| (A_{\Tset_{yy,t}}^\dag A_{\Delta_{\add}}x_{t,\Delta_{\add}} + A_{\Tset_{\add,t}}^\dag w_t\|_{\infty}\nn\\
&&\leq \|({A_{\Tset_{\add,t}}}'A_{\Tset_{\add,t}})^{-1}\|_{\infty}\|A_{\Tset_{\add,t}}A_{\Delta_{\add}}x_{t,\Delta_{\add}}\|_{\infty}+\|A_{\Tset_{\add,t}}^\dag w_t\|_{\infty}\nn\\
&&\approx \|({A_{\Tset_{\add,t}}}'A_{\Tset_{\add,t}})^{-1}\|_{\infty}C_1\theta_{|\Tset_{\add,t}|,|\Delta_{\add}|} C_2\hat{x}_{\min}+\|A_{\Tset_{\add,t}}^\dag \hat{w}_t\|_{\infty},\nn
\eea
where $C_1, C_2$ are some constant larger than 1. Here we use the fact that for any matrix $B$, $\|B\|_{\infty}\leq C_1\|B\|$ for some constant $C_1$ and that only small elements are missed and hence we can approximate $\|x_{t,\Delta_{\add}}\|_\infty$ by $C_2$ times $\xhat_{\min,t}$ where $C_2$ is a small constant larger than 1. We cannot compute $\theta_{|\Tset_{\add,t}|,\Delta_{\add}}$, but it is fair to assume that it is small (significantly smaller than one). If we assume that $$C_1C_2\|({A_{\Tset_{\add,t}}}'A_{\Tset_{\add,t}})^{-1}\|_{\infty}\theta_{|\Tset_{\add,t}|,|\Delta_{\add}|}\leq 0.3,$$
then the above bound simplifies to $0.3\xhat_{\min,t}+\|A_{\Tset_{\add,t}}^\dag \hat{w}_t\|_{\infty}$. We can approximate $\hat{w}_t$ by $y_t-A\xhat_{t,\modcs}$. Thus, we set $\alpha_{\del,t}=0.7\hat{x}_{\min,t}-\|A_{\Tset_{\add,t}}^\dag (y_t-A\xhat_{t,\modcs})\|_{\infty}$.

For Algorithm \ref{modcsalgo}, we set $\alpha_t$ as follows. If $\|x_t-\xhat_{t,\modcs}\|_{\infty}\leq Cx_{\min,t}$ for some $C<1$, then setting $\alpha_t =(1-C)x_{\min,t}$ will ensure that there are no misses. If this bound holds for most entries $i$, then most entries will be correctly recovered, i.e., there will be few misses. If we ensure $\sigma_{\min}(A_{\tT_t})\geq 0.4$ then the number of extras will be bounded. To try to ensure that both the above hold, we let $\alpha_t$ to be the smallest value such that $\min_{j\in\tT_t} |(\xhat_{t,\modcs})_j|_{j}\geq (1-C)\xhat_{\min,t}=0.5\xhat_{\min,t}$ (we pick $C=0.5$), and $\sigma_{\min}(A_{\tT_t})\geq 0.4$.

To get a more robust estimate of the minimum nonzero value of $x_t$, we use a short-time average of $\{\xhat_{\min,\tau}, t-t_0\leq \tau\leq t\}$ as the estimate of $x_{\min,t}$. In our experiments, $t_0 =10$.

\subsection{Simulation Results}
In the discussion so far, we only compared sufficient conditions required by different algorithms. The general conclusion obtained by comparing the sufficient conditions was that modified-CS-add-LS-del is the best algorithm followed by modified-CS and then noisy $\ell_1$. In this section, we use simulations to demonstrate the same thing. We compared noisy $\ell_1$ (simple CS), i.e. solution of (\ref{simpcs}) at each time instant, modified-CS(mod-CS) as given in Algorithm \ref{modcsalgo}, and modified-CS-add-LS-del (mod-CS-Add-LS-Del) as given in Algorithm \ref{modcsalgo_2}. The parameters for the algorithms were set as explained in Sec \ref{param} above.

The data was generated as follows.
We used Signal Model \ref{sigmodgen} generated as explained in Appendix \ref{sigmodgengen} with $m=200$, $S=20$, $d_{\min}=3$, $a_{\min}=r_{\min}(d_{\min})=r$, $S_a=2$, $b=3$, $\ell=a_{\min}+d_{\min}r_{\min}(d_{\min})=4r$ and $r$ was varied. The measurement matrices $A_t$ were zero mean random Gaussian $n_t\times m$ matrices with columns normalized to unit  norm. We used $n_0=160$ and $n_t=n=57$ for $t>0$. The measurement noise, $(w_t)_j \sim^{i.i.d.} uniform(-c_t,c_t)$ for $1\leq j\leq m$. For $t=0$, $c_t=0.01266$; for $t\geq 1$, $c_t=c=0.1266$. Here $\sim^{i.i.d.}$ means that $(w_t)_j$ are independent and identically distributed (i.i.d.) both for different $j$'s and for different $t$'s.

\begin{figure*}[!htb]
   \centerline{\subfigure[]{\includegraphics[width=6.5cm]{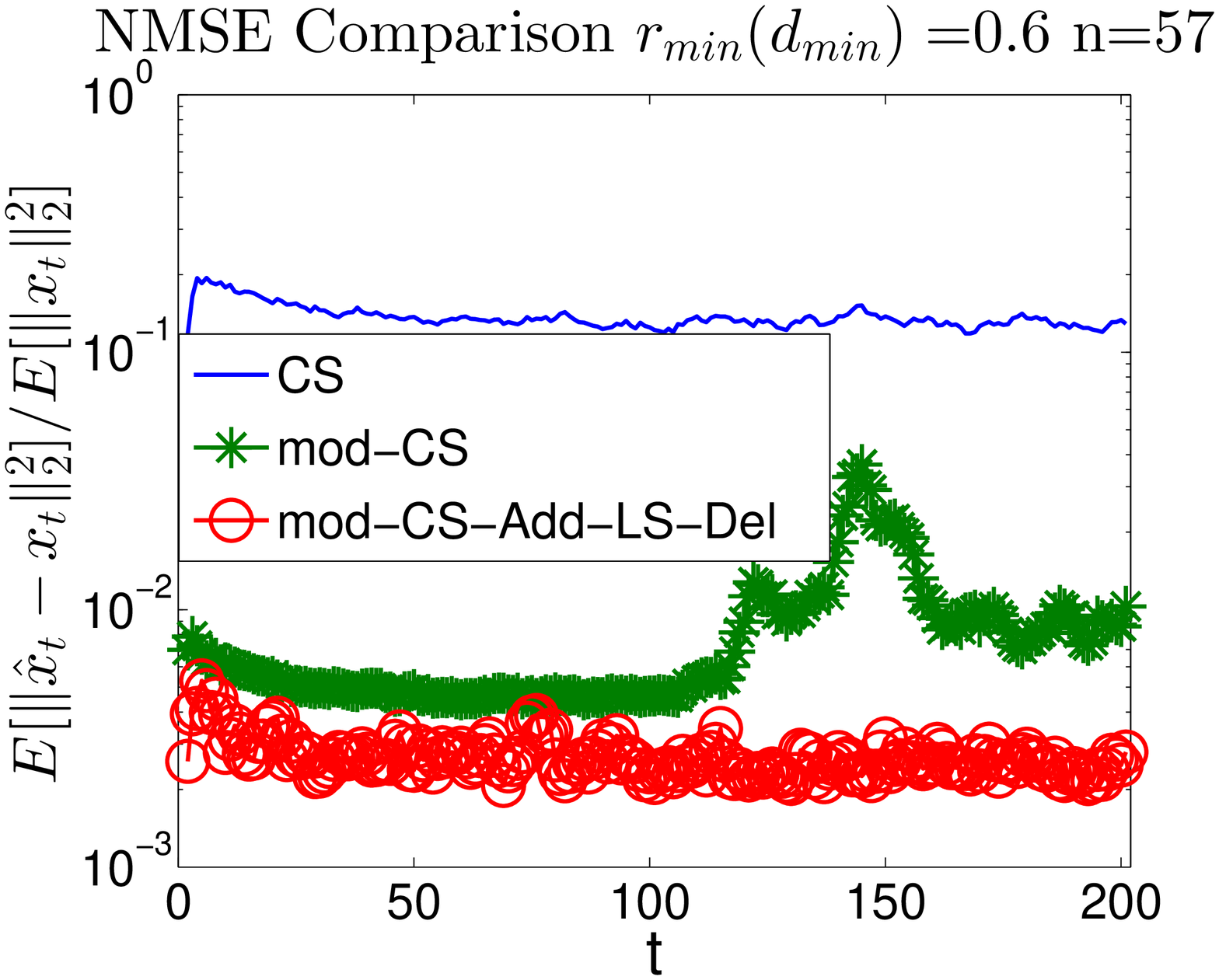}
   \label{err_gen_A_r_06}}
     \hfil
     \subfigure[]{\includegraphics[width=6.5cm]{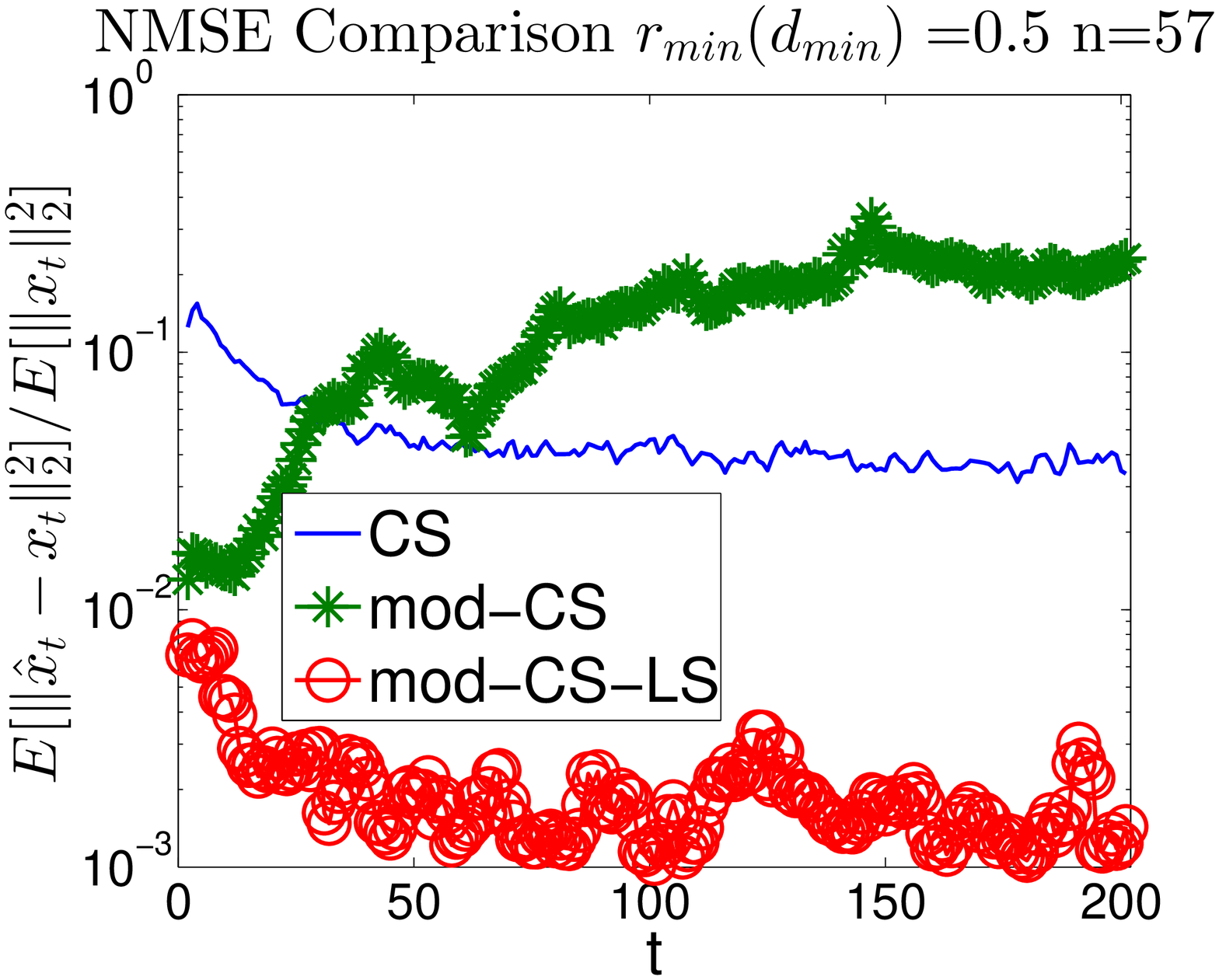}
     \label{err_gen_A_r_06}}}
   \caption{Error Comparison with Fixed Measurement Matrix. ``CS" in the figures refers to noisy $\ell_1$, i.e. the solution of (\ref{simpcs}) at each time.}
   \label{err_comp_general_model}
\end{figure*}

\begin{figure*}[!htb]
\centerline{
\subfigure{
\begin{tabular}{lcr}
\epsfig{file = 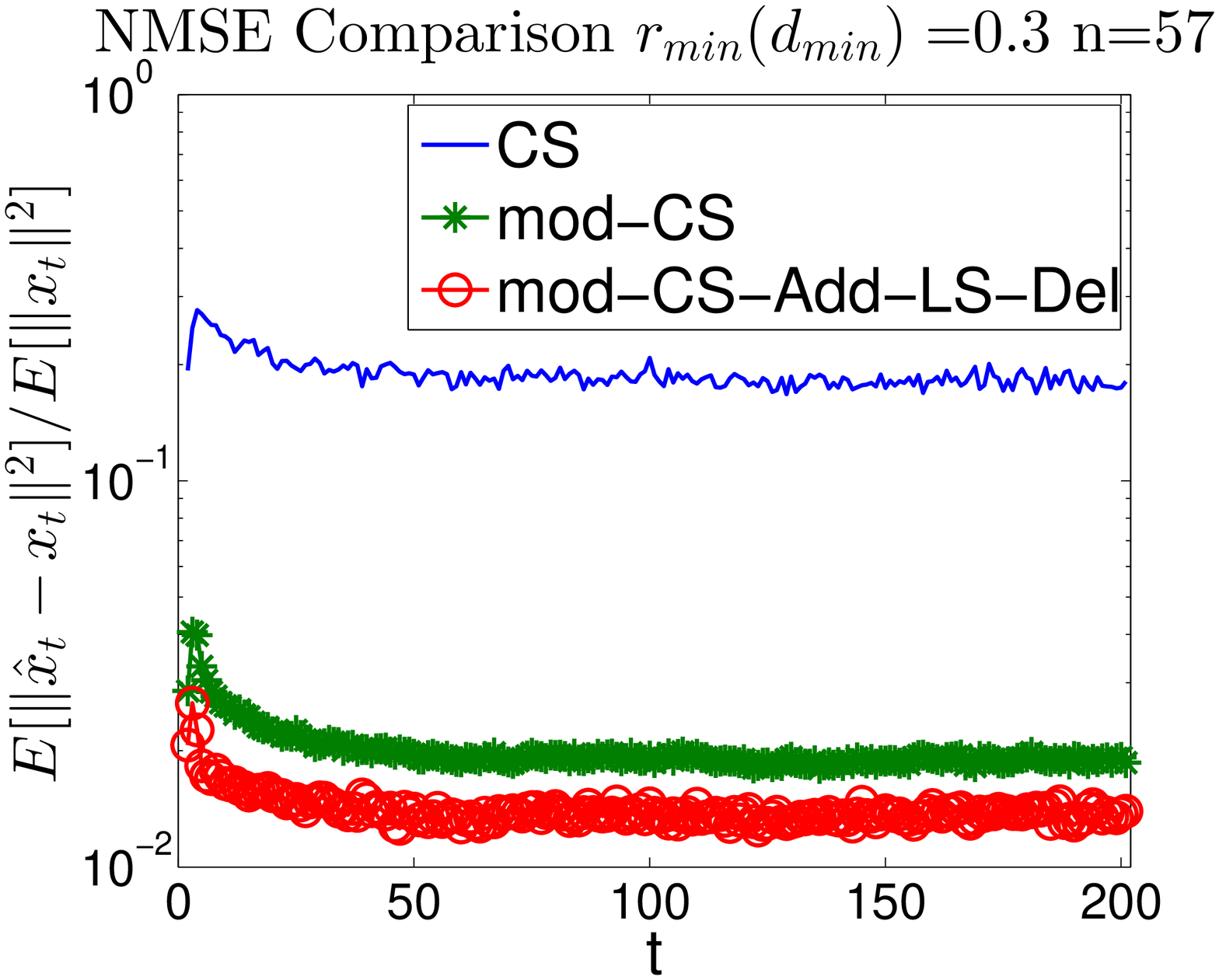, width=6.5cm}&\epsfig{file = 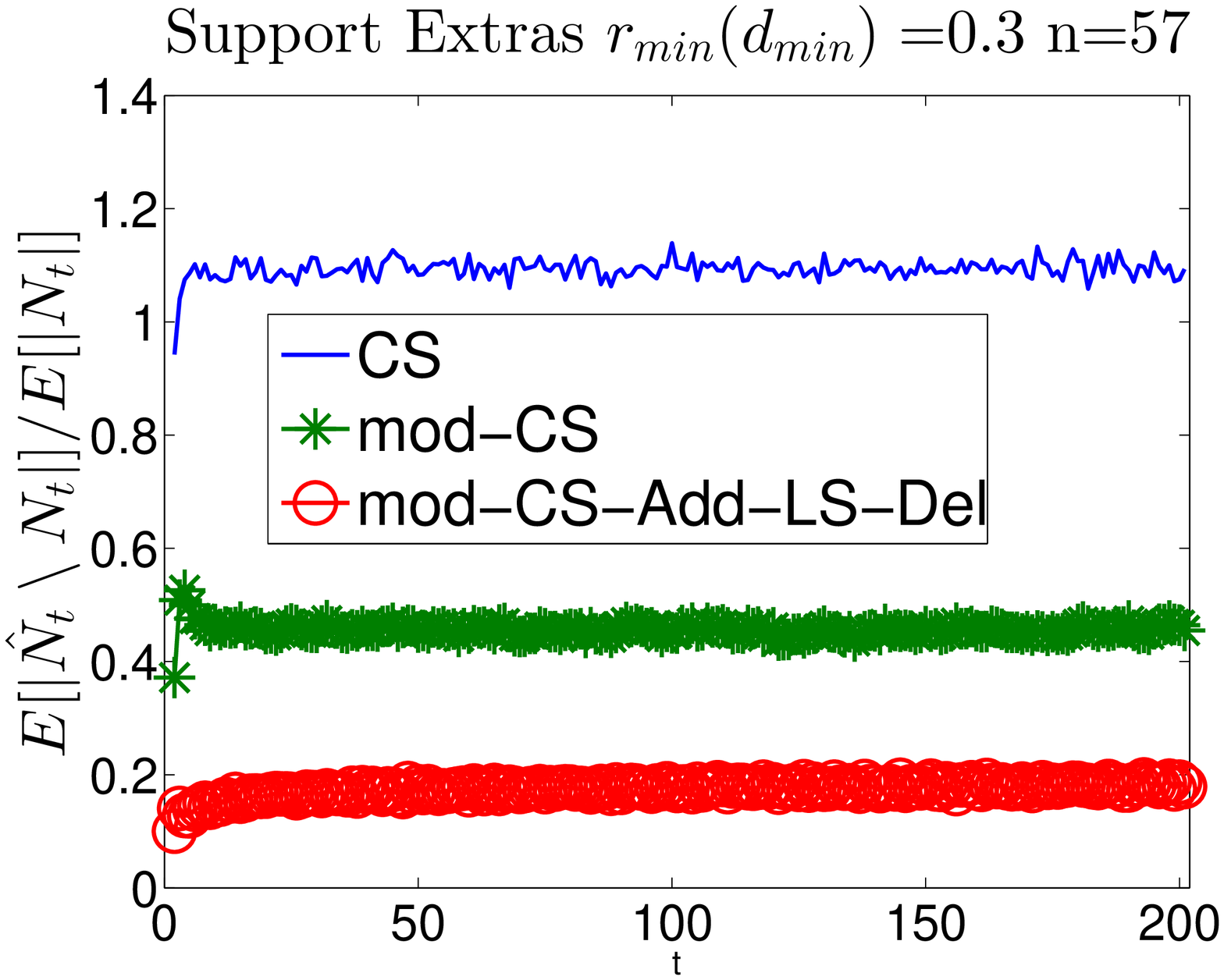, width=6.5cm}&\epsfig{file = 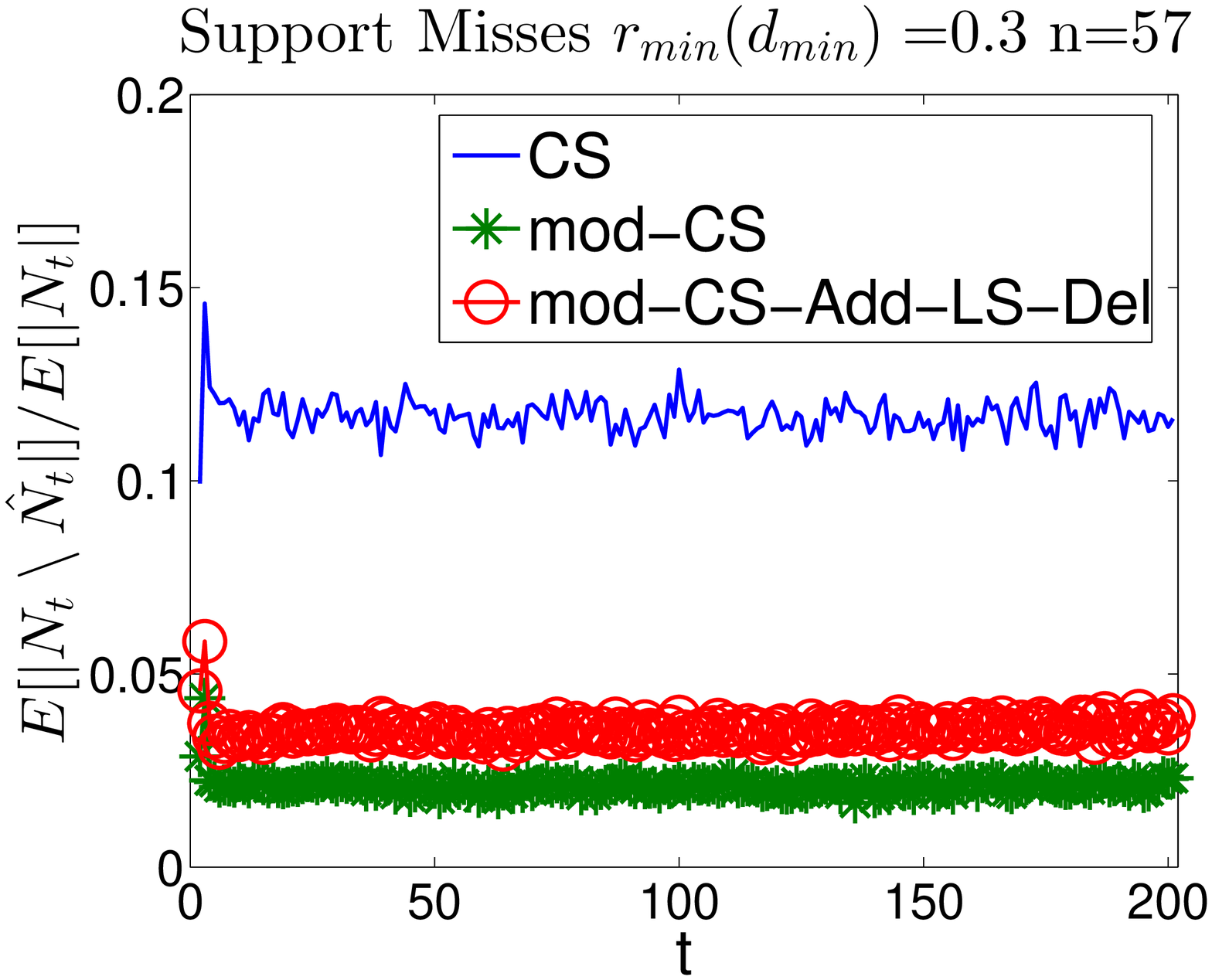, width=6.5cm}
\end{tabular}
}
}
\caption{Error Comparison with Time Varying Measurement Matrices. ``CS" in the figures refers to noisy $\ell_1$, i.e. the solution of (\ref{simpcs}) at each time.}
\label{err_comp_general_model_vA}
\end{figure*}

In the first set of experiments shown in  Fig. \ref{err_comp_general_model}, we used the same measurement matrix $A_t=A$ for all $t\geq 1$. In the second experiment shown in Fig. \ref{err_comp_general_model_vA}, $A_t$ was time varying. 

The normalized mean squared error (NMSE), $\frac{\E[\|x_t - \xhat_t\|^2]}{\E[\|x_t\|^2]}$, the normalized mean extras, $\frac{\E[|\tilde{\Nset_t} \setminus \Nset_t|]}{\E[|\Nset_t|]}$, and the normalized mean misses, $\frac{\E[| \Nset_t \setminus \tilde{\Nset_t}|]}{\E[|\Nset_t|]}$ are used to compare the reconstruction performance. Here $\E[.]$ denotes the empirical mean over the 500 realizations.
Consider the results of  Fig \ref{err_comp_general_model}. Clearly, both mod-CS and mod-CS-Add-LS-Del significantly outperform noisy $\ell_1$ (simple CS). This is because for $t>0$, the number of measurements, $n_t=57$ is too small for a 200 length 20 sparse signal. When $a_{\min} = r_{\min}(d_{\min}) = r$ is large enough, both mod-CS and mod-CS-Add-LS-Del are stable at 5\% error or less. When $r$ is reduced, mod-CS becomes unstable. Of course when $r$ is reduced even further to $r = 0.2$, both become unstable (not shown).
If Fig \ref{err_comp_general_model_vA}, we show results for the case when $A_t$ changes with time and all other parameters are the same as Fig \ref{err_comp_general_model} (a). Clearly in this case, the performance of both mod-CS and mod-CS-add-LS-del has improved significantly.

\begin{figure}[!htb]
\centering
\includegraphics[width=7.5cm]{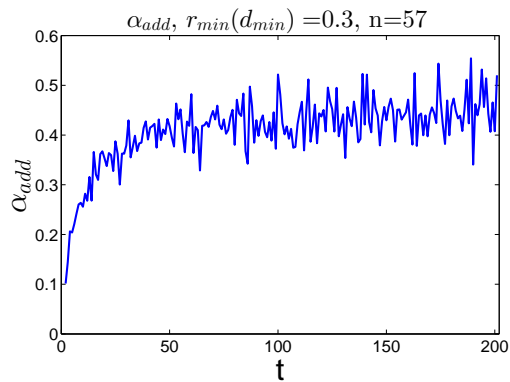}
\caption{Mean of $\alpha_{\add}$ over time.}
\label{thresh_add}
\end{figure}

In Fig. \ref{thresh_add}, we plot the average value of  $\alpha_{\add,t}$ for the simulations corresponding to Fig \ref{err_comp_general_model_vA}. As can be seen, this threshold is close to $4c = 4 \eps/\sqrt{n}$ at all times.


For solving the minimization problems given in (\ref{simpcs}) and (\ref{modcs}), we used the YALL1 software, which is provided in http://yall1.blogs.rice.edu/.
Both the modified-CS algorithms and noisy $\ell_1$ took roughly the same amount of time.
For the results of Fig. \ref{err_comp_general_model_vA}, when running the code in MATLAB on the same server, noisy $\ell_1$ needed 0.0466 seconds per frame; mod-CS needed  0.0432 seconds per frame and mod-CS-Add-LS-Del needed 0.0517 seconds. These numbers are computed by averaging over all 500 realizations and over the 200 time instants per realization.

\section{Conclusions and Future Work} \label{conc}
In this work we obtained performance guarantees for recursive noisy modified-CS which has been shown in earlier work to be a practically useful algorithm \cite{modcsjp,fmrimodcs,regmodbpdn}. We show that, under mild assumptions -- a lower bound on either the initial nonzero magnitude or on the magnitude increase rate, or an upper bound on the maximum number of nonzero entries with magnitude below a certain threshold; mild RIP conditions (which imply conditions on the required number of measurements); appropriately set algorithm parameters; and a special start condition  -- the support and signal recovery error of modified-CS and its improvement, modified-CS-add-LS-del can be bounded by time-invariant and small values. 

The special start condition is a possible limitation of our analysis. This can be removed in various ways. If some prior knowledge about signal support is available, that can be used at $t=0$ as suggested and demonstrated in \cite{modcsjp}. Or, one can solve a batch problem (multiple measurement vector (MMV) problem) for the first set of $k$ frames. If we let ${\cal N} = \cup_{t=1}^k {\cal N}_t$, then we have an MMV problem with row support ${\cal N}$ that can be solved using mixed norm minimization \cite{tropp2006algorithms2}, simultaneous-OMP \cite{chen2006theoretical,tropp2006algorithms1}, compressive MUSIC \cite{kim2012compressivemusic}, iterative MUSIC \cite{lee_bresler}, block sparsity approaches \cite{eldar2009compressed} or M-SBL (Sparse Bayesian Learning) \cite{wipf2007empirical}.  In this case one could adopt guarantees for the chosen batch method for the initialization. 

In this work, we used a deterministic set of assumptions on signal change. Notice however that one can assume any probabilistic model that ensures that $a_{j,t} \ge a_{\min}$ and $r_{j,\tau}$ is anything larger than $r_{\min}(d_0)$ for for the first $d_0$ frames after a new addition; and at later times, $r_{j,\tau}$ can be anything between zero and infinity. Similarly, any probabilistic model for coefficient decrease that ensures removal within at most $b$ frames after decrease begins will suffice. We can fix $d_0$ to be any integer between zero and $d_{\min}$ and our result will then hold for that particular value of $d_0$. 

Other ongoing and future work includes designing and analyzing better support prediction techniques rather than just using the previous support estimate as the prediction for the current support. Some initial ideas are presented in \cite{rrpcp_isit}.

\appendix

\subsection{Proof of Lemma \ref{modcsbnd_2}}
\label{modcserrorbound_2}
We provide the proof here for the sake of completion and for ease of review. This will be removed later.
In this proof, we use $\Tset, \Delta, \Nset$ instead of $\Tset_t, \Delta_t, \Nset_t$ respectively for simplicity. Let $h : = \xhat_{modcs} - x$. We adapt the approach of \cite{candes_rip} to bound the reconstruction error, $\|h\| := \|\xhat_{modcs}-x\|$. A similar result was obtained in \cite{jacques}.
Let $\Delta_1$ denote the set of  indices of $h$ with the $|\Delta|$ largest values outside of $\Tset\cup \Delta$, let $\Delta_2$ denote the indices of the next $|\Delta|$ largest values and so on. Then using the same approach as that of \cite{candes_rip}, i.e., $\|h_{\Delta_j}\|\leq\frac{1}{\sqrt{\Delta}}\|h_{\Delta_{j-1}}\|_1$,
\bea
\|h_{(\Tset \cup \Delta \cup \Delta_1)^c}\| \sle \sum_{j \ge 2} \|h_{\Delta_j}\|
                                          \le \frac{1}{\sqrt{|\Delta|}} \|h_{(\Tset \cup \Delta)^c}\|_1
\label{h1}
\eea
Since $\xhat_{modcs} = x+h$ is the minimizer of (\ref{modcs}) and since both $x$ and $\xhat_{modcs}$ are feasible; and since $x$ is supported on $\Nset \subseteq \Tset \cup \Delta$,
\bea
\|x_\Delta\|_1=\|x_{\Tset^c}\|_1 \sge  \|(x+h)_{\Tset^c}\|_1 \nn \\
\sge \|x_\Delta\|_1 - \|h_\Delta\|_1 + \|h_{(\Tset \cup \Delta)^c}\|_1  
\eea
Thus,
\bea
\|h_{(\Tset \cup \Delta)^c}\|_1 \le \|h_\Delta\|_1    
\eea
Combining this with (\ref{h1}), and using $\frac{\|h_\Delta\|_1}{\sqrt{|\Delta|}} \le \|h_\Delta\|$, we get
\bea
\|h_{(\Tset \cup \Delta \cup \Delta_1)^c}\| \sle \sum_{j \ge 2} \|h_{\Delta_j}\| \le \|h_\Delta\|
\label{h2}
\eea

Next, since both $x$ and $\xhat_{modcs}$ are feasible,
\bea
\|Ah\| \se \|A(x - \xhat_{modcs})\| \nn \\
\sle \|y - Ax\| + \|y - A \xhat_{modcs}\| \le 2 \eps
\label{feas}
\eea
In this proof, let
\bea
\delta \sdefn \delta_{|\Tset| + 3|\Delta|} 
\eea

Now, we upper bound $\|h_{\Tset \cup \Delta \cup \Delta_1}\|$. By $\delta_{|\Tset|+2|\Delta|}\le \delta$, we have
\bea
(1-\delta) \|h_{\Tset \cup \Delta \cup \Delta_1}\|^2 \le  \|Ah_{\Tset \cup \Delta \cup \Delta_1}\|^2
\eea
To bound the RHS of the above, notice that $Ah_{\Tset \cup \Delta \cup \Delta_1} = Ah - \sum_{j \ge 2} A h_{\Delta_j}$ and so
\bea
&& \|Ah_{\Tset \cup \Delta \cup \Delta_1}\|^2  = \langle Ah_{\Tset \cup \Delta \cup \Delta_1}, Ah \rangle - \sum_{j \ge 2}\langle A h_{\Tset \cup \Delta \cup \Delta_1}, A h_{\Delta_j} \rangle \nn
\eea
Using (\ref{feas}) and the definition of $\delta_S$ given in (\ref{def_delta}) and $\delta_{|\Tset|+2|\Delta|}\le \delta$,
\bea
|\langle A h_{\Tset \cup \Delta \cup \Delta_1}, Ah \rangle | \le 2\eps \sqrt{1+\delta}\|h_{\Tset \cup \Delta \cup \Delta_1}\|
\eea
Using the definition of $\theta_{S_1,S_2}$ given in (\ref{def_theta}); equation (\ref{h2}); and the fact that $\|h_{\Tset}\| + \|h_{\Delta \cup \Delta_1}\| \le \sqrt{2}\|h_{\Tset \cup \Delta \cup \Delta_1}\|$, we get the following. Using $\theta_{|\Tset|,|\Delta|} \le \delta_{|\Tset|+|\Delta|} \le \delta_{|\Tset|+3|\Delta|}, \theta_{2|\Delta|,|\Delta|} \le \delta_{3|\Delta|} \le \delta_{|\Tset|+3|\Delta|}$ \cite{decodinglp},
\bea
&& |\sum_{j \ge 2} \langle A h_{\Tset \cup \Delta \cup \Delta_1}, A h_{\Delta_j} \rangle | \nn \\
&&  \le \theta_{|\Tset|+2|\Delta|,|\Delta|} \|h_{\Tset \cup \Delta \cup \Delta_1}\| \sum_{j \ge 2} \|h_{\Delta_j}\| \nn \\
&& \le  \delta \|h_{\Tset \cup \Delta \cup \Delta_1}\| \ \|h_\Delta\|    
\label{b1}
\eea
Combining the last six equations above, 
using $\|h_\Delta\| \le \|h_{\Tset \cup \Delta \cup \Delta_1}\|$, we can simplify the above to get
\bea
\|h\| 
\sle 2\|h_{\Tset \cup \Delta \cup \Delta_1}\|  \le \frac{4 \sqrt{1+\delta}}{1-2\delta} \eps \nn\\
&\le& \frac{4 \sqrt{1+\delta}}{1-2\delta} \eps
\eea

Clearly, all of the above discussion holds only if the RHS is positive which is true only if $2\delta_{|\Tset|+3|\Delta|} < 1$. Thus, we can get Lemma \ref{modcsbnd_2}.

\subsection{Proof of Theorem \ref{theorem_modcs_simplified}}
\label{proof_modcs_simplified}

We prove the first two claims by induction. Using condition \ref{initass_simplified} of the theorem, the claim holds for $t=0$. This proves the base case. For the induction step, assume that the claims hold at $t-1$, i.e. $|\tDelta_{e,t-1}| =0$, $|\tT_{t-1}| \le S$, and $|\tDelta_{t-1}| \leq S_a$, so $|\Tset_t| \le S$. At $t$, there are at most $S_a$ new support, so $|\Delta_t| \leq |\tDelta_{t-1}|+S_a \leq 2S_a$; there are at most $S_a$ removed support at time $t$, so $|\Delta_{e,t}| \leq |\tDelta_{t-1}|+S_a =S_a$. Thus the second claim holds.

Next we bound $|\tDelta_t|$, $|\tDelta_{e,t}|$, $|\tT_t|$. Consider the support estimation step. Since condition 1 of the theorem holds, we can apply Lemma \ref{modcsbnd_2} with $S_{\Tset_t}=S$, $S_{\Delta_t} = 2S_a$. This gives $\|x_t - \xhat_{t,modcs}\| \le 7.5 \eps$. Using Proposition \ref{prop0}, this, along with conditions 2 and 3 implies that all elements of $\Nset_t\setminus \Bset_t$ will get detected and all zero elements will get deleted, i.e., there will be no false detections. Thus, $|\tDelta_t| \leq |\Bset_t| \leq S_a$ and $|\tDelta_{e,t}|=0$ and so $|\tT_t| \leq |\Nset_t|+|\tDelta_{e,t}| \leq S$. Thus the first claim holds.

The third claim follows using the second claim and Lemma \ref{modcsbnd_2}.

\subsection{Proof of Theorem \ref{theorem_modcsls_simplified}}
\label{proof_modcsls_simplified}

We prove the first three claims of the theorem by induction. Using condition \ref{initass_simplified} of the theorem, the claim holds for $t=0$. This proves the base case. For the induction step, assume that the claim holds at $t-1$, i.e. $|\tDelta_{e,t-1}| =0$, $|\Tset_{t-1}| \le S$, and $|\tDelta_{t-1}| \leq S_a$. Using this, we prove the first three claims holds at $t$. 

The bounding of $|\Tset_t|, |\Delta_t|, |\Delta_{e,t}|$ is exactly as in the proof of Theorem \ref{theorem_modcs_simplified}. 

Consider the detection step. There are at most $f$ false detects (from condition \ref{addthresh_simplified}) and thus $|\tDelta_{e,\add,t}| \le |\Delta_{e,t}| + f \le S_a + f$. Thus $|\Tset_{\add,t}| \le |\Nset_t| + |\tDelta_{e,\add,t}| \le S+S_a+f$. So the third claim holds.

Next, consider $|\Delta_{\add,t}|$. Applying Lemma \ref{modcsbnd_2} with condition \ref{measmodel_modcsls_simplified}, i.e., $\delta_{|\Tset_t|+3|\Delta_t|} \leq \delta_{S+6S_a} \leq 0.207$, we have $\|x_t - \xhat_{t, modcs}\| \leq 7.50 \eps$. Thus, all elements of $\{i: |(x_t)_i| > \alpha_\add + 7.50\eps\}$ will definitely get detected at time $t$ and so $\Delta_{\add,t} \subseteq \{i: |(x_t)_i| \leq  \alpha_\add + 7.50\eps\}$. Since condition \ref{minmag_modcsls_simplified} holds, $\{i: |(x_t)_i| \leq  \alpha_\add + 7.50\eps\} \subseteq \Bset_t$, and so $|\Delta_{\add,t}| \leq |\Bset_t| \leq S_a$.

Consider the deletion step. As $\Delta_{\add,t} \subseteq \Bset_t$, and $|(x_t)_i| \leq \alpha_\add + 7.50\eps$ for $i\in \Delta_{\add,t}$, we have $\|(x_t)_{\Delta_{\add,t}}\| \leq \sqrt{S_a}(\alpha_\add + 7.50\eps)$. Applying Lemma \ref{errls1} with condition \ref{measmodel_modcsls_simplified}, i.e., $\delta_{|\Tset_{\add,t}|+|\Delta_{\add,t}|} = \delta_{S+2S_a+f} \leq 0.207$, we have $\|(x_t - x_{t, \add})_{\Tset_{\add,t}}\| \leq 1.12 \epsilon + 0.261 \sqrt{S_a}(\alpha_\add + 7.50\eps)$. Thus, using these facts and condition \ref{delthresh_simplified}, all elements of $\tDelta_{e,\add,t}$ will get deleted and elements of $\{i: |(x_t)_i| > 2\alpha_\del\}$ will not be deleted. Thus $|\tDelta_{e,t}|=0$, and since condition \ref{minmag_modcsls_simplified} holds,  $\tDelta_{t} \subseteq \{i: |(x_t)_i| \leq  2 \alpha_\del \} \subseteq \Bset_t$, i.e., $|\tDelta_t| \leq S_a$. Thus $|\tT_t| \le |\Nset_t| + |\tDelta_{e,t}| \le S$. So the first claim holds.%


The fourth claim follows using the previous claims and Lemma \ref{modcsbnd_2}. The fifth claim follows using previous claims, Lemma \ref{errls1}.

\subsection{Proof of Theorem \ref{stabres_simple_modcs}}
\label{proof_simple_modcs}

We prove the first claim by induction. Using condition \ref{initass_simple} of the theorem, the claim holds for $t=0$. This proves the base case. For the induction step, assume that the claim holds at $t-1$, i.e. $|\tDelta_{e,t-1}| =0$, $|\tT_{t-1}| \le S$, and $\tDelta_{t-1} \subseteq \Sset_{t-1}(d_0)$ so that $|\tDelta_{t-1}| \le 2(d_0-1)S_a$. Using this we prove that the claim holds at $t$. In the proof, we use the following facts often: (a) $\Rset_t \subseteq \Nset_{t-1}$ and $\Aset_t \subseteq \Nset_{t-1}^c$, (b) $\Nset_t = \Nset_{t-1} \cup \Aset_t \setminus \Rset_t$, and (c) if two sets $B,C$ are disjoint, then, $D \cup C \setminus B :=(D \cup C) \setminus B =(D \cap B^c) \cup C$ for any set $D$.%

We first bound $|\Tset_t|$, $|\Delta_{e,t}|$, $|\Delta_t|$. Since $\Tset_t = \tT_{t-1} = \Nhat_{t-1}$, so $|\Tset_t| \le S$. Also, $\Delta_{e,t} = \Nhat_{t-1} \setminus \Nset_t =  \Nhat_{t-1} \cap [(\Nset_{t-1}^c \cap \Aset_t^c) \cup \Rset_t] \subseteq \tDelta_{e,t-1} \cup \Rset_t = \Rset_t$. The last equality follows since $|\tDelta_{e,t-1}| =0$. Thus $|\Delta_{e,t}| \le |\Rset_t| = S_a$.

Consider $|\Delta_t|$. Notice that $\Delta_t = \Nset_t \setminus \Nhat_{t-1}  = (\Nset_{t-1} \cap \Nhat_{t-1}^c \cap \Rset_t^c) \cup (\Aset_t \cap \Nhat_{t-1}^c) = (\tDelta_{t-1} \cap  \Rset_t^c) \cup ( \Aset_t \cap \Nhat_{t-1}^c) \subseteq  (\Sset_{t-1}(d_0)  \cap  \Rset_t^c) \cup  \Aset_t = \Sset_{t-1}(d_0) \cup  \Aset_t \setminus \Rset_t$. Here we used $\tDelta_{t-1} \subseteq \Sset_{t-1}(d_0)$.
When $d_0 \geq 2, \Rset_t \subseteq \Sset_{t-1}(d_0)$ and $\Aset_t$ is disjoint with $\Sset_{t-1}(d_0)$. Thus $|\Delta_t| \le |\Sset_{t-1}(d_0)| + |\Aset_t| - |\Rset_t| = 2(d_0-1)S_a + S_a - S_a$. When $d_0=1, \Sset_{t-1}(d_0)=\emptyset$, and $\Aset_t$ is disjoint with $\Rset_t$. Thus $|\Delta_t| \le |\Aset_t \setminus \Rset_t| = |\Aset_t|= S_a$. Thus, $|\Delta_t| \le k_1S_a$.

Next we bound $|\tDelta_t|$, $|\tDelta_{e,t}|$, $|\tT_t|$. Consider the support estimation step. Apply the first claim of Lemma \ref{lemma_modcs} with $S_\Nset=S$, $S_{\Delta e}=S_a$, $S_\Delta = k_1S_a$, and $b_1 = d_0r$. Since conditions \ref{measmodel_simple} and \ref{add_del_simple} of the theorem hold, all elements of $\Nset_t$ with magnitude equal to or greater than $d_0r$ will get detected. Thus, $\tDelta_t \subseteq \Sset_t(d_0)$. Apply the second claim of the lemma. Since conditions \ref{measmodel_simple} and \ref{threshes_simple} hold, all zero elements will get deleted and there will be no false detections, i.e. $|\tDelta_{e,t}|=0$. Finally, $|\tT_t| \le |\Nset_t| + |\tDelta_{e,t}| \le S + 0$.

The second claim for time $t$ follows using the first claim for time $t-1$ and the arguments from the paras above. The third claim follows using the second claim and  Lemma \ref{modcsbnd_2}.

\subsection{Proof of Theorem \ref{gencase}}
\label{proof_addLSdel_modcs}

We prove the first claim of the theorem by induction. Using condition \ref{initass} of the theorem, the claim holds for $t=0$. This proves the base case. For the induction step, assume that the claim holds at $t-1$, i.e. $|\tDelta_{e,t-1}| =0$, $|\Tset_{t-1}| \le S$, and $\tDelta_{t-1} \subseteq \Sset_{t-1}(d_0)$ so that $|\tDelta_{t-1}| \le 2(d_0-1)S_a$. Using this, we prove that the claim holds at $t$. We will use the following facts often: (a) $\Rset_t \subseteq \Nset_{t-1}$,  (b) $\Aset_t \subseteq \Nset_{t-1}^c$, (c) $\Nset_t = \Nset_{t-1} \cup \Aset_t \setminus \Rset_t$, and (d) if two sets $B,C$ are disjoint, then, $D \cup C \setminus B :=(D \cup C) \setminus B =(D \cap B^c) \cup C$ for any set $D$.%

The bounding of $|\Tset_t|, |\Delta_t|, |\Delta_{e,t}|$ is exactly as in the proof of Theorem \ref{stabres_simple_modcs}. Since $\Tset_t = \tT_{t-1}$, so $|\Tset_t| \le S$. Also, $\Delta_{e,t} = \Nhat_{t-1} \setminus \Nset_t =  \Nhat_{t-1} \cap [(\Nset_{t-1}^c \cap \Aset_t^c) \cup \Rset_t] \subseteq \tDelta_{e,t-1} \cup \Rset_t = \Rset_t$. 
Thus $|\Delta_{e,t}| \le |\Rset_t| = S_a$.
Finally, $\Delta_t = \Nset_t \setminus \Nhat_{t-1}  =  (\tDelta_{t-1} \cap  \Rset_t^c) \cup ( \Aset_t \cap \Nhat_{t-1}^c) \subseteq  (\Sset_{t-1}(d_0)  \cap  \Rset_t^c) \cup  \Aset_t$. Thus,
\bea
\Delta_t \subseteq \Sset_{t-1}(d_0) \cup  \Aset_t \setminus \Rset_t
\label{Delta_sub_0}
\eea
When $d_0\geq 2, \Rset_t \subseteq \Sset_{t-1}(d_0)$ and $\Aset_t$ is disjoint with $\Sset_{t-1}(d_0)$, so $|\Delta_t| \le |\Sset_{t-1}(d_0)| + |\Aset_t| - |\Rset_t| = 2(d_0-1)S_a + S_a - S_a$. When $d_0=1, \Sset_{t-1}(d_0)=\emptyset$, and $\Aset_t$ is disjoint with $\Rset_t$, so $|\Delta_t| \le |\Aset_t \setminus \Rset_t| = |\Aset_t| = S_a$. Thus, $|\Delta_t| \le k_1S_a$.

Consider the detection step. There are at most $f$ false detects (from condition \ref{addthresh}) and thus $|\tDelta_{e,\add,t}| \le |\Delta_{e,t}| + f \le S_a + f$. Thus $|\Tset_{\add,t}| \le |\Nset_t| + |\tDelta_{e,\add,t}| \le S+S_a+f$.

Next, consider $|\Delta_{\add,t}|$. Notice that
\bea
\Delta_t &\subseteq& \Sset_{t-1}(d_0) \cup \Aset_t\setminus\Rset_t\nn \\
&\subseteq& \Sset_t(d_0) \cup \Iset_t(d_0) \setminus \Dset_t(d_0-1).
\label{Delta_sub}
\eea
The first $\subseteq$ is from (\ref{Delta_sub_0}), the second one follows by using (\ref{sseteq_2}) for $j=d_0$. Now, apply Lemma \ref{detectcond_modcs} with $S_{\Nset_t} = S$, $S_{ \Delta_{e,t}}=S_a$, $S_{\Delta_t} = k_1S_a$, and with $b_1 = d_0r$. Using (\ref{Delta_sub}), $\{i\in\Delta_t:|(x_t)_i|\geq b_1\} = \Delta_t \cap \Iset_t(d_0) $.
Since conditions \ref{measmodel} and \ref{add_del} hold, by Lemma \ref{detectcond_modcs}, all elements of $\{i\in\Delta_t:|(x_t)_i|\geq b_1\}$ will definitely get detected at time $t$. Thus $\Delta_{\add,t} \subseteq \Delta_t \setminus \{i\in\Delta_t:|(x_t)_i|\geq b_1\} \subseteq  \Delta_t \setminus \Iset_t(d_0)$. But from (\ref{Delta_sub}), $\Delta_t \setminus \Iset_t(d_0) \subseteq \Sset_t(d_0) \setminus \Dset_{t}(d_0-1)$.
Since when $d_0\geq 2$, $\Dset_{t}(d_0-1) \subseteq \Sset_{t}(d_0)$, then $|\Delta_{\add,t}| \le |\Sset_t(d_0)| - |\Dset_{t}(d_0-1)|= 2(d_0-1)S_a - S_a$; when $d_0=1, \Dset_{t}(d_0-1) = \Sset_{t}(d_0) = \emptyset$, then $|\Delta_{\add,t}|=0$. Thus, $|\Delta_{\add,t}|\le k_2S_a$

Consider the deletion step. Apply  Lemma \ref{nofalsedels_truedels_cond} with $S_{\Tset_{\add,t}} = S$, $S_{\Delta_{\add,t}} = k_1S_a$. Since condition \ref{measmodald_delta2} holds, $\delta_{S+S_a+f} < 1/2$ holds. Since $\Delta_{\add,t} \subseteq \Sset_t(d_0) \setminus \Dset_{t}(d_0-1)$, $\Delta_{\add,t}$ contains only $2S_a$ elements of magnitude $\{r, 2r, \cdots, (d_0-2)r\}$ and $S_a$ elements of magnitude $(d_0-1)r$. Thus,  $\|(x_t)_{\Delta_{\add,t}}\| \le k_3\sqrt{S_a}r$. Using these facts and condition \ref{delthresh}, by Lemma \ref{nofalsedels_truedels_cond}, all elements of $\tDelta_{e,\add,t}$ will get deleted. Thus $|\tDelta_{e,t}|=0$. Thus $|\tT_t| \le |\Nset_t| + |\tDelta_{e,t}| \le S$.%

To bound $|\tDelta_{t}|$, apply Lemma \ref{nofalsedels_truedels_cond} with $S_{\Tset_{\add,t}} = S+S_a+f$, $S_{\Delta_{\add,t}} = k_2S_a$, $b_1 = d_0r$. 
By Lemma \ref{nofalsedels_truedels_cond}, to ensure that all elements of $\{i\in\Tset_{\add,t}:|(x_t)_i|\geq b_1\}$ do not get falsely deleted, we need $\delta_{S_0+S_a+f} < 1/2$ and $d_0r > \alpha_{\del} + \frac{\zeta_L}{\sqrt{S_a}}(\sqrt{2} \eps + 2 \theta_{S_0+S_a+f, k_2S_a} k_3\sqrt{S_a} r)$. From condition \ref{delthresh}, $\alpha_{\del} =  \sqrt{\frac{2}{S_a}}\zeta_L \eps +  2  k_3  \theta_{S+S_a+f,k_2S_a}\zeta_L r $. Thus, we need $\delta_{S_0+S_a+f} < 1/2$ and $d_0r > 2(\sqrt{\frac{2}{S_a}}\zeta_L \eps +  2  k_3  \theta_{S+S_a+f,k_2S_a}\zeta_L r  )$. $\delta_{S_0+S_a+f} < 1/2$ holds since condition \ref{measmodald_delta2} holds. The second one holds since condition \ref{measmodald_theta} and $r \ge G_2$ of condition \ref{add_del} hold. Thus, we can ensure that all elements of $\{i\in\Tset_{\add,t}:|(x_t)_i|\geq b_1\}$, i.e. all elements of $\Tset_{\add,t}$ with magnitude greater than or equal to $b_1=d_0r$ do not get falsely deleted.
But nothing can be said about the elements smaller than $d_0r$ (in the worst case all of them may get falsely deleted). Thus, $\tDelta_t \subseteq \Sset_t(d_0)$ and so $|\tDelta_t| \le 2(d_0-1) S_a$.

This finishes the proof of the first claim. To prove the second and third claims for any $t>0$: use the first claim  for $t-1$ and the arguments from the paragraphs above to show that the second and third claim hold for $t$. The fourth claim follows using the previous claims and Lemma \ref{modcsbnd_2}. The fifth claim follows using previous claims, Lemma \ref{errls1} and a bound on $\|(x_t)_{\tDelta_t}\|_2$. It is easy to see that $\|(x_t)_{\tDelta_t}\|_2 \le k_3\sqrt{S_a}r$.
\subsection{Proof of Theorem \ref{modcsthm}}
\label{modcs_sta_proof}
Recall from the signal model that $|\Nset_t|\le S$ for all $t$, and that $|\SDset_t| \le  \frac{(b+1)}{2} S_d$. Also $\Nset_t=\cup_{\tau=t-d_{\min}+1}^t \Aset_{\tau}\cup \Lset_t\cup \SDset_t$, noting that the first two sets might not be disjoint.

The proof follows using induction. The base case is easy. Assume that the result holds at $t-1$.
At $t$, at most $S_a$ new elements get added to the support, thus $|\Delta_t| \le |{\tDelta}_{t-1}| + S_a \le \frac{(b+1)}{2} S_d + d_0 S_a + S_a$. Also, since $\Tset_t = \tT_{t-1}$, thus $|\Tset_t| \le S$. And $\Delta_{e,t}={\tDelta}_{e,t-1}\cup R_t$, indicating $|\Delta_{e,t}|\le |{\tDelta}_{e,t-1}|+|R_t|\le S_r$. The second condition of the theorem ensures that $\delta_{|\Tset_t| + 3|\Delta_t|} \le (\sqrt{2}-1)/2$. Thus using Lemma \ref{modcsbnd_2},  $||x_t - \xhat_t|| \le 7.50 \eps$.

Consider the support detection step. Consider an $i \notin \Nset_t$, i.e. $(x_t)_i = 0$. Since $\alpha =  \frac{\zeta_M}{\sqrt{S_a}} 7.50 \eps \ge \frac{\zeta_M}{\sqrt{S_a}} ||x_t - \xhat_t|| \ge ||x_t - \xhat_t||_\infty \ge |(\xhat_t)_i|$, thus $i$ will never get detected into the support estimate. Thus, $|{\tDelta}_{e,t}|= 0$. Thus $|\tT_t| \le |\Nset_t| + |{\tDelta}_{e,t}| \le S$.

The third condition ensures that any newly added element exceeds $\alpha + \frac{\zeta_M}{\sqrt{S_a}} 7.50 \eps$ within $d_0$ time units and any element of $\Lset_t$ exceeds $\alpha + \frac{\zeta_M}{\sqrt{S_a}} 7.50 \eps$ as $\ell>\alpha + \frac{\zeta_M}{\sqrt{S_a}} 7.50 \eps$. Consider any such element $j$. This means that $|(\xhat_t)_j| \ge |(x_t)_j| - |(x_t - \xhat_t)_j| \ge |(x_t)_j| - ||x_t - \xhat_t||_\infty \ge |(x_t)_j| - \frac{\zeta_M}{\sqrt{S_a}}||x_t - \xhat_t|| \ge |(x_t)_j| -\frac{\zeta_M}{\sqrt{S_a}}7.50 \eps  \ge \alpha$. Thus such an element will definitely get detected into the support. This means that the only nonzero elements that are missed are either those that got added in the last $d_0$ frames or those that are currently decreasing. The maximum number of elements that got added in the last $d_0$ time units is $d_0 S_a$. The maximum number of decreasing elements at $t$ is less than or equal to  $\frac{(b+1)}{2} S_d$. Thus,  $|{\tDelta}_t| \le \frac{(b+1)}{2} S_d + d_0 S_a$. This finishes the proof of the induction step and hence of the theorem.

\subsection{Proof of Theorem \ref{modcsaldwkthm}}
\label{modcsaldthmproof}

\begin{proposition}[simple facts] Consider Algorithm \ref{modcsalgo_2}.%
\ben
\item An $i \in \Nset_t$ will definitely get detected  if $|(x_t)_i|  > \alpha_{\add} + \frac{\zeta_M}{\sqrt{S_a}} \|x_t - \xhat_{t,modcs}\|$.
\label{det1}

\item An $i \in \Nset_t$ will definitely not be deleted if  $|(x_t)_i|  > \alpha_{\text{del}} + \frac{\zeta_L}{\sqrt{S_a}}\|x_t - \xhat_{t,\add}\|$.


\item All $i \in \Delta_{e,t}$ (the zero elements of $\Tset_t$) will definitely get deleted if $\alpha_{\text{del}} \ge \|x - \xhat_{t,\add}\|_{\infty}$. 

\een
\label{prop2}
\end{proposition}

Recall from the signal model that $\Nset_t=\cup_{\tau=t-d_{\min}+1}^t \Aset_{\tau}\cup \Lset_t\cup \SDset_t$, noting that the first two sets might not be disjoint.
By the induction assumption, $|\tT_{t-1}| \le S$. Since $\Tset_t = \tT_{t-1} = \Nhat_{t-1}$, thus,
\bea
|\Tset_t| \le S
\eea
Also, by the induction assumption,
\bea
\tDelta_{t-1} \subseteq \SDset_{t-1} \cup \Aset_{t-1} \dots \Aset_{t-d_0}
\eea
Recall that $\Nset_t = \Nset_{t-1} \cup \Aset_t \setminus \Rset_t$. Also, $\SDset_{t-1} \subseteq \SDset_{t} \cup \Rset_t$. Thus, $\SDset_{t-1} \cap \Rset_t^c \subseteq \SDset_t$.
Thus,
\bea
\Delta_t = \Nset_t \cap \Nhat_{t-1}^c \se (\Nset_{t-1} \cap \Rset_t^c \cap \Nhat_{t-1}^c) \cup (\Aset_t \cap \Nhat_t^c) \nn \\
& \subseteq & (\tDelta_{t-1} \cap \Rset_t^c) \cup \Aset_t \nn \\
& \subseteq & \SDset_{t} \cup \Aset_{t-1} \dots \cup \Aset_{t-d_0} \cup \Aset_t
\label{Deltat1}
\eea
Thus,
\bea
|\Delta_t| \le\frac{(b+1)}{2} S_a + d_0S_a + S_a
\eea

Using the above bounds on $|\Tset_t|$ and $|\Delta_t|$ and the RIP condition of the theorem, we can apply Lemma \ref{modcsbnd_2} to show that
\bea
\| x_t - \xhat_{t,modcs}\| \le 7.50 \eps
\eea
Thus, using the Proposition \ref{prop2} and condition 3, all elements of $\Aset_{t-d_0}$ are definitely detected in the add step at $t$, i.e.
\bea
\Aset_{t-d_0} \subseteq \Ahat_t
\label{A1}
\eea
Also since $\ell$ satisfies condition 3, all elements of $\Lset_t$ will be detected in the add step at $t$.

Using (\ref{A1}), 
\bea
\Delta_{\add,t}  = \Delta_t \setminus \Ahat_t &=& \SDset_{t}\cup \Aset_t \cup \Aset_{t-1} \dots \cup \Aset_{t-d_0}\setminus\Ahat_t \nn\\
& \subseteq & \SDset_{t} \cup \Aset_t \cup \Aset_{t-1} \dots \cup \Aset_{t-d_0+1} 
\label{Deltaadd1}
\eea
Thus,
\bea
|\Delta_{\add,t}| \le\frac{(b+1)}{2} S_a + d_0 S_a
\label{Tdett}
\eea
Also, $\Tset_{\add,t} \subseteq \Nset_t \cup \Delta_{e,\add,t}$ and
\bea
\Delta_{e,\add,t} = \Delta_{e,t} \cup (\Ahat_t \setminus \Nset_t) \subseteq \tDelta_{e,t-1} \cup \Rset_t \cup (\Ahat_t \setminus \Nset_t)
\eea
Thus, $|\Delta_{e,\add,t}| \le S_a + f$ and so
\bea
|\Tset_{\add,t}| \le S + |\Delta_{e,\add,t}| \le S + S_a + f
\eea
By Lemma \ref{errls1} and condition 2c of the Theorem, we have
\bea
&&\|(x_t - \xhat_{t,\add})\| \le 1.12 \eps + (1+ 1.261{\theta_{|\Tset_{\add,t}|,|\Delta_{\add,t}|}} ) \|(x_t)_{\Delta_{\add,t}}\|\nn\\
&&\le 1.12 \eps + 1.261\|(x_t)_{\Delta_{\add,t}}\|
\eea
Recall that, by Proposition \ref{prop2}, any element of $x_{\Delta_{\add,t}}$ will have magnitude smaller than $\alpha_{\add} +\frac{\zeta_M}{\sqrt{S_a}} 7.50\eps$. By (\ref{Tdett}), we have
\bea
\|x_{\Delta_{\add,t}}\|&\leq&\sqrt{|\Delta_{\add,t}| (\alpha_{\add} + \frac{\zeta_M}{\sqrt{S_a}}7.50 \eps)}\nn\\
&\leq&\sqrt{(\frac{(b+1)}{2} S_a + d_0 S_a)(\alpha_{\add} + \frac{\zeta_M}{\sqrt{S_a}}7.50 \eps)}
\eea
Let $h=\sqrt{(\frac{(b+1)}{2}  + d_0 )(\alpha_{\add} +\frac{\zeta_M}{\sqrt{S_a}}7.50 \eps)}$. Combining this with the bound on $|\Tset_{\add,t}|$ and $|\Delta_{\add,t}|$ we can bound the LS step error by a time-invariant quantity,
\bea
\|(x_t - \xhat_{t,\add})_{\Tset_{\add,t}}\| \le 1.12 \eps + 1.261 h\sqrt{S_a}
\eea
Using Assumption \ref{spread2}, we have,
\bea
\|(x_t - \xhat_{t,\add})_{\Tset_{\add,t}}\|_\infty \le 1.12 \frac{\zeta_L}{\sqrt{S_a}} \eps + 0.261 \zeta_L h
\label{infnorm}
\eea
Using the fact that $\alpha_{\text{del}}$ is equal to the RHS of the above equation and proposition fact 3, if $(x_t)_j = 0$, then $j \in \Rhat_t$. Thus,
\bea
\Nset_t^c \subseteq \Rhat_t
\eea
Next, using (\ref{eqdel}), (\ref{infnorm}), fact 2 of Proposition \ref{prop2} and the value of $\alpha_{\del}$, we can conclude the following: if $j\in \Lset_t, j$ will not get falsely deleted; the same is true if $j\in \Aset_{\tau}, \tau\leq t-d_0$. Thus, 
\bea
\Rhat_t \subseteq \Nset_t^c \cup \SDset_t \cup \Aset_t \cup \Aset_{t-1} \dots \cup \Aset_{t-d_0+1}
\eea

%
Recall that $\Nhat_t = \Nhat_{t-1} \cup \Ahat_t \setminus \Rhat_{t}$.
Thus
\bea
\tDelta_t &=& \Nset_t \setminus \Nhat_t = (\Nset_t \cap \Nhat_{t-1}^c \cap \Ahat_t^c) \cup (\Nset_t \cap \Rhat_t)  \nn \\
   & \subseteq &  (\Delta_t \cap \Ahat_t^c) \cup (\SDset_t \cup \Aset_t \cup \Aset_{t-1} \dots \Aset_{t-d_0+1})
\label{tDeltat1}
\eea
Since $\Aset_{t-d_0} \subset \Ahat_t$, using (\ref{Deltat1}), we get
\bea
\Delta_t \cap \Ahat_t^c  \subseteq \SDset_{t} \cup \Aset_t \cup \Aset_{t-1} \dots \cup \Aset_{t-d_0+1}
\eea
Thus, using (\ref{tDeltat1}),
\bea
\tDelta_t \subseteq \SDset_{t} \cup \Aset_t \cup \Aset_{t-1} \dots \cup \Aset_{t-d_0+1}
\eea
Thus,
\bea
|\tDelta_t| \le \frac{(b+1)}{2} S_a + d_0 S_a
\label{tdelta_size}
\eea

Now consider $\tDelta_{e,t}$.
\bea
\tDelta_{e,t} \se \Nhat_t \setminus \Nset_t \nn \\
\se (\Nhat_{t-1} \cap \Rhat_t^c \cap \Nset_t^c) \cup (\Ahat_t \cap \Rhat_t^c \cap \Nset_t^c) \nn
\eea
As $\Nset_t^c \subseteq \Rhat_t $, we have $\Rhat_t^c \subseteq \Nset_t $.
Thus,
\bea
\tDelta_{e,t} = \emptyset
\eea
Thus,
\bea
|\tDelta_{e,t}| = 0
\eea
Since $|\Nset_t| \le S$ and since $|\tT_t| \le |\Nset_t| + |\tDelta_{e,t}|$, thus
\bea
|\tT_t| \le S
\eea
By condition 2,
\bea
&&\theta_{|\tT_{t}|,|\tDelta_{t}|} \leq \theta_{S,\frac{b+1}{2}S_a+d_0S_a+S_a}\nn\\
&&\leq \delta_{S+3(\frac{b+1}{2}S_a+d_0S_a+S_a)}\leq 0.207
\eea
and $$\delta_{|\tT_t|}\leq \delta_{S}\leq \delta_{S+S_a+f}\leq 0.207$$
Using the same way as getting $\|(x_t - \xhat_{t,\add})\|$, we have
$$\|(x_t - \xhat_{t})\|\le 1.12 \eps + 1.261 \|x_{\tDelta_{t}}\|$$
Also, using Proposition \ref{prop2}, any element of $x_{\tDelta_{t}}$ will have magnitude smaller than $\alpha_{\del}+1.12 \frac{\zeta_L}{\sqrt{S_a}} \eps $. By (\ref{tdelta_size}), we have
$$\|x_{t,\tDelta_t}\|\leq \sqrt{(\frac{(b+1)}{2}S_a+d_0S_a)(\alpha_{\del}+1.12 \frac{\zeta_L}{\sqrt{S_a}} \eps )}$$ 
Thus, the final claim is proved.

\subsection{Proof of Remark \ref{modcs_noise_free_rip}: Necessary and Sufficient conditions}
\label{modcs_noise_free_rip_proof}

Necessity: Consider the noise-free case, i.e. $\eps=0$ and Algorithm \ref{modcsalgo}.  We claim that $\delta_{S+S_a,\text{left}}<1$ at all times $t>0$ is necessary to ensure exact recovery of all sparse signal sequences with support size at most $S$, and number of support additions and removals at most $S_a$. We prove this here. Assume exact recovery at $t-1$. Assume also that the support size at $t-1$ is $S$, there are $S_a$ new additions and $S_a$ new removals at time $t$. Thus support size at time $t$ is also $S$.

Suppose that $\delta_{S+S_a,\text{left}}<1$ does not hold. This means there is a set, $R$, of size $S+S_a$ for which $rank((A_t)_R) < S+S_a$. Pick a $z$ so that $z_R \in null((A_t)_{R})$  (i.e. $(A_t)_R z_R=0$) and $z_{R^c}=0$. Partition $R$ into three sets $R=D \cup D_1 \cup D_2$ s.t. all are disjoint;  $|D|=S-S_a$, $|D_1|=S_a = |D_2|$ and $\|z_{D_2}\|_1 \le \|z_{D_1}\|_1$. Create two sparse vectors $x^1$ and $x^2$ supported on $D \cup D_1$ and $D \cup D_2$ respectively as follows. Let $(x^1)_{D} = z_D/2$,  $(x^1)_{D_1} = z_{D_1}$, $(x^1)_{(D \cup D_1)^c} = 0$. Let $(x^2)_{D} = -z_D/2$,  $(x^2)_{D_2} = -z_{D_2}$, $(x^2)_{(D \cup D_2)^c} = 0$. Then both $x^1$ and $x^2$ have support size $S$.

Suppose that the signal at time $t$ is $x^1$, i.e. $x_t=x^1$ so that $y_t = A_t x^1$,  and suppose that the support (equal to support estimate) from $t-1$ is $T = D \cup \Delta_e$ where $\Delta_e$ is a subset of $(D \cup D_1 \cup D_2)^c$ of size $S_a$. Consider the solution of modified-CS with $\eps=0$. In this case, both $x^1$ and $x^2$ are feasible since $A_t (x^1-x^2) = (A_t)_D z_D/2 + (A_t)_{D_1} z_{D_1} - (A_t)_D (-z_D/2) - (A_t)_{D_2} (-z_{D_2}) = (A_t)_R z_R$. But, $\|(x^1)_{D^c}\| = \|(x^1)_{D_1}\|_1 = \|z_{D_1}\|_1 \ge \|z_{D_2}\|_1 = \|(x^2)_{D^c}\|_1$. Thus, clearly $x^1$ will not be the unique solution to modified-CS with $\eps=0$. This proves that $\delta_{S+S_a,\text{left}}<1$ is necessary.

Sufficiency: Assume exact recovery at $t-1$, i.e., $\Tset_t=\tT_{t-1}=\Nset_{t-1},$ $\Delta_{t}=\Nset_{t}\setminus \Tset_t = \Nset_{t} \setminus \Nset_{t-1}$, i.e., $|\Tset_t|\leq S$, $|\Delta_{t}| \leq S_a$, thus by Lemma \ref{modcsbnd_2} and $\delta_{S+3S_a}< 0.5$, we have $\|x_t-\xhat_t\|\leq 0$, i.e., $\xhat_t=x_t$.

\subsection{Generative model for Signal Model \ref{sigmodgen}:}
\label{sigmodgengen}
This model requires that when a new element $j$ gets added to the support, its magnitude keeps increasing at rate $r_{j,t}$ until it reaches large set, and that an element $i$ of the large set starts to decrease at rate $r_{i,t}$ until it reaches 0. The sign is selected as +1 or -1 with equal probability when the element gets added to the support, but remains the same after that. We can choose values for $a_{\min}, d_{\min}, r_{\min}(d_{\min}), S_a, m, b$ during simulation.

Mathematically, it can be described as follows. Let $(x_t)_j = (M_t)_j(s_t)_j$ where $(M_t)_j$ denotes the magnitude and $(s_t)_j$ denotes the sign of $(x_t)_j$ at time t.
$x_t$ is a $m\times 1$ vector; $S_0=[\mu_1S]$, here $\mu_1$ is a random number between 0.9 and 1.

For $1\leq t\leq b$, let $S_{a,t}=0$, $S_{r,t}=0$, $S_{d,t}=S_a$;
%
For any $t>b$, do the following.
\ben
\item Generate 
\ben
\item the new addition set, $\Aset_t$, of size $S_{a,t}=[\mu_2(\Sigma^{t-b}_{\tau=1}S_{d,\tau}-\Sigma^{t-1}_{\tau=1}S_{a,\tau})]$ (here $\mu_2$ is a random number between 0.9 and 1) uniformly at random from ${\Nset_{t-1}}^c$,
\item the new decreasing set, $\Bset_t$, of size $S_{d,t}=[\mu_3S_a]$ (here $\mu_3$ is a random number between 0.5 and 1) uniformly at random from $\Lset_{t-1}$, and%
\item the new deleted set, $\Rset_t$, of size $S_{r,t}=[\mu_4|\SDset_{t-1}|]$ (here $\mu_4$ is a random number between 0.1 and 0.3), as the smallest $S_{r,t}$ elements of $\SDset_{t-1}$.
\een


\item Update the coefficients' magnitudes as follows.
\bea
&&(M_t)_i = \nn\\
&&\left\{ \begin{array}{ll}
              (M_{t-1})_i + r_{i,t}, & \ i \in \Aset_{t-d_{\min}}\cup\Lset_{t-1}\setminus \Bset_t, r_{j,t}=\mu_5;  \\
              (M_{t-1})_i + r_{i,t}, & \ i \in \cup_{\tau=t-d_{\min}+1}^{t}\Aset_{\tau}, r_{i,t}=\mu_6 r_{\min}(d_{\min});\\
              (M_{t-1})_i - r_{i,t}, & \ i \in \SDset_{t-1}\setminus \Rset_t, r_{i,t}=\mu_7 \frac{\ell}{b};\\
              (M_{t-1})_i - r_{i,t}, & \ i \in \Bset_t, r_{i,t}=\mu_8 (M_{i,t-1}-\ell);\\
              0,     & \ i \in \Nset_t^c.
              \end{array}\nn
              \right.
\eea
where $\mu_6$, $\mu_7$ and $\mu_8$ are random numbers between 1 and 1.44; $\mu_5$ is a random number larger than $-((M_{t-1})_i-\ell)$.
\item Update the signs as follows.
\bea
(s_t)_i \se \left\{ \begin{array}{ll}
              (s_{t-1})_i, & \ i \in \Nset_t \setminus \Aset_t  \\
              iid(\pm 1), & \ i \in \Aset_t \\
              0,     & \ i \in \Nset_t^c
              \end{array}
              \right.
\eea
where $iid(\pm 1)$ refers to generating the sign as +1 or -1 with equal probability and doing this independently for each element $i$.

\item Set $(x_t)_i = (M_t)_i (s_t)_i$ for all $i$.
\item Update 
\bea
\Lset_t&=&\Aset_{t-d_{\min}}\cup\Lset_{t-1}\setminus\Bset_t,\nn \\
\SDset_{t}&=&\SDset_{t-1}\cup \Bset_t\setminus \Rset_{t}.\nn
\eea
\een

\bibliographystyle{IEEEbib}
\bibliography{tipnewpfmt_kfcsfullpap}

\section*{Biographies}

{\bf Jinchun Zhan} received a B.S. from University of Science and Technology of China in 2011 in Electronic Engineering and Information Science. He is currently a Ph.D. candidate in Electrical Engineering at Iowa State University. His research interests are sparse recovery, robust PCA and their applications in video and medical imaging.

{\bf Namrata Vaswani} received a B.Tech. from Indian Institute of Technology (IIT), Delhi, in 1999 and a Ph.D. from University of Maryland, College Park, in 2004, both in Electrical Engineering. During 2004-05, she was a research scientist at Georgia Tech. Since Fall 2005, she has been with the Iowa State University where she is currently an Associate Professor of Electrical and Computer Engineering. She has held the Harpole-Pentair Assistant Professorship at Iowa State during 2008-09. During 2009 to 2013, she served as an Associate Editor for IEEE Transactions on Signal Processing. She is the recipient of the 2014 Iowa State Early Career Engineering Faculty Research Award and the 2014 IEEE Signal Processing Society Best Paper Award for her Modified-CS paper in IEEE Transactions on Signal Processing (jointly with her former graduate student).

Vaswani's research interests lie at the intersection of signal and information processing and machine learning for high dimensional problems. She also works on applications in video and big-data analytics and in bio-imaging. In the last several years her work has focused on developing provably accurate online algorithms for various high-dimensional structured data recovery problems such as online sparse matrix recovery (recursive recovery of sparse vector sequences) or dynamic compressed sensing, online robust principal components' analysis (PCA) and online matrix completion; and on demonstrating their usefulness in dynamic magnetic resonance imaging (MRI) and video analytics.

\end{document}